\documentclass[12pt]{article}
\usepackage{supercent}
\usepackage{titlesec}
\usepackage{soul}
\usepackage{setspace}
\usepackage{footmisc}

\titlespacing\section{0pt}{12pt plus 4pt minus 2pt}{0pt plus 2pt minus 2pt}
\titlespacing\subsection{0pt}{12pt plus 4pt minus 2pt}{0pt plus 2pt minus 2pt}
\newtheoremstyle{exampstyle}
  {0pt plus 2.0pt minus 4.0pt} 
  {0pt plus 2.0pt minus 4.0pt} 
  {} 
  {} 
  {\bfseries} 
  {.} 
  {.5em} 
  {} 

\theoremstyle{exampstyle}

\usepackage{lineno}

\newcommand{\blind}{1}

\makeatletter
\newcommand*{\addFileDependency}[1]{
  \typeout{(#1)}
  \@addtofilelist{#1}
  \IfFileExists{#1}{}{\typeout{No file #1.}}
}
\makeatother

\newcommand*{\myexternaldocument}[1]{%
    \externaldocument{#1}%
    \addFileDependency{#1.tex}%
    \addFileDependency{#1.aux}%
}
\usepackage{xr,xr-hyper}
\myexternaldocument{\detokenize{R2_supplement}}
\RequirePackage[colorlinks,citecolor=blue,linkcolor=blue,urlcolor=blue,pagebackref]{hyperref}

\makeatletter
\long\def\@makecaption#1#2{
        \vskip 0.8ex
        \setbox\@tempboxa\hbox{\small {\bf #1:} #2}
        \parindent 1.5em  
        \dimen0=\hsize
        \advance\dimen0 by 0em
        \ifdim \wd\@tempboxa >\dimen0
                \hbox to \hsize{
                        \parindent 0em
                        \hfil 
                        \parbox{\dimen0}{\def\baselinestretch{0.96}\small
                                {\bf #1.} #2
                                } 
                        \hfil}
        \else \hbox to \hsize{\hfil \box\@tempboxa \hfil}
        \fi
        }
\makeatother

\newif\ifdoublespacing

\begin{document}



\def\spacingset#1{\renewcommand{\baselinestretch}%
{#1}\small\normalsize} \spacingset{1}


\if1\blind
{
  \title{Network Regression and \\ Supervised Centrality Estimation}
  \date{}
  \author{Junhui Cai
    \\
    {\normalsize University of Notre Dame} \vspace{.5em} \\ 
    Dan Yang \\
    {\normalsize The University of Hong Kong} \vspace{.5em} \\
    Ran Chen
    \\    
    {\normalsize Washington University in St. Louis} \vspace{.5em} \\ 
    Wu Zhu \\ 
    {\normalsize Tsinghua University} \vspace{.5em} \\
    Haipeng Shen \\
    {\normalsize The University of Hong Kong} \vspace{.5em} \\
    Linda Zhao \\
    {\normalsize University of Pennsylvania} \vspace{.5em}}
  \maketitle
} \fi

\if0\blind
{
  \bigskip
  \bigskip
  \bigskip
  \begin{center}
    {\large\bf Network Regression and Supervised Centrality Estimation}
\end{center}
  \medskip
} \fi

\begin{abstract}
The centrality in a network is often used to measure nodes' importance and model network effects on a certain outcome. Empirical studies widely adopt a two-stage procedure, which first estimates the centrality from the observed noisy network and then infers the network effect from the estimated centrality, even though it lacks theoretical understanding. 
{We propose a unified modeling framework to study the properties of centrality estimation and inference and the subsequent network regression analysis with noisy network observations.}
Furthermore, we propose a supervised centrality estimation methodology, which aims to simultaneously estimate both centrality and network effect. 
{We showcase the advantages of our method compared with the two-stage method both theoretically and numerically via extensive simulations and a case study in predicting currency risk premiums from the global trade network.}

\end{abstract}

\noindent%
{\it Keywords:}  
Hub centrality, Authority centrality, 
{Network effect,}
Global trade network, Currency risk premium 




\spacingset{1.8} 

\setlength{\parskip}{0pt}
\setlength{\abovedisplayskip}{0pt}
\setlength{\belowdisplayskip}{0pt}
\setlength{\abovedisplayshortskip}{0pt}
\setlength{\belowdisplayshortskip}{0pt}

\newdimen\origiwspc%
\newdimen\origiwstr%
\origiwspc=\fontdimen2\font
\origiwstr=\fontdimen3\font
\if1\blind
{
\pagebreak
}
\fi
\section{Introduction}
\label{sec:intro}

In many disciplines such as economics, finance, and sociology, there has been great interest in studying the network effect, that is, the effect of a network on certain outcomes of interest due to relationships among agents (e.g., individuals, firms, industries, and countries). One popular approach 
is to bridge the outcome and network via an intermediary or a sufficient statistics -- the centrality of the network. 

As a low-rank summary of a network, centrality is a common metric to measure agents' importance in the network, which in turn induces a wide range of agent behaviors that consequently shapes certain outcomes of {theirs}. A strong motivation for centrality is that many real-world networks exhibit a low-rank structure, i.e., the leading singular value dominates the rest in magnitude~\citep{allen2019ownership,zhu2020networks1,liu2020dynamical}. Centrality itself has rich implications for studying human capital investment \citep{jackson2017economic}, information sharing and advertising \citep{banerjee2019using, breza2019social}, 
firms' investment decision-making \citep{allen2019ownership}, the identification of banks that are too-connected-to-fail \citep{gofman2017efficiency}, 
and stock returns \citep{ahern2013network,  richmond2019trade1},
among many others.

To be specific, researchers often regress the outcome of interest on the network centrality to study the network effect.
This approach has been implemented in many fields including portfolio management, finance, and social networks. In portfolio management, \citet{hochberg2007whom, ahern2013network} and \citet{richmond2019trade1} demonstrated that, for a trade network of firms or countries, a strategy that shorts portfolios with high centralities and longs those with low centralities yields a significant excess return, and regressing risk metrics on the centrality of the financial institutions 
{helps to} understand the amplification of severe adversarial shocks to the central institutions in the network. \citet{liu2019industrial} examined the effect of centrality in the production network on the government's investment in strategic industries to illustrate the effectiveness of industrial policies. For social networks, \citet{ozsoylev2014investor} and \citet{rossi2018network} regressed the excess returns of investment managers on the centrality of their social networks to study trading behaviors; 
\citet{kornienko2018peer} and \citet{mojzisch2021interactive}
 studied the network effect on mental health by regressing the stress level on the network centrality.

\fontdimen3\font=0.15em
Network centrality, however, is not directly observable. In practice, researchers often follow a two-stage procedure: in Stage 1, they compute the centrality from a given network adjacency matrix using an algorithm; in Stage 2, the computed centrality is then used as an input in the regression analysis. Such a practice will be referred to as the \emph{two-stage} procedure throughout.
\fontdimen3\font=\origiwspc

The validity of the two-stage procedure, however, hinges upon one critical assumption that the centrality is computed from a \emph{noiseless}  observed adjacency matrix in Stage 1 so that it is accurate. In reality, a network is often observed with noise due to the cost of data collection \citep{lakhina2003sampling}.
There are numerous examples of such noise: the friendship network on Facebook or Twitter is far from a perfect measure of real-life social connections; using self-reported friendships to measure social ties suffers from  subjective biases \citep{banerjee2013diffusion}; using patent citations to measure the knowledge flow between companies neglects the communication among workers or executives \citep{zhu2020networks1}. Overlooking noise in networks has demonstrable consequences for network analysis \citep{borgatti2006robustness, frantz2009robustness, wang2012measurement, martin2019influence, candelaria2022identification}.


Given a \emph{noisy} observed network, one has two goals in understanding the network effect:
\begin{enumerate}[align=left,label={(G\arabic*)},itemsep=-.5em, topsep=0pt]
\item \label{g1} Estimate centrality accurately from the observed noisy network.
\item \label{g2} Estimate and conduct valid inference of the network effect through the centrality. 
\end{enumerate}

The two-stage procedure attempts to achieve these two goals in a sequential manner, yet it has the following drawbacks. First, Stage 1 only uses the information from the noisy network to estimate centrality without incorporating the auxiliary information from the regression on the centrality, which can result in inaccurate estimation of the centrality due to large observational errors in the network. Second, Stage 2 is contingent upon Stage 1 -- regressing the outcome on the inaccurately estimated centrality exacerbates an inaccurate estimation of the regression coefficients, thereby invalidating the follow-up statistical inference. 

To remedy the shortcomings of the two-stage procedure, 
we first propose a \emph{unified} framework that fuses two models to achieve the two goals: 
 one network generation model based on the centralities for \ref{g1} and 
 one network regression model for the dependency of the outcome on the centralities for \ref{g2}.
We then propose a novel \emph{supervised network centrality estimation} (SuperCENT) methodology 
that accomplishes both \ref{g1} and \ref{g2} \emph{simultaneously}, instead of sequentially.

SuperCENT exploits information from the two models -- the network regression model contains auxiliary information on the centrality in addition to the network, and thus provides \emph{supervision} to the centrality estimation. 
The supervision effect improves the centrality estimation, which in turn benefits the network regression. Therefore, the centrality estimation and the network regression complement and empower each other. Under the unified framework, we derive the theoretical convergence rates and asymptotic distributions of the centralities and regression coefficients estimators, for both the two-stage and SuperCENT methods, which can be used to construct confidence intervals.

\fontdimen3\font=0em
We summarize our contributions as follows.
First, to the best of our knowledge, despite the popular adoption of the two-stage procedure, we are the first to provide a unified framework to study properties of centrality estimation and inference, and the subsequent network regression analysis when the observed network is noisy. 

{
Second, we are the first to study the properties of the common practice of the two-stage procedure and 
demonstrate that it can be problematic when the network noise is large.
The accuracy of the two-stage centrality estimates in Stage 1 depends on the network noise.
When the network noise is large, the centrality estimates are inaccurate, which results in 
\emph{inaccurate} centrality coefficient estimates with \emph{invalid} ad-hoc inference in Stage 2.
}

\fontdimen3\font=\origiwspc

Thirdly, we show theoretically and empirically that the proposed SuperCENT dominates the two-stage procedure universally. 
    Specifically, for \ref{g1}, SuperCENT yields a more accurate centrality estimation, especially under large network noise;
    {for \ref{g2}, SuperCENT boosts the accuracy of the regression coefficient estimation and provides confidence intervals that are \emph{valid} and \emph{narrower} than the ad-hoc two-stage confidence intervals.}

\fontdimen3\font=0.1em
Lastly, we apply both SuperCENT and the two-stage procedure to predict the currency risk premium,
based on an economic theory that links a country's currency risk premium with its importance within the global trade network \citep{richmond2019trade1}. We show that a long-short {trading} strategy based on SuperCENT centrality estimates {yields} a return {\emph{double}} that of the two-stage procedure. Furthermore, SuperCENT can verify the economic theory via a rigorous statistical test while the two-stage  fails.
\fontdimen3\font=\origiwspc

Our paper contributes to several strands of literature, including network modeling, network regression with centralities, covariate-assisted network modeling, and network effect modeling. First, the proposed unified framework bridges the gap between research on noisy networks and network regression with centralities. Most of the existing network literature focuses on one of these two aspects.  On one hand, in studies involving noisy networks, many empirical works have estimated the true network without incorporating centrality measures (e.g., \cite{lakhina2003sampling}, \cite{handcock2010modeling}, \cite{banerjee2013diffusion}, \cite{le2018estimating}, \cite{rohe2019critical}, \cite{breza2020using}). On the other hand, numerous work, including those mentioned earlier, have focused on the network regression model with centralities while ignoring the estimation error of the centralities inherited from the noise of the network.

Our unified framework also relates to the line of research on networks with covariates supervision~\citep{zhang2016community, li2016supervised, fan2016projected, binkiewicz2017covariate, yan2019statistical, ma2020universal}. 
One major difference is that SuperCENT uses both the covariates and the response to supervise the estimation, instead of only the covariates. In addition, the existing literature has focused mostly on network formation or community detection. 

In econometrics, there has been significant effort to model the network effect on an outcome of interest through regression~\citep{de2017econometrics}. One popular approach follows the pioneering work of \citet{manski1993identification} and his ``reflection model'' 
\citep{lee2007identification, bramoulle2009identification, lee2010specification, hsieh2016social,zhu2017network}. 
This approach models the network effect through the observed adjacency matrix itself, not through the centralities like ours. 
There has also been a recent surge of literature in network recovery based on the reflection model \citep{de2019identifying, battaglini2021endogenous}.
This literature focuses on the issue of identifiability of the network effect, while our work attends to both estimation and inference of the network effect. 
Another popular approach assumes that the outcome depends on individual fixed effects, 
and casts the role of the network through the Laplacian matrix, such that connected nodes share similar individual fixed effects
\citep{li2019prediction, le2020linear}.
This approach emphasizes network homophily, while ours concentrates on the nodes' position or importance in the network using the centralities.

{
It is worth mentioning that there is an extensive body of literature discussing the concept of centrality in networks with negative-weight edges. The foundational work on these networks stems from social balance theory in sociology \citep{harary1953notion, cartwright1956structural}. \cite{bonacichCalculatingStatusNegative2004} extends the concept of centrality to such networks, providing interpretations grounded in balance theory. Several subsequent studies have built on this foundation \citep{chiang2014prediction, everettNetworksContainingNegative2014, singh2019eigenvector, ma2019clusters, gromov2025social}. Empirical research has also explored these networks, including studies on workplace dynamics \citep{labiancaExploringSocialLedger2006} and alliance-enemy networks in wartime \citep{konig2017networks}. Our method is applicable to these networks, and centrality can be interpreted within the framework established by the literature.
}

The rest of this article is organized as follows. Section \ref{sec:model} provides the background and formally introduces the unified framework. 
Descriptions of the two-stage procedure and SuperCENT are given in Section \ref{sec:method}. 
Theoretical properties are studied in Section \ref{sec:theory} and the simulation study is shown in Section \ref{sec:sim}. 
Section \ref{sec:case} presents the case study of the relationship between currency risk premiums and the global trade network centralities.
Section \ref{sec:conclusion} concludes with a summary and future work. 
The supplementary materials contain additional background information on network and centralities, detailed descriptions of the algorithms for undirected networks, more simulation results, additional information of the case study, some concrete mathematical expressions, and the proofs.
We developed an R package, \texttt{SuperCENT}, that implements the methods 
\if0\blind
(\url{https://cccfran.github.io/SuperCENT}).
\fi
\if1\blind
(\url{https://jh-cai.com/SuperCENT}).
\fi

\vspace{-1.5em}
\section{A unified framework}
\label{sec:model}

\vspace{-1em}
\subsection{Set-up and background of network}
\label{ssec:model-setup}

We observe a sample of $n$ observations $(\x_1, y_1), \, (\x_2, y_2),\ldots, (\x_n, y_n)$ where $y_i \in \mathbb{R}$ is the response and $\x_i\in\mathbb{R}^{p-1}$ is the vector of $p-1$ covariates for the $i$-th observation as in the multivariate regression setting.
Let $\by \in \mathbb{R}^n$ denote the column vector of outcome
and $\bX \in \mathbb{R}^{n\times p}$ denote the design matrix including the intercept, which is assumed to be fixed.

In {a} network, the nodes are agents 
and the edges represent relationships between the agents.
The edges can be directed or undirected depending on whether 
the relationships are reciprocal. This article focuses on directed networks; the Supplement provides the results for undirected ones.
A weighted directed network with $n$ nodes can be represented by an \emph{asymmetric} adjacency matrix $\bA \in \mathbb{R}^{n \times n}$ where $a_{ij}$'s represent the weighted edges.


Researchers have used multiple versions of network centrality.
We refer to Chapter 2 of \cite{jackson2010social}
for a comprehensive introduction to centrality. 
We focus on the \textit{hub} and \textit{authority centralities} \citep{kleinberg1999authoritative}, which extend the well-known {eigenvector centrality} associated with the undirected network to the directed network. 

For directed networks, there is a distinction between the giver and the recipient, such as the citee-citor in citation networks or web-page networks, the exporter-importer in trade networks, and the investor-investee in investment networks. 
The hub and authority centralities take into account the different roles of the giver and the recipient, and thus measure the importance of nodes from these two different perspectives.
The concept of ``hubs and authorities'' 
{originated}
from web searching. Intuitively, the {hub centrality} of a web page depends on the total level of authority centrality of the web pages it links to, while the {authority centrality} of a web page depends on the total level of hub centrality of the web pages it receives links from. 
Supplement \ref{app:centrality-toy-exmaples} provides an example to further illuminate this intuition.

Let $u_i$ denote the hub centrality and $v_i$ denote the authority centrality for node $i$, and let $\bu=(u_1,u_2,\ldots,u_n)^\top,~~\bv=(v_1,v_2,\ldots,v_n)^\top$. 
Their relationship hence satisfies
$
\bu =\A \bv,~
\bv=\A^\top \bu.
$
Given $\A$, to calculate the centralities, \cite{kleinberg1999authoritative} proposes iterating with proper normalization as follows until convergence, for $k = 1, 2, 3, \ldots,$ 
\be
\bu^{(k)} \leftarrow	\A \bv^{(k-1)}, ~~\bv^{(k)} \leftarrow \A^\top \bu^{(k)}.
\ee
This iterative algorithm is also well known as the power method to compute the singular value decomposition (SVD) of $\bA$ \citep{van1996matrix}. 
Therefore, the hub and authority centralities are the leading left and right singular vectors of $\bA$ respectively. It is worth mentioning that such definition of centrality and the algorithm essentially assume that the adjacency matrix $\bA$ is noiseless.

\ifdoublespacing \vspace{-.1in} \fi 
\vspace{-1em}
\subsection{A unified framework}
\label{sec:supercent-model}

\fontdimen3\font=0em
We propose the following unified modelling framework
that encapsulates \ref{g1}-\ref{g2},
{
\begin{subequations}
\label{eq:model-supercent}
  \begin{align}[left ={\empheqlbrace}]
	\bA &= \bA_0 + \bE = \bU\bD\bV^\top + \bE = d\bu\bv^\top + \sum_{l=2}^r d_l \bu_l \bv_l^\top + \bE, \label{eq:model-supercent1} \\
	\by &= \bX\bbeta_x + \bu\beta_u + \bv\beta_v + \bepsilon, \label{eq:model-supercent2}
  \end{align}
\end{subequations}
where $\bD$ is a diagonal matrix of dimension $r\times r$ with the singular values $d > d_2 \geq  \ldots \geq d_r \geq 0$ as the diagonal entries, and $\bU = (\bu, \bu_2, \ldots, \bu_r)$ and $\bV = (\bv, \bv_2, \ldots, \bv_r)$ are two matrices of size $n\times r$ with orthogonal columns of length $\sqrt{n}$.
}
The intuitions of the unified framework are as follows. The hub and authority centralities are calculated as the leading left and right singular vectors of the observed adjacency matrix. As such, it is natural to consider the generative model \eqref{eq:model-supercent1} for the observed adjacency matrix, where $\bA_0$ is the true adjacency matrix, the true centralities $\bu,\bv\in \mathbb{R}^n$ are the parameters of interest to be estimated, 
{$(\bu_2, \ldots, \bu_r)$ and $(\bv_2, \ldots, \bv_r)$ are the non-leading singular vectors orthogonal to $\bu, \bv$,}
and $\bE$ is the additive noise of mean zero. Then,~\eqref{eq:model-supercent2} naturally models the relationship between the centralities and the response variable.
Here, $\bbeta_x \in \mathbb{R}^p$ 
is the vector of the regression coefficients, $\betau, \betav \in \mathbb{R}$ are the coefficients of the hub and authority centralities, and the regression error $\bepsilon$ has mean zero. Note that in  \eqref{eq:model-supercent2} it is the true centralities, not the estimated ones, that have direct impacts on the response 
{and only $\bu$ and $\bv$ are included instead of the entire $\bU$ and $\bV$ because it is common practice to consider the network effect via only the centralities.}

Under the unified framework \eqref{eq:model-supercent} with observed data $\{\bA, \bX, \by\}$, our original two goals \ref{g1}-\ref{g2} become concrete: 
(i) estimate the true centralities $\bu,\bv$; (ii) estimate the regression coefficients $\bbeta_x,\beta_u,\beta_v$; and (iii) construct \textit{valid} confidence intervals (CIs) for the centralities and the regression coefficients.

\fontdimen3\font=0em
The low-rank mean plus noise model \eqref{eq:model-supercent1} has been commonly adopted for matrix estimation or denoising~\citep{shabalin2013reconstruction1, yang2016rate, cai2018rate1},
matrix completion~\citep{candes2010matrix},
and network community detection with slight modifications~\citep{rohe2011spectral, zhao2012consistency, lei2015consistency, le2016optimization, gao2021minimax}.
There is a strand of literature on latent variables network models
that can be rewritten as \eqref{eq:model-supercent1}
\citep{hoff2009multiplicative, soufiani2012graphlet, fosdick2015testing}. 
\fontdimen3\font=\origiwstr%


The unified framework unites our estimation goals and provides a theoretical framework to study the behaviors of the two-stage procedure
and motivates our new methodology.
Under Model \eqref{eq:model-supercent1} and some extra assumptions on the noise, \citet{shabalin2013reconstruction1} proves that if the noise-to-signal ratio is large,
the leading singular vector of $\bA$ and that of $\bA_0$ converge to orthogonal as $n$ goes to infinity.
This implies that the naive estimation of the centralities by implementing SVD on the observed network will fail in the presence of large noise, which invalidates the common practice of two-stage.
Furthermore, unifying the two models motivates our supervised network centrality estimation (SuperCENT) methodology, which we will describe formally in the next section. 
We name it the ``supervised'' centrality estimation because $(\bX, \by)$ in the regression \eqref{eq:model-supercent2} can be thought of as the supervisors that offer additional supervision to the centrality estimation. 
It is expected that if the centralities indeed have strong predictive power (that is, the centrality regression coefficients $\betau,\betav$ are large compared with the regression noise level), the estimation of the centralities will be better when considering both \eqref{eq:model-supercent1} and \eqref{eq:model-supercent2} instead of only \eqref{eq:model-supercent1}. 
With the improved estimation of the centralities, 
{SuperCENT can further improve the estimation and inference of the regression model.}
Note that $\bu,\bv$ are only identifiable up to a scalar. SVD assumes $\bu$ and $\bv$ have unit length. However, we assume $\|\bu\|_2=\|\bv\|_2=\sqrt{n}$, because the network can grow and consequently the centralities should roughly be on the same scale with the network. This prevents the centrality regression coefficients from exploding as the network grows. 

\vspace{-2em}
\section{Methodology}
\label{sec:method}

\fontdimen3\font=0em
\jcM{
Sections \ref{sec:ts} and \ref{sec:supercent-est} formally introduce
the two-stage procedure and SuperCENT, respectively. 
In Supplement \ref{app:algo-supercent}, we will derive SuperCENT algorithm and prove its algorithmic convergence, 
discuss the prediction procedure and tuning parameter selection, 
and provide an algorithm for undirected networks with eigenvector centrality.
}
\fontdimen3\font=\origiwstr%

\vspace{-1.5em}
\subsection{The two-stage procedure}
\label{sec:ts}

As mentioned in the introduction, given 
the unified framework \eqref{eq:model-supercent} and the observed data $\{\bA, \bX, \by\}$, a natural and ad-hoc procedure is the two-stage estimator, which can serve as a benchmark. 
In view of \eqref{eq:model-supercent1}, the first stage is to perform SVD on the observed adjacency matrix $\bA$ and take its leading left and right singular vectors and rescale them to have length $\sqrt{n}$, denoted as $\huts$ and $\hvts$, as the estimates for the centralities $\bu$ and $\bv$, respectively. The superscript \textit{ts} stands for {\textbf{t}wo-\textbf{s}tage}.
In view of \eqref{eq:model-supercent2}, given the estimates $\huts$ and $\hvts$, the second stage performs {the} ordinary least square (OLS) regression of $\by$ on $\bX$ and $\huts,\hvts$, treating $\huts,\hvts$ as fixed covariates.

Hence, the two-stage procedure solves the following two optimizations \emph{sequentially},
\begin{subequations}
\label{eq:two-stage}
 \begin{align}[left ={\empheqlbrace}]
	(\hdts,\huts,\hvts) &:= \argmin_{d,\|\bu\|_2=\|\bv\|_2=\sqrt{n}}\|\bA - d\bu\bv^\top\|_F^2, \label{eq:ah-def2} \\
	\hbbetats := ((\hbetaxts)^\top,\hbetauts,\hbetavts)^\top &:= \argmin_{\bbeta_x,\beta_u,\beta_v} \|\by - \bX\bbeta_x - \huts\beta_u - \hvts\beta_v\|_2^2. \label{eq:reg-obj}
 \end{align}
\end{subequations}
It follows that $\hbbetats = (\hW^\top \hW)^{-1}\hW^\top\y$, where $\hW = (\X, \huts, \hvts)$.



\begin{rmk}
\vspace{-1em}
\label{rmk:ts-inference}
(Two-stage ``ad-hoc'' CI)
Besides the estimation of the unknown parameters, 
valid inference is necessary to evaluate the network effect.
Empirical studies usually
construct CIs of the regression coefficients from the second-stage regression by assuming that $\huts$ and $\hvts$ are fixed and noiseless. This assumption simplifies the inferential statement, because it follows that $\cov(\hbbetats) = \sigma_y^2(\hW^\top \hW)^{-1}$, where $\hW = (\X, \huts, \hvts)$. 
However, the observed network $\bA$ is one realization from $\bA_0+\bE$ as in Model \eqref{eq:model-supercent1}, which makes its singular vectors $\huts, \hvts$ random. If one ignores this randomness, 
the inference becomes invalid.
We refer to such ``ad-hoc'' CI as the  ``two-stage-adhoc'' method. To account for the randomness of the estimated singular vectors $\huts, \hvts$ and obtain valid inference, 
{Section \ref{sec:theory} derives the asymptotic distribution of the two-stage estimators, which depends on the network noise $\bE$ as well as the singular values and singular vectors of $\bA_0$,}
and discusses the theoretical property of the naive two-stage-adhoc CI. Section \ref{sec:sim} shows that the two-stage-adhoc CI is either conservative or invalid, depending on the network noise level.
\end{rmk}

\vspace{-1.5em}
\subsection{SuperCENT methodology}
\label{sec:supercent-est}
In the two-stage procedure, the estimation of the regression model in Step 2 depends on the centrality estimation in Step 1. The more accurate the centrality estimates are, the better we are able to make inference in the regression model. 
On the other hand, the centralities are incorporated into the regression model as regressors, 
so $(\bX, \by)$ can supervise centrality estimation and thus boost the estimation accuracy.

\fontdimen3\font=\origiwspc
Motivated by the intuition above, we propose to optimize the following objective function $\mathcal L(\bu,\bv,\bbeta, d)$, where $\bbeta = (\bbeta_x^\top,\betau,\betav)^\top$, to obtain the SuperCENT estimates,
\be
\label{eq:supercent-obj}
  (\hu,\hv, \hbbeta,\hd) :=
   \argmin_{\substack{\|\bu\|_2=\|\bv\|_2=\sqrt{n}, ~\bbeta, d}}
   \frac{1}{n}\|\by - \bX\bbeta_x - \bu\beta_u - \bv\beta_v\|_2^2 +
   \frac{\lambda}{n^2} \|\bA - d\bu\bv^\top\|_F^2,
\ee
where $\hbbeta = (\hbetax^\top,\hbetau,\hbetav)^\top$ and $\|\cdot\|_F$ is the Frobenius norm of a matrix.
The above objective function combines the residual sum of squares in \eqref{eq:reg-obj} and the rank-one approximation error of the observed network in \eqref{eq:ah-def2}.
The connection between the two terms is the centralities. The trade-off between them can be tuned through {a} proper selection of the hyper-parameter $\lambda$.

To solve \eqref{eq:supercent-obj}, we use a block gradient descent algorithm by updating $(\hu,\hv, \hbbeta,\hd)$
iteratively until convergence. 
The initialization 
is based upon the two-stage estimation, $(\bu^{(0)}, \bv^{(0)}) = (\huts, \hvts)$.
{For any matrix $\bA$, let} $\bP_\bA$ 
{denote} the projection matrix that projects onto the column space of $\bA$, and $\|\bA\|_2$ 
{be} the matrix operator norm. 
We use 
{$(\bu^{(t)}, \bv^{(t)}, \bbeta^{(t)}, d^{(t)})$} to denote the estimation in the $t$-th iteration. 
The complete algorithm with a given tuning parameter $\lambda$ is shown in Algorithm \ref{algo:supercent}.

\begin{algorithm}
\SetAlgoLined
\jcM{
\KwResult{{$\hd$, $\hu$, $\hv$, and $\hbbeta$}.}
 \textbf{Input:} $\A\in\mathbb{R}^{n\times n}$,  $\X\in\mathbb{R}^{n\times p}$, $\y\in\mathbb{R}^{n}$,  tuning penalty parameter $\lambda$,  tolerance parameter $\rho>0$,  maximum number of iteration $T$ \;
 Initiate $t=0$\; 
 \quad \quad \quad $(\bu^{(0)}, \bv^{(0)}) = (\huts, \hvts)$\;
 \quad \quad \quad $\W^{(0)} = (\X, \bu^{(0)}, \bv^{(0)})$\;
 \quad \quad \quad $\bbeta^{(0)} = (\W^{(0){^\top}}\W^{(0)})^{-1}\W^{(0){^\top}}\by$\;
 \quad \quad \quad $d^{(0)} = {\bu^{(0)}}^\top \A \bv^{(0)} / (\|\bu^{(0)}\|_2^2 \|\bv^{(0)}\|_2^2)$\;
	\While{$t\leq 1$ or $(\max\left(\| \bP_{\bu^{(t-1)}} - \bP_{\bu^{(t)}}\|_2, \|\bP_{\bv^{(t-1)}} - \bP_{\bv^{(t)}} \|_2 \right) > \rho$ and $t<T)$ }{
 	\begin{enumerate}[topsep=0pt,itemsep=-1ex,partopsep=1ex,parsep=1ex]
 		\item $t \leftarrow t+1$; 
		\item $\bu^{(t)} = \left((\beta_u^{(t-1)})^2 + \frac{1}{n} \lambda (d^{(t-1)})^2 \|\bv^{(t-1)}\|_2^2 \right)^{-1}$ \label{alg-u}\\
\hspace{8em} $\times \left[\beta_u^{(t-1)}(\y - \X\bbeta_x^{(t-1)} - \bv^{(t-1)}\beta_v^{(t-1)})+ \frac{1}{n}\lambda d^{(t-1)}\A\bv^{(t-1)}\right]$;
		\item $\bv^{(t)} = \left((\beta_v^{(t-1)})^2 + \frac{1}{n} \lambda (d^{(t-1)})^2 \|\bu^{(t)}\|_2^2 \right)^{-1} $ \label{alg-v}\\
\hspace{8em} $\times \left[\beta_v^{(t-1)}(\y - \X\bbeta_x^{(t-1)} - \bu^{(t)} \beta_u^{(t-1)})+ \frac{1}{n}\lambda d^{(t-1)}\A^\top\bu^{(t)}\right]$;
        \item Normalize $\bu^{(t)}, \bv^{(t)}$ to have norm $\sqrt{n}$: \label{alg-norm}\\
        \hspace{1in} $\bu^{(t)} = \sqrt{n}\bu^{(t)} / \|\bu^{(t)}\|_2,  \bv^{(t)} = \sqrt{n} \bv^{(t)} / \|\bv^{(t)}\|_2$;
    \item
    $\W^{(t)} = (\X, \bu^{(t)}, \bv^{(t)})$; \label{alg-w}
 		\item
        $\bbeta^{(t)} = (\W^{(t){^\top}}\W^{(t)})^{-1}\W^{(t){^\top}}\by$;\label{alg-beta}
 		\item $d^{(t)} = {\bu^{(t)}}^\top \A \bv^{(t)} / (\|\bu^{(t)}\|_2^2 \|\bv^{(t)}\|_2^2) $; 	\label{alg-d}\end{enumerate}
 }
 $\hu = \bu^{(t)}, \hv = \bv^{(t)}$, $\hbbeta=\bbeta^{(t)}$, $\hd=d^{(t)}$.
}
 \caption{SuperCENT algorithm with a given tuning parameter. }
 \label{algo:supercent}
\end{algorithm}

\begin{rmk}[Identifiability of $\hbetau,\hbetav,\hbu,\hbv$]
\label{rmk:identifiability}
Note that although $\hu$ and $\hv$ with length $\sqrt{n}$ are only identifiable up to the sign, $\hu\hv^\top$, $\hu\hbetau$, $\hv\hbetav$ are uniquely identifiable. 
{
Without additional structure, the sign is irrelevant since with proper flipping of $(\betau,\betav,\bu,\bv)$, the objective function will have the same value and the flipped sequence will remain a valid sequence generated by the algorithm.  When additional information is available (e.g., positive entries), one can determine the sign as follows: 
identify the entry that has the largest magnitude in $(\hu^\top,\hv^\top)$,
}
make that entry positive, and adjust the signs of the remaining entries in $\hu, \hv$, as well as $\hbetau, \hbetav$ accordingly.
\end{rmk}
\vspace{-1.5em}


\jcM{
\begin{rmk}[Full-rankness of $(\X, \bu^{(t)}, \bv^{(t)})$]
\label{rmk:algo-full-rank-t}
Further note that Step \ref{alg-beta} requires $(\X, \bu^{(t)}, \bv^{(t)})$ to be full-rank
and this condition is satisfied with high probability under assumptions to be introduced in Section \ref{sec:theory}. Please refer to Remark \ref{rmk:full-rank-t} in Supplement {\ref{sec:thm-algo}} for more discussions.
\end{rmk}
\vspace{-1.5em}

\begin{rmk}[Algorithmic convergence of Algorithm \ref{algo:supercent}]
\label{rmk:supercent-convergence}
Note that Algorithm \ref{algo:supercent} will converge to stationary points whose objective function value is smaller than that of the initialization. Precisely, 
$\lim_{t\to \infty} \| \partial \mathcal L (\bu^{(t)}, \bv^{(t)} ,\bbeta^{(t)}, d^{(t)} ) \|_2 =0$ and $ \sup_{t\ge 1}\mathcal L(\bu^{(t)}, \bv^{(t)}, \bbeta^{(t)}, d^{(t)} ) \le \mathcal L(\bu^{(0)}, \bv^{(0)}, \bbeta^{(0)}, d^{(0)})$. Details of the proof are deferred to Supplement \ref{app:supercent-rankone}. The theoretical results of the output from this algorithm is given in Supplement \ref{sec:thm-algo},
which depends on the stationarity of the estimators.
\end{rmk}

The tuning parameter can be chosen via cross-validation, for which prediction procedure has to be introduced. The prediction is a non-trivial task, since it involves the network expansion and justification to use the same regression coefficients for centralities with more nodes introduced into the model. Given the length constraint, the detailed prediction procedure and why it works are given in Supplement \ref{app:prediction-algo}, and the cross-validation procedure is provided in Supplement \ref{sec:cv}.
}

\vspace{-1.5em}
\section{Theoretical properties}
\label{sec:theory}
We investigate the statistical properties of SuperCENT and compare it with the two-stage procedure in this section.
\jcM{The proofs are deferred to Supplement \ref{sec:proof}.}
We start with introducing notations and assumptions.

Denote SuperCENT estimators,
\jcM{i.e., the minimizer of the objective function \eqref{eq:supercent-obj} with a given tuning parameter $\lambda$,}
as $\hd$, $\hu$, $\hv$, and $\hbbeta = ((\hbetax)^\top,\hbetau,\hbetav)^\top$ and the two-stage counterparts as
$\hdts$, $\huts$, $\hvts$, and $\hbbetats$.
We further denote
$\bAp = \bUp \bDp \bVp^\top$ where $\bUp = (\bu_2, \ldots, \bu_r)$, $\bVp = (\bv_2, \ldots, \bv_r)$ and 
$\bDp = diag(d_2, \ldots, d_r)$,
$\bm\Omega = \left(
       \begin{array}{cc}
        \sigma_y^2\bI_n & \zero_{n\times n^2} \\
        \zero_{n^2\times n} & \sigma_a^2 \bI_{n^2} \\
       \end{array}
      \right)$,
$\pxuv$ as the projection matrix that projects onto the column space of $(\bX, \bu,\bv)$, 
and similarly for $\bP_{\bX}$, $\pu$ and $\pv$. 
Define $\tu = (\bI-\bP_{\bX})\bu$,
$\tv = (\bI-\bP_{\bX})\bv$, which are the centralities projected onto the orthogonal space of $\bX$,
and $
\bCuv =
 \left(\tu,~ \tv\right)^\top
 \left(\tu,~ \tv\right)
$. 
In the following, the theorems are for $\argmax_{\mathbf{h}\in \{\hu,-\hu\}} \mathrm{sign}(\mathbf{h}^\top\bu) $ and $\argmax_{\mathbf{g}\in \{\hv,-\hv\}} \mathrm{sign}(\mathbf{g}^\top\bv) $, and we continue to use $\hu,\hv$ to denote them. While these notions are a bit of an abuse of notation, it is reasonable since both the objective function and algorithm are sign-invariant (i.e., proper flipping of signs gives the same value or another valid iteration sequence). The same notation applies in the simulations as well.




\vspace{-.8em}
\begin{assumption}
\label{assump:normal}
The network noise $\bE$ and regression noise $\bepsilon$ have independent normal entries with mean 0 and variance $\sigma_a^2$ and $\sigma_y^2$ respectively, and they are independent.
\end{assumption}
\vspace{-1.5em}
\begin{assumption}
\label{assump:x}
The fixed design matrix in the regression $\bX \in \mathbb{R}^{n\times p}$ {satisfies} $n > p+2$, and the dimension $p$ is non-diverging. 
{We further assume $(\bX, {\bu}, {\bv})$ is full rank
and the condition number of $(\bX, {\bu}, {\bv})^\top (\bX, {\bu}, {\bv})$ is smaller or equal to $1/\tau^2$ for some positive constant $\tau$. 
}
\end{assumption}
\vspace{-1.5em}
\begin{assumption}
\label{assump:consistent}
The scaled network noise-to-signal ratio 
{$\netsnr: = \frac{\sigma_a^2}{(d-d_2)^2 n} \rightarrow 0$.}
\end{assumption}

In Assumption \ref{assump:normal}, the independence is assumed for simplicity. If the network noises $e_{ij}$'s or the regression noises $\epsilon_i$'s are dependent with known covariance, the theorems and the corollaries below still hold with slight modifications {by simply plugging their covariance matrices into appropriate places}; if they are dependent with unknown covariance, extra assumptions on the covariance structure need to be made and new methodologies and theories should be developed.
Assumption \ref{assump:x} simply states that the regression is in the conventional low-dimensional fixed-design regime.
Assumption \ref{assump:consistent} is required for the consistency of the two-stage and SuperCENT, which essentially requires the signal-to-noise ratio (SNR) of the network 
and the gap between the leading and second singular values of the network to be large enough.

\jcM{
\begin{rmk}[Verification of the full-rank assumption of $(\bX, {\bu}, {\bv})$]
\label{rmk:full-rank1}
Unlike classical regression settings where the rank of $\bX$ can be easily checked, determining the rank of $(\bX, \bu, \bv)$ is more complex because the true centralities $(\bu, \bv)$ are not directly observable. However, this full-rank assumption can be verified as follows.
Under certain high signal-to-noise ratio assumptions for the observed network, a simple SVD of the noisy adjacency matrix $\bA$ generates consistent estimates $(\hbu^{ts}, \hbv^{ts})$ of their population counterparts, and the error bound for $(\hbu^{ts}, \hbv^{ts})$ is easily obtainable.
One can bound the quantities $|\bu^\top \opx \bv|$, $\|\opx \bu\|_2$, and $\|\opx \bv\|_2$ by their sample counterparts and the error bound for $(\hbu^{ts}, \hbv^{ts})$. These bounds collectively lead to an upper bound on $\frac{\left|\bu^\top \opx \bv\right|}{\left\|\opx \bu\right\|_2 \left\|\opx \bv\right\|_2}$, thereby proving the full-rank assumption of $(\bX, \bu, \bv)$.
\end{rmk}

\begin{rmk}[Practical implication of the full-rank assumption of $(\bX, {\bu}, {\bv})$]
\label{rmk:full-rank2}
The widespread use of two-stage procedures suggests that researchers and practitioners find centralities useful for predicting responses, even after accounting for the effects of predictors $\bX$. This is because centralities typically provide additional information beyond what the covariates offer\footnote{\jcM{
When all nodes are equally important, the centralities will not provide additional information and thus should not be included in the regression. In such cases, $\bu$ or $\bv$ is constant and the full-rank condition of $(\bX, \bu, \bv)$ will be violated due to collinearity with the intercept. This scenario occurs when the row sums or column sums of $\bA_0$ are equal, a condition often modeled by a homogeneous Erd\H{o}s-R\'enyi model.
In practice, one can test whether a network is generated from a homogeneous Erd\H{o}s-R\'enyi model 
\citep{bubeck2016testing, bickel2016hypothesis, lei2016goodness, zhang2017hypothesis, banerjee2017optimal, gao2017testing, ouadah2020degree, hu2021using, brune2024goodness}.
}}.  This widespread use implies that they consider $(\bX, \hbu^{ts}, \hbv^{ts})$ to be full rank. Often, the noise in the network is overlooked, as noted in the literature in Section \ref{sec:intro}, leading to an implicit assumption that the estimated centralities are the true centralities, i.e., $\hbu^{ts} = \bu$ and $\hbv^{ts} = \bv$. Therefore, researchers and practitioners implementing two-stage procedures essentially assume the full-rankness of $(\bX, \bu, \bv)$.
\end{rmk}
}

\begin{theorem}
\label{thm:supercent-normality}
Under the unified framework
\eqref{eq:model-supercent} and Assumptions \ref{assump:normal}-\ref{assump:consistent}, 
\jcM{
suppose $
\jcM{\sqrt{\kappa}}\frac{\sigma_y}{\sqrt{\beta_{u}^{2}+\beta_{v}^{2}}} = o(1)$,
$\sigma_y = o\left(\sqrt{\frac{n}{\log n}}\right)$,
and 
$\left|\frac{\beta_{u}}{\beta_{v}}\right| \in[\underline{\alpha}, \bar{\alpha}]$
for positive constants $\bar{\alpha}>\underline{\alpha}>0$,
then the SuperCENT estimators,
defined as the minimizer of the objective function \eqref{eq:supercent-obj} with a given tuning parameter $\lambda$ satisfying
$\frac{1}{\lambda} \kappa \frac{n \sigma_y^2}{\sigma_a^2} \left(1+\frac{\sigma_y}{\sqrt{\beta_u^2 + \beta_v^2}}\right)^2 = o(1)$, have the following asymptotic distributions,}
\begin{enumerate}[nolistsep]
\item Centralities: 
\jcM{
\be
\hu - \bu = \etau + o(\etau)
\quad\mbox{and}\quad
\hv - \bv = \etav + o(\etav),
\ee
}
\item Network effect: 
\jcM{
\be
\hbbeta - \bbeta = \etabeta + o(\etabeta)
= \left(\eta_\betax^\top, \etabetau, \etabetav\right)^\top + o\left(\left(\eta_\betax^\top, \etabetau, \etabetav \right)^\top \right),
\ee
}
\end{enumerate}
{
\setstretch{1}
\jcM{
where $\left(
 \begin{array}{c}
  \etau \\
  \etav \\
  \etabeta
 \end{array}
\right) \sim 
N\Big(
\zero_{(2n+2+p)\times 1}, 
\bC\left(
              \begin{array}{cc}
                \sigma_y^2\bI_n & \zero_{n\times n^2} \\
                \zero_{n^2\times n} & \sigma_a^2 \bI_{n^2} \\
              \end{array}
              \right){\bC}^\top
\Big)$,
$\frac{\|o\left(\etau\right)\|}{\|\etau\|} \overset{P}{\longrightarrow} 0$,
$\frac{\|o\left(\etav\right)\|}{\|\etav\|} \overset{P}{\longrightarrow} 0$,
$\frac{\left\|o\left(\etabetax\right)\right\|}{\left\|\etabetax\right\|} \overset{P}{\longrightarrow} 0$,
$\frac{|o\left(\etabetau\right)|}{|\etabetau|} \overset{P}{\longrightarrow} 0$,
$\frac{|o\left(\etabetav\right)|}{|\etabetav|} \overset{P}{\longrightarrow} 0$,
and 
$\bC = \left(
  \begin{array}{cc}
    \bC_{11} & \bC_{12} \\
    \bC_{21} & \bC_{22} \\
    \bC_{31} & \bC_{32} \\
    \bC_{41} & \bC_{42} \\
    \bC_{51} & \bC_{52} 
  \end{array}
\right)$ whose specific forms are as follows.}

The matrices related to $\hu$ and $\hv$ are
{\small
\be
\begin{multlined}[t]
\left(
 \begin{array}{cc}
  \bC_{11} & \bC_{12} \\
  \bC_{21} & \bC_{22} \\
 \end{array}
\right)
=
\left[
\frac{\lambda d}{n}
\left(
  \begin{array}{cc}
    dn \bI & -\bAp \\
    (-\bAp)^\top & dn \bI \\
  \end{array}
\right)
+
\left(
  \begin{array}{cc}
    \beta_u^2 & \betau\betav \\
    \betau\betav & \beta_v^2 \\
  \end{array}
\right)
\otimes \opxuv
\right]^{-1} \\
\left(
  \begin{array}{cc}
    \betau\opxuv & \lambda d \bv^\top \otimes (\bI-\pu)/n \\
    \betav\opxuv & \lambda d \left(\bu^\top \otimes (\bI-\pv)/n\right)\bK \\
  \end{array}
\right), 
\label{du-dv-final}
\end{multlined}
\ee
}
the matrices related to $\hbetau$ and $\hbetav$ are
{\small
\be
\left(
 \begin{array}{cc}
  \bC_{41} & \bC_{42} \\
  \bC_{51} & \bC_{52} \\
 \end{array}
\right)
=
\bCuvi
\left(
 \begin{array}{c}
  \tu^\top \\
  \tv^\top \\
 \end{array}
\right)
\left(-\betau\bI_n~~ -\betav\bI_n ~~ \bI_n\right)
\left(
 \begin{array}{cc}
  \bC_{11} & \bC_{12} \\
  \bC_{21} & \bC_{22} \\
  \bI_n & \zero_{n\times n^2}
 \end{array}
\right),
\ee
}
and the matrices related to $\hbetax$ are
{
\small
\be
 \left(
 \begin{array}{cc}
  \bC_{31}, & \bC_{32} \\
 \end{array}
\right) =
(\bX^\top\bX)^{-1}\bX^\top
 \left(
 -\betau\bI_n ~~ -\betav\bI_n ~~ -\bu ~~ -\bv ~~ \bI_n
\right)
\left(
 \begin{array}{cc}
  \bC_{11} & \bC_{12} \\
  \bC_{21} & \bC_{22} \\
  \bC_{41} & \bC_{42} \\
  \bC_{51} & \bC_{52} \\
  \bI_n & \zero_{n\times n^2}
 \end{array}
\right).
\ee
}

} 
 \vspace{-1em}
\end{theorem}

\jcM{
\begin{rmk}[Asymptotic results for SuperCENT estimators from Algorithm \ref{algo:supercent}]
\label{rmk:two-vers-consistent}
Theorem \ref{thm:supercent-normality} presents the asymptotic  distribution of
the SuperCENT estimators as the global minimizer of the objective function \eqref{eq:supercent-obj}.
However, this global minimizer may or may not be achievable using Algorithm \ref{algo:supercent}.
The same asymptotic distribution holds for the estimators produced by Algorithm \ref{algo:supercent}, as long as these estimators are consistent. As a matter of fact, they are indeed consistent under slightly more {conditions}; see Supplement {Sections \ref{sec:thm-algo} and} \ref{app:consistency-algo} for the details.
\end{rmk}
}

{Similarly, we derive the asymptotic distribution of the two-stage estimator in Theorem \ref{thm:two-stage-normality}.}
Comparing the covariance matrices with those of the two-stage in Theorem \ref{thm:two-stage-normality}, 
all $\bSigma_\bu, \bSigma_\bv$ and $ \bSigma_\bbeta$  involve both 
$\sigma_a^2$ and $\sigma_y^2$ due to the simultaneous estimation,
while $\bsigmats_\bu$ and $\bsigmats_\bv$ of the two-stage only involve $\sigma_a^2$ and $\bsigmats_\bbeta$ involves both.
Specifically, $\bC_\bu$ and $\bC_\bv$ are functions of $(\sigma_a, \bD, \bU, \bV, \sigmay, \bX, \betau, \betav, \lambda)$,
while the two-stage counterparts $\bCts_\bu$ and $\bCts_\bv$ only involve  $(\sigma_a, \bD, \bU, \bV)$.
Therefore, the difference between SuperCENT and the two-stage estimators of $\bu$ and $\bv$ lies in the tuning parameter $\lambda$ as well as the SNRs of the network and regression.
Following Theorem \ref{thm:supercent-normality},
Proposition \ref{prop:supercent-rate1} provides the convergence rates of $\hu$ and $\hv$.
We focus on the rank-one scenario where $\bA_0 = d\bu\bv^\top$ in model \eqref{eq:model-supercent1} to provide clearer insights for understanding the difference between SuperCENT and the two-stage estimators.

\begin{proposition}(Convergence rates of $\hu$, $\hv$)
\label{prop:supercent-rate1}
\jcM{
Under the same assumptions and conditions as in Theorem \ref{thm:supercent-normality} and further assume $\bA_0$ to be rank-one,
}
the SuperCENT estimators satisfy the following, 
\be
\frac{1}{n}\E\|\hu-\bu\|_2^2
&=&
\left(
\frac{\sigma_a^2(n-1)}{d^2n^2} -
\frac{n-p-2}{n}\beta_u^2 \delta_{ts,sc}\right)(1+o(1)) \\
&=& \netsnr (1+o(1)) - \beta_u^2 \delta_{ts,sc}(1+o(1)),
\label{eq:supercent-pu} \\
\frac{1}{n}\E\|\hv-\bv\|_2^2
&=&
\left(\frac{\sigma_a^2(n-1)}{d^2n^2} -
\frac{n-p-2}{n}\beta_v^2 \delta_{ts,sc}\right)(1+o(1)) \\
&=& \netsnr (1+o(1)) - \beta_v^2 \delta_{ts,sc}(1+o(1)),
\label{eq:supercent-pv} 
\ee
where
\be 
\label{eq:delta-ts-sc}
\delta_{ts,sc} = 
(\lambda d^2 + \beta_u^2 + \beta_v^2)^{-2}
\left[
\frac{2\lambda d^2 + \beta_u^2 + \beta_v^2}{d^2 n}
\sigma_a^2
- \sigma_y^2
\right].
\ee 
\end{proposition}

{
\begin{rmk}(The role of $\delta_{ts,sc}$)
\label{rmk:role-of-delta}
Comparing Corollaries \ref{prop:two-stage-rate1} and \ref{prop:supercent-rate1}, the discrepancies between the two-stage and SuperCENT estimators of the centralities are all proportional to $\delta_{ts,sc}$ since
$\E\|\huts-\bu\|_2^2/n- \E\|\hu-\bu\|_2^2/n = \beta_u^2\delta_{ts,sc}$. 
It can be seen that, whenever $\delta_{ts,sc} > 0$, SuperCENT always outperforms the two-stage. 

The positiveness of $\delta_{ts,sc}$
requires $\frac{2\lambda d^2 + \beta_u^2 + \beta_v^2}{d^2 n}
\sigma_a^2 - \sigma_y^2 > 0$, which depends on the interplay of $(\sigma_a, d, \sigmay, \betau, \betav, n, \lambda)$.
Specifically, $\delta_{ts,sc}$ is positive, when
the signal of the regression $\betau,\betav$ is large, the regression noise $\sigmay$ is small, the signal of the network $d$ is small, or the network noise $\sigmaa$ is large. This exactly verifies our intuition: 
when the regression SNR
is high, we gain information from the regression to assist centrality estimation; and the advantage is more pronounced when 
the network SNR is weak.
Moreover, $\delta_{ts,sc}$ involves a tuning parameter $\lambda$, and is positive when $\lambda$ is large enough. This is especially true when $\lambda$ takes the optimal value $\lambda_0=n\sigma_y^2/\sigma_a^2$ given in the remark below.

\end{rmk}
}



\vspace{-1em}
\begin{rmk} (Optimal $\lambda$)
\label{rmk:opt-lambda}
Minimizing the convergence rates \eqref{eq:supercent-pu} or \eqref{eq:supercent-pv} {with respect to $\lambda$} leads to the optimal tuning parameter
$
\lambda_0 = \frac{n\sigma_y^2}{\sigma_a^2}.
$
With the optimal $\lambda_0$, SuperCENT achieves its best performance and obtains the most improvement over the two-stage. Plugging the optimal $\lambda_0$ into \eqref{eq:delta-ts-sc}, we obtain the discrepancy  $\delta_{ts,sc}=\frac{\frac{\netsnr^2}{\sigma_y^2}}{1 + \netsnr \left(\frac{\beta_u^2}{\sigma_y^2} + \frac{\beta_v^2}{\sigma_y^2}\right)}$, which is always positive. This implies that as long as the tuning parameter is properly selected, SuperCENT will always be superior over the two-stage.
\jcM{Note that $\lambda_0$ satisfies the condition for $\lambda$ because when plugging $\lambda_0$ into the condition, we still have $\kappa \left(1+\frac{\sigma_y}{\sqrt{\beta_u^2 + \beta_v^2}}\right)^2 =o(1)$ under Assumption \ref{assump:consistent}.}
\end{rmk}


\vspace{-1em}
\begin{rmk} ($\supercentplugin$ and $\supercentcv$) 
\label{rmk:supercent-plugin-lambda}
The benefit of the optimal value $\lambda_0$ is two-fold: 1) to benchmark the cross-validation (CV) procedure in Algorithm \ref{algo:supercent-cv}; 
2) to provide a candidate for the tuning parameter $\lambda$ by plugging in the two-stage estimates of $\sigma_y^2$ and $\sigma_a^2$, i.e., $\hat\lambda_0 = n(\hat\sigma_y^{ts})^2/(\hat\sigma_a^{ts})^2$, instead of the time-consuming cross-validation.
We refer to SuperCENT using $\hat\lambda_0$ as $\supercentplugin$. 
Furthermore, $\hat\lambda_0$ can be used as a guide to lay out the cross-validation grid points in Algorithm \ref{algo:supercent-cv}, to obtain $\hat\lambda_{cv}$ and $\supercentcv$.
\end{rmk}

\jcM{
\begin{rmk}(Estimation comparison)
\label{rmk:inconsistent}
We further compare the estimation of $\bu, \bv$ of two-stage and SuperCENT. 
Plugging in the optimal $\lambda_0$, the standardized MSEs $\E\|\hu-\bu\|_2^2/n$ in \eqref{eq:supercent-pu} and $\E\|\hv-\bv\|_2^2/n$ in \eqref{eq:supercent-pv} respectively become
\vspace{.5em}\be
\label{eq:supercent-rate-consistency}
\netsnr\frac{1 + \netsnr\frac{\beta_v^2}{\sigma_y^2}}{1 + \netsnr \left(\frac{\beta_u^2}{\sigma_y^2} + \frac{\beta_v^2}{\sigma_y^2}\right)}(1+o(1)) 
\quad \mbox{and} \quad 
\netsnr\frac{1 + \netsnr\frac{\beta_u^2}{\sigma_y^2}}{1 + \netsnr \left(\frac{\beta_u^2}{\sigma_y^2} + \frac{\beta_v^2}{\sigma_y^2}\right)}(1+o(1)).
\ee
Note that the standardized MSEs of two-stage are $\E\|\huts-\bu\|_2^2/n=\E\|\hvts-\bv\|_2^2/n=\netsnr(1+o(1))$ in \eqref{eq:two-stage-mse-uv-exact}. 
Given~\eqref{eq:supercent-rate-consistency}, the performance of SuperCENT boils down to how the regression SNRs regarding $\bu$ and $\bv$  compare with the network SNR, i.e., $\kappa\frac{\beta_u^2}{\sigma_y^2}$ and $\kappa\frac{\beta_v^2}{\sigma_y^2}$.
Intuitively, when $\kappa\frac{\beta_u^2}{\sigma_y^2}$ is larger than $\kappa\frac{\beta_v^2}{\sigma_y^2}$, $\hu$ converges faster and $\hbetau$ also performs better due to the supervision effect; and vice versa.
\vspace{-1em}
\end{rmk}
}

\begin{rmk}({Two-stage CI versus ``two-stage-ad-hoc'' CI of $\betau$})
\label{rmk:inference}
Based on Theorem \ref{thm:two-stage-normality}, we can construct a valid confidence interval for $\betau$.
Specifically, the asymptotic variance of $\hbetauts$ when $\bA_0$ is rank-one has the following two-terms: \eqref{eq:two-stage-betau-exact-term1}-\eqref{eq:two-stage-betau-exact} where the first term is the same as the variance in the classical regression results 
and the second term is due to the randomness nature of $\huts,\hvts$.
Compared with the ``two-stage-ad-hoc'' CI, i.e., the CI that obtained via software directly from the regression in Stage 2, this ``two-stage-ad-hoc'' CI uses \eqref{eq:two-stage-betau-exact-term1} alone and is thus invalid unless $\sigma_a = 0$. 


\end{rmk}

\vspace{-2.5em}
\section{Simulation}
\label{sec:sim}

In this section, we investigate the empirical performances, including the \emph{estimation} and \emph{inference properties} of the two-stage and SuperCENT estimators under various settings. 
Section  \ref{sec:sim-setup} describes the simulation setups and 
Section \ref{sec:sim-incons} shows the results. 
Additional simulations, including a phase-transition experiment, are deferred to Supplement \ref{app:more-sim}. 

\vspace{-1em}
\subsection{Simulation setup}
\label{sec:sim-setup}

We generate the network following model \eqref{eq:model-supercent1}.
We consider the case of $r = 10$ where the leading singular value $d = 1$ and the non-leading ones as $d_2 = \ldots = d_r = 2^{-1}$.
All entries of $\bU$ are first generated from i.i.d. $N(0,1)$ and $\bV = 0.5\bU+\bepsilon_{\bV}$  where $\bepsilon_{\bV}$ are generated from i.i.d. $N(0,1)$.
We then apply Gram–Schmidt to ensure orthogonality between columns of $\bU$ and $\bV$, and finally rescale each column to have length $\sqrt{n}$.
For the regression model \eqref{eq:model-supercent2}, 
the regression coefficients are $\bbeta_x = (1,3,5)^\top$, the design matrix $\X$ consists of a column of 1's and $p-1$ columns whose entries follow $N(0,1)$ independently.

\fontdimen3\font=0.15em
For the properties of the estimators and inference, only the network SNR $\netsnr$
and the regression SNR $(\frac{\beta_u}{\sigma_y}, \frac{\beta_v}{\sigma_y})$ matter. Hence,  
we fix $n=2^{8},~d=1$, {$d_2=\ldots=d_r=2^{-1}$,} and $\betav=1$ and vary $\sigma_a, \sigma_y$, and $\beta_u$. 
To study the effect of the regression SNR, we consider $\sigmay \in 2^{-4, -2, 0 }$ and $\betau \in 2^{0, 2, 4}$, while ensuring 
{$\frac{\beta_u^2}{\sigma_y^2} \geq \frac{{\beta_v^2}}{\sigma_y^2}$.}
As the network SNR is controlled by $\sigmaa$,
we vary $\sigmaa\in 2^{ 0, 2 }$.
{We study the effects of the non-leading singular values $d_2, \ldots, d_r$ in additional simulations in Supplement \ref{app:more-sim}.}
\fontdimen3\font=\origiwstr%

For estimation property, we compare the following procedures: 1. \textbf{Two-stage}; 2. \textbf{$\supercentoracle$}, which implements Algorithm \ref{algo:supercent} with oracle $\lambda_0 = n\sigma_y^2/\sigma_a^2$ using the true $\sigmay,\sigmaa$ and serves as the benchmark; 3. \textbf{$\supercentplugin$} is SuperCENT with estimated tuning parameter $\hat\lambda_0 = n(\hat\sigma_y^{ts})^2/(\hat\sigma_a^{ts})^2$, where $(\hat\sigma_y^{ts})^2 = \frac{1}{n-p-2}\|\hy^{ts}-\by\|_2^2$ and $(\hat\sigma_a^{ts})^2 = \frac{1}{n^2} \|\hAts - \bA_0\|_F^2$ are estimated from the two-stage procedure; and 4.  \textbf{$\supercentcv$} is SuperCENT with tuning parameter $\hat\lambda_{cv}$ chosen by cross-validation 
as in Algorithm \ref{algo:supercent-cv}.

For inference property, we consider the following procedures to construct the confidence intervals (CIs) for the regression coefficient: 1.  \textbf{$\ts$-{adhoc}}: 
	$\hbeta^{ts} \pm z_{1-\alpha/2} \hat\sigma^{OLS}(\hbeta^{ts})$, where $z_{1-\alpha/2}$ denote the $(1-\alpha/2)$-quantile of the standard normal distribution, $\hbeta^{ts}$ is the two-stage estimate of $\beta$ and $\hat\sigma^{OLS}(\hbeta^{ts})$ is the standard error from OLS,  assuming $\huts,\hvts$ are fixed predictors; 2. \textbf{$\ts$-oracle}: 
	$\hbeta^{ts}  \pm z_{1-\alpha/2} \sigma(\hbeta^{ts})$, where $\sigma(\hbeta^{ts})$ is the standard error of $\hbeta^{ts}$, whose mathematical expressions are given in \eqref{eq:two-stage-betau-exact-term1}-\eqref{eq:two-stage-betau-exact} or \eqref{eq:two-stage-betav-exact-term1}-\eqref{eq:two-stage-betav-exact} with the true parameters plugged in; 
	3. \textbf{$\ts$-plugin}: $\hbeta^{ts} \pm z_{1-\alpha/2} \hat\sigma(\hbeta^{ts})$, where $\hat\sigma(\hbeta^{ts})$ is the standard error of $\hbeta^{ts}$ by plugging all the two-stage estimators into \eqref{eq:two-stage-betau-exact-term1}-\eqref{eq:two-stage-betau-exact} or \eqref{eq:two-stage-betav-exact-term1}-\eqref{eq:two-stage-betav-exact};
	4. \textbf{$\supercentoracle$-oracle}: $\hbetaoracle \pm z_{1-\alpha/2}  \sigma(\hbetaoracle)$, where $\hbetaoracle$ is the estimate of $\beta$ by $\supercentoracle$ and $\sigma(\hbetaoracle)$ follows \eqref{eq:supercent-betauv-exact} 
    with the true parameters plugged in;
	and 5. \textbf{$\supercentcv$}: $\hbetacv \pm z_{1-\alpha/2} \hat\sigma(\hbetacv)$, where $\hbetacv$ is the estimate of $\beta$ by $\supercentcv$ and $\hat\sigma(\hbetacv)$ is obtained by plugging the $\supercentcv$ estimates into 
    \eqref{eq:supercent-betauv-exact}.
 {Note that for the  \textbf{$\ts$-plugin} and \textbf{$\supercentcv$}, $\hat\sigma(\hbeta^{ts})$ and $\hat\sigma(\hbetacv)$ involve estimation for $\bAp = \bUp \bDp \bVp^\top$.
 For the \textbf{$\ts$-plugin}, we plug in the estimate from SVD; for \textbf{$\supercentcv$}, we perform SVD on $\bA - \hd \hu \hv^\top$ and then plug in the estimates.}
	The experiments are repeated 500 times.

\vspace{-1em}
\subsection{Simulation results}
\label{sec:sim-incons}

From the perspective of estimation, we compare the following metrics: the estimation accuracy for the centralities, the network, and the regression coefficients. Let $\bP$ denote the projection matrix.
Figure \ref{fig:u-betau-2} shows the loss $l(\hu,{\bu}) = \| \bP_{\hu} - \bP_{\bu}\|^2_2$ and $\hbetau-\betau$, respectively, across different $\sigmaa$, $\sigmay$ and $\betau$ with $d = 1$ and $\betav = 1$. Losses such as 
$l(\hv,{\bv}) = \| \bP_{\hv} - \bP_{\bv}\|^2_2$,  $l(\hA,{\bA_0}) = \|\hA - \bA_0\|_F^2/\|\bA_0\|_F^2$, $l(\hbetau,{\betau}) = (\hbetau - \betau)^2/\beta_u^2$, $l(\hbetav,{\betav}) = (\hbetav - \betav)^2/\beta_v^2$, and $\hbetav-\betav$ are given in Supplement \ref{sec:more-sim-inconsistent}.

\begin{figure}[!t]
\centering
\begin{subfigure}{\textwidth}
  \centering
  \includegraphics[width=.9\linewidth]{{\simpathtwo u.pdf}}
  \caption{Boxplot of $\log_{10}(l(\hu,\bu))$.}
  \label{fig:u-2}
\end{subfigure}
\begin{subfigure}{\textwidth}
  \centering
  \includegraphics[width=.9\linewidth]{{\simpathtwo bias_betau.pdf}}
  \caption{Boxplot of $\hbetau - \beta_u$. The dashed lines correspond to 0.}
  \label{fig:betau-bias-2}
\end{subfigure}
\caption{Boxplot of $\log_{10}(l(\hu,\bu))$ for the four estimators across different $\sigmaa$, $\sigmay$ and $\betau$ with fixed $d = 1,~\betav = 1$. The super-imposed red symbols show the theoretical rates of the two-stage {and SuperCENT calculated from Theorems \ref{thm:two-stage-normality} and \ref{thm:supercent-normality} respectively} 
in Figure \ref{fig:u-2} and the median of $\hbetau - \beta_u$ in Figure \ref{fig:betau-bias-2} respectively.
}
\label{fig:u-betau-2}
\vspace{-1em}
\end{figure}



Figure \ref{fig:u-2} shows
the boxplot of $\log_{10}(l(\hu,\bu))$.
The rows correspond to $\log_2(\sigmaa)$
and the columns correspond to $\log_2(\betau)$.
For each panel, the x-axis is $\log_2(\sigmay)$ and the y-axis is $\log_{10}(l(\hu,\bu))$.
The super-imposed red symbols show the theoretical rates of $\huts$
{and $\hu$ calculated from Theorems \ref{thm:two-stage-normality} and \ref{thm:supercent-normality} respectively.}
As expected, three SuperCENT-based methods estimate $\bu$ much more accurately than the two-stage procedure. In particular, the supervision effect of $(\bX, \by)$ is more pronounced when the noise of the outcome regression ($\sigmay$) is small, or when the signal of the outcome regression ($\betau$) is large, or when the network noise-to-signal ($\frac{\sigmaa}{d}=\sigmaa$) is large. The numerical comparison validates Remarks \ref{rmk:role-of-delta} and \ref{rmk:inconsistent} on the theoretical comparison of the estimators. Comparing the three SuperCENT-based methods, 
$\supercentcv$ and $\supercentplugin$ are sometimes worse than the benchmark $\supercentoracle$, but still better than the two-stage.
$\supercentplugin$ is typically comparable to or worse than $\supercentcv$, because $\supercentplugin$ fails to locate the optimal $\lambda_0$ due to inaccurate estimate of $\sigma_a$ and $\sigma_y$ from the two-stage procedure.

Figure \ref{fig:betau-bias-2} shows $\hbetau - \betau$.
With large $\sigmaa$ or large $\betau$, the two-stage estimates are inaccurate,
while SuperCENT estimates remain accurate.
{
In particular, the two-stage estimates becomes more inaccurate as $\sigmaa$ or $\betau$ increases.
The three SuperCENT-based methods all outperform the two-stage and $\supercentcv$ and $\supercentplugin$ are comparable with the benchmark $\supercentoracle$.
}

\begin{figure}[!t]
\centering
\begin{subfigure}{\textwidth}
  \centering
  \includegraphics[width=.85\linewidth]{{\confintpathtwo cov_u.pdf}}
  \caption{Empirical coverage of $CI_{\betau}$. The dashed lines show the nominal confidence level 0.95.}
  \label{fig:confint-u-2}
\end{subfigure}
\begin{subfigure}{\textwidth}
  \centering
  \includegraphics[width=.85\linewidth]{{\confintpathtwo width_u.pdf}}
  \caption{$\log_{10}$ of the width of $CI_{\beta_u}$.}
  \label{fig:width-u-2}
\end{subfigure}
\caption{Empirical coverage and $\log_{10}$ of the width of $CI_{\betau}$ across different $\sigmaa$, $\sigmay$ and $\betau$ with $d = 1$ and $\betav = 1$.
$\supercent$ variants are labelled as circles ($\circ\;\bullet$) and the two-stage variants are labelled as triangles ($\vartriangle\mathlarger{\mathlarger{\mathlarger{\blacktriangledown}}}\;\blacktriangle$). The hollow ones are for oracles and the solid ones are for non-oracles.
}
\label{fig:CI-2}
\vspace{-1em}
\end{figure}

From the perspective of inference property, Figure \ref{fig:CI-2} shows the empirical coverage probability (CP) and the average width of the 95\% confidence interval for $\betau$ respectively. The CP and width for the centralities, the network, and $\betav$ are given in the Supplement.

Figure \ref{fig:confint-u-2} shows how the inaccurate estimation of $\betau$ by the two-stage further affects its confidence interval. 
Regarding empirical coverage, when $\betau$ is small (leftmost column), all methods are above the nominal level. As $\betau$ increases and $\sigmaa$ remains small (top right two panels), most methods (except for two-stage-oracle) remain valid, but for different reasons: the two SuperCENT-based methods remain valid due to the accurate estimation of both $\betau$ and the standard error, whereas two-stage and two-stage-{ad-hoc} remain valid mainly because they over-estimate ${\sigma_y^2}$, 
and this conservativeness masks the issue of the inaccurate estimation. Two-stage-oracle uses the true ${\sigma_y^2}$ and the issue of the inaccurate estimate cannot be hidden,
hence the corresponding intervals undercover.
When $\betau$ increases and $\sigmaa$ gets large too (bottom right panel), the over-estimation of ${\sigma_y^2}$ can no longer hide the issue of inaccurate estimation, causing all two-stage-related methods to become invalid. On the other hand, the empirical coverage of SuperCENT remains closer to the nominal level. 

As for the width of  $CI_\betau$,
Figure \ref{fig:width-u-2} shows that the confidence intervals by the SuperCENT-based methods have better coverage and are narrower than those by the two-stage methods. 
The improvement in width becomes more pronounced with larger $\betau$ and $\sigmaa$.

\vspace{-1em}
\section{Global trade network and currency risk premium}
\label{sec:case}


In this case study,
we demonstrate that SuperCENT can provide a more accurate estimation of the centralities using the global trade network.
This has a profound and lucrative implication on portfolio management because the centrality is closely related to currency risk premium, i.e., the excess return from holding foreign currency compared to the US dollar.
We further show the advantage of SuperCENT over the two-stage in the inference of regression coefficients, 
and thus strengthens a related economic theory.

In international finance literature, economists have studied extensively the currency risk premium
and remain puzzled by its driving forces.
One recent theory, developed by \cite{richmond2019trade1} using a general equilibrium, shows that countries' positions in the trade network can explain the difference in currency premiums and countries that are central in the trade network exhibit lower currency risk premiums.
This theory has two implications: 
{(i)}
the regression coefficients for the centralities should be negative; 
and {(ii)}
international investors can leverage and profit from a long-short strategy for foreign exchange by taking a long position in currencies of countries with low centralities and a short position in currencies of countries with high centralities.
Therefore, if the centralities can be estimated accurately, one can yield a significant investment return based on the strategy.

Motivated by \cite{richmond2019trade1}, we investigate how the global trade network drives the currency risk premium by regressing the currency risk premium on the centrality of the international trade network. 
To be specific, we consider a triplet of $\{\bA, \bX, \by\}$, where $\bA$ is the country-level trade network, $\by$ is the currency risk premium, and $\bX$ is 
the share of the world's GDP.
Since all these quantities are not directly available, we compute them following \citet{richmond2019trade1}.
It is worth mentioning that the trade linkage in $\bA$ is defined as the trade volume normalized by the pair-wise total GDP, which represents the relative trade (export/import) intensity between two countries.
We use a five-year moving average:
when considering year $t$, the average is taken from year $t-4$ to year $t$.
More details are provided in Supplement \ref{app:case-data}.
We focus on the period between 1999 and 2013 and include the 24 countries/regions whose exchange rates are available during this period.\footnote{The euro was first adopted in 1999. The bilateral trade data is available untill 2013.}\footnote{The list of country abbreviations is provided in Supplement \ref{app:case}.} 
In Figure \ref{fig:trade-hub}, the dotted line 
shows the time series plot of the rank of the five-year moving average of risk premium from 2003 to 2012 for the 24 countries/regions.\footnote{We leave the last available year 2013 for the validation purposes.}
In each year, we rank the 24 countries/regions' risk premiums from the largest to the smallest as the 1st to 24th.
We show a circular plot to visualize the average trade volume (2003-2012) in 
Figure \ref{fig:trade-circular}.
\paragraph{Centrality estimation.}
Since neither the two-stage nor SuperCENT is applicable for panel data, we will repeat the analysis for each year from 2003 to 2012. 
Besides the network and the response variable, we also include the GDP share as a predictor, which is defined as the percentage of country/region GDP among 
the total GDP of all available countries in the sample for that year.
In summary, the unified framework is, for each $t$,
\be
{a_{ijt}}
&=& d \cdot\text{Hub}_{it} \times \text{Authority}_{jt} + e_{ijt},\nonumber\\
{y_{it}}
&=& \alpha + {\beta_{ut}} \cdot \text{Hub}_{it} + {\beta_{vt}} \cdot \text{Authority}_{it} + {\beta_{xt}} \cdot \text{GDP share}_{it} + \epsilon_{it}.\nonumber
\ee

{In Sections \ref{sec:theory} and \ref{sec:sim}, we have demonstrated that the two-stage is problematic under large network noise. 
In this case study, the observational error of the network comes from two sources: GDPs and the trade volumes, because 
each entry of the observed network $a_{ijt}$ is defined as the trade volume normalized by their GDPs.
The accounting of GDP has been a challenge in macroeconomics \citep{landefeld2008taking}. 
For the trade volume, measurement errors are mostly due to (i) underground or illegal import and export; (ii) excluding service trade; (iii) trade cost like transportation or taxes \citep{lipsey20091}.
Consequently, the observed trade network can be very noisy and the two-stage will perform badly.
}

\plotfig[.94]{\trade}{plot/hub_gap5}{
Time series of ranking of risk premium in descending order and ranking of hub centrality estimated by two-stage and SuperCENT in ascending order from 2003 to 2012.
The vertical dashed line indicates 2008, the year of the financial crisis. 
}{fig:trade-hub}

On the other hand, SuperCENT can significantly improve over the two-stage when the network noise is large.
In what follows, we focus on $\supercentcv$ using $10$-fold cross-validation.
We will refer to $\supercentcv$ as SuperCENT for simplicity and use the superscript $sc$ for all the $\supercentcv$-related estimates.
We determine the signs of the centrality estimates by the empirical rule described at the end of Section \ref{sec:supercent-est}.
Figure \ref{fig:trade-hub} shows the time series plots of the ranking of the hub centrality estimated by two-stage and SuperCENT for the 24 countries/regions, together with the ranking of the currency risk premium. 
Figure \ref{fig:trade-authority} is for the authority centrality.
{We rank the centrality in ascending order and the risk premium in descending order. 
Based on the negative relationship between centralities and risk premium established in \cite{richmond2019trade1},
the closer the trends of rankings between centralities and risk premium are, the better the centralities capture the time variation in the risk premium.}
The centrality estimated by the two-stage procedure is relatively more stable over time compared to SuperCENT. This is because SuperCENT incorporates information of both the GDP share and the currency risk premium, which is more volatile than the trade network itself. Asian trade hubs such as Hong Kong (HKG) and Singapore (SGP) are the most central; while countries like South Africa (ZAF) and New Zealand (NZL) are peripheral. Comparing the ranking of risk premium,
the time variation is not reflected in the centrality estimated by the two-stage procedure, while it is well captured by SuperCENT. For the 2008 financial crisis, the SuperCENT centralities fluctuate together with the risk premium while the two-stage centralities mostly
{remain unchanged}.

\plotfig[.9]{\trade}{plot/excess_return_next_gap5_top3}{Time series of the next-year return 
{from 2004 to 2013}
based on a strategy
{that takes a long position on the currencies with the lowest 3 centralities and a short position on the currencies with the highest 3 centralities {estimated from 2003 to 2012 respectively}.}
}{fig:excess-return}

To emphasize the importance of accurate centrality estimation for portfolio management, we examine whether a long-short strategy based on SuperCENT's estimated centrality can significantly boost investment performance based on two-stage. 
For either two-stage or SuperCENT, we take a long position on the currencies with the lowest 3 centralities (bottom 10\%) and a short position on the currencies with the highest 3 centralities (top 10\%). We obtain a return based on the estimated centrality of the period between year $t-4$ and $t$. 
{Similarly, we include a naive long-short strategy based on the return of year $t-1$ as a baseline.}
Figure \ref{fig:excess-return} shows the year $t+1$ return based on this strategy. 
{The centrality-based portfolios both outperform the naive strategy.}
The return based on the centrality estimated by SuperCENT is much higher than that of the two-stage procedure. 
Table \ref{tab:tbl:excess-return-5} shows the 10-year average {annualized return and Sharpe ratio \citep{sharpe1994sharpe}} based on this strategy with the top and bottom 3, 4, and 5 currencies, respectively. The 10-year average return based on SuperCENT centralities {doubles} that of the two-stage procedure. SuperCENT achieves Sharpe ratios ranging from 0.27 to 0.39, compared to the Sharpe ratios of 0.28 for the Dow Jones, 0.42 for the S\&P 500, and 0.39 for the NASDAQ over the sample period (2004-2013).


{\small
\begin{table}
\small
\caption{\label{tab:tbl:excess-return-5}The 10-year average annualized return.}
\centering
\begin{tabular}[t]{lllllll}
\toprule
\multicolumn{1}{c}{} & \multicolumn{2}{c}{Top/Bottom 3} & \multicolumn{2}{c}{Top/Bottom 4} & \multicolumn{2}{c}{Top/Bottom 5} \\
\cmidrule(l{3pt}r{3pt}){2-3} \cmidrule(l{3pt}r{3pt}){4-5} \cmidrule(l{3pt}r{3pt}){6-7}
 & Hub & Authority & Hub & Authority & Hub & Authority\\
\midrule
Naive & -0.13\% & -0.13\% & -0.86\% & -0.86\% & -1.86\% & -1.86\%\\
SuperCENT CV & 3.49\% & 0.03\% & 2.17\% & 0.65\% & 3.12\% & 0.28\%\\
Two-stage & 0.53\% & -1.56\% & 1.28\% & -1.16\% & 1.50\% & -0.65\%\\
\hline
SuperCENT Sharpe ratio & 0.38 & -0.02 & 0.27 & 0.07 & 0.39 & 0.02\\
Relative difference of return: \\ SuperCENT vs Two-stage & 554.5\% & 102.0\% & 70.3\% & 156.5\% & 108.1\% & 142.6\%\\
\bottomrule
\end{tabular}
\end{table}

}

\vspace{-1.5em}
\paragraph{Inference of regression.}
We further demonstrate the superiority of SuperCENT in inference.
{Again since our method is not directly applicable to longitudinal data, 
we take the 10-year average of trade volume and GDP to construct a 10-year trade network
and GDP share. Similarly, we take the 10-year average of risk premium as the response.}

To better understand the behavior of the two-stage and SuperCENT estimators and 
how much improvement SuperCENT can potentially achieve,
we compare the noise-to-signal ratio $\netsnr$ of the trade network and the SNR of the regression.
Since both quantities are unknown, we estimate using results from SuperCENT.
For the noise-to-signal ratio, $\hat\netsnr^{\lambdacv} = 0.36 \approx 2^{-1.5}$, which is larger than
$\netsnr=2^{-8}$ in the simulation where the accuracy of the two-stage estimators is already low.
For the SNR of the regression: 
$(\hbetau^\lambdacv/\widehat{\sigma}_y^{\lambdacv})^2 = 1.8 \times 10^{7} \approx 2^{24}$ and 
$(\hbetav^\lambdacv/\widehat{\sigma}_y^{\lambdacv})^2  =  6.1\times 10^{5}  \approx 2^{19}$.
Compared with the simulation settings where $\netsnr = 2^{-4}$, $\beta_u^2/\sigma_y^2 \leq 2^{16} $ and $\beta_v^2/\sigma_y^2 \leq 2^{8}$,
We expect SuperCENT to significantly outperform the two-stage method in both the estimation and inference of $\betau$, due to the large value of $(\hbetau^\lambdacv/\widehat{\sigma}_y^{\lambdacv})^2$ and the fact that $|\hbetau^\lambdacv| \gg |\hbetav^\lambdacv|$ under a relatively large $\hat\netsnr^{\lambdacv}$, while the improvement of $\betav$ is less pronounced.



\begin{table} \centering 
  \caption{The summary table of the regression comparing three methods in terms of coefficient estimation, standard error (in parenthesis) and the significant level (by asterisks).   } 
  \label{tab:rx-summary} 
\begin{tabular}{@{\extracolsep{5pt}}lllllll} 
\hline 
& \multicolumn{2}{c}{Two-stage-adhoc} & \multicolumn{2}{c}{Two-stage} & \multicolumn{2}{c}{$\supercentcv$} \\ 
\hline 
 GDP share & $-$0.0159$^{*}$ & (0.0083) & $-$0.0159$^{*}$  & (0.0083) & $-$0.0157$^{***}$ & (0.0039) \\ 
 Hub & $-$0.0011 & (0.0006) & $-$0.0011$^{*}$ & (0.0007) & $-$0.0020$^{***}$  & (0.0001) \\ 
 Authority & $-$0.0005  & (0.0006) & $-$0.0005 & (0.0006) & $-$0.0004$^{*}$   & (0.0003)    \\ \hline
\textit{Note:}  & \multicolumn{6}{r}{$^{*}$p$<$0.1; $^{**}$p$<$0.05; $^{***}$p$<$0.01} 
\end{tabular} 
\vspace{-1em}
\end{table} 

Table \ref{tab:rx-summary} shows the coefficient estimation, the standard error, and the significant level for the two-stage-adhoc, two-stage, and $\supercent$, respectively.
{
The standard errors of two-stage and $\supercent$ are based on the trade network being rank-one as we tested using the rank inference by \cite{han2023universal} in Supplement \ref{app:net-exmaples}.
}
For the hub centrality $\betau$, 
(i) the estimate from the two-stage methods is $-0.0011$, while the estimate from $\supercent$ is $-0.0020$, 
{which is consistent with the inaccuracy we observed in the simulation;}
{(ii) the standard errors from the two-stage methods are close to $0.0007$, much larger than $0.0001$ from $\supercent$, which reinforces the problem of overestimation of $\sigma_y^2$ in two-stage; }
(iii) 
{the above two facts combined}
make the confidence intervals by two-stage-adhoc and two-stage unnecessarily wide, yet still invalid: 
{consequently the hub centrality $\betau$ is barely significant at level $0.1$ using two-stage and is insignificant using two-stage-adhoc;}
(iv) the two facts in (i) and (ii) also lead to a valid but narrower confidence interval for $\supercent$, 
making the hub centrality a significant factor 
{at}
level $0.01$ for the currency risk premium;
and (v) 
{conclusions drawn from the}
two-stage-adhoc and two-stage methods contradict the theory in \cite{richmond2019trade1}, while $\supercent$ supports the theory. 
Other regression coefficients' significance can be also explained by Remark \ref{rmk:inference}; the details are given in Supplement \ref{app:case-aut}.

%

\section{Conclusion and discussion}
\label{sec:conclusion}

Motivated by the rising use of centrality in empirical literature, we examined centrality estimation and inference on a noisy network \ref{g1} as well as network effect through the centralities in the subsequent network regression \ref{g2}. We proposed a unified framework that incorporates the network generation model and the network regression model to achieve both goals. 
{
Under the unified framework, we showed that the properties of the commonly used two-stage procedure and that it could yield inaccurate centrality estimates and regression coefficient estimates, as well as invalid inference when the noise-to-signal ratio of the network is large.
}
We proposed SuperCENT which incorporates the two models and simultaneously estimates the centralities and the effects of the centralities on the outcome. We further derived the convergence rate and the distribution of the SuperCENT {estimator} and provided valid confidence intervals for all the parameters of interest. 
{
We showed that SuperCENT dominates the two-stage universally and improves over the two-stage in terms of centrality estimation, regression coefficient estimations, and inference.
}
The theoretical results are corroborated with extensive simulations and a real case study in predicting currency risk premiums from the global trade network.

The {unified framework and SuperCENT methodology} 
can be extended in multiple directions. One can consider a generalized linear model for the outcome model and 
\jcM{a generalized network model for networks with noncontinuous edges via link functions to generalize SuperCENT. } 
In the case when only a subset of covariates and outcomes are observed, semi-supervised SuperCENT can be developed. In the case when the network is partially observed, we can perform matrix completion with supervision. SuperCENT can also be extended to a longitudinal model with additional assumptions by using techniques from tensor decomposition as well as functional data analysis to obtain centralities that are smooth over time. For ultra-high-dimensional problems, sparsity can be imposed on centralities due to the existence of abundant peripheral nodes.

\if1\blind
\section*{Acknowledgement}
Shen's research is supported in part by Hong Kong CRF C7162-20GF, the Ministry of Science and Technology Major Project of China 2017YFC1310903, University of Hong Kong (HKU) Stanley Ho Alumni Challenge Fund, and HKU BRC Grant.
Yang's research is supported in part by NSF grant IIS-1741390, Hong Kong GRF 17301620 and CRF C7162-20GF.
Zhao's research is supported in part by Wharton Global Initiative Fund.
Zhu's research is supported in part by Tsinghua University Initiative Scientific Research Program and Tsinghua University School of Economics and Management Research Grant.
\fi

{\linespread{1}\selectfont{}
\small

\setlength{\bibsep}{.6em plus 0.3ex}

\fontdimen3\font=0em
\fontdimen2\font=0.7ex

\bibliographystyle{apalike}
\bibliography{biblio.bib}

\begin{thebibliography}{}

\bibitem[Ahern, 2013]{ahern2013network}
Ahern, K.~R. (2013).
\newblock Network centrality and cross section of stock returns.
\newblock {\em SSRN 2197370}.

\bibitem[Allen et~al., 2019]{allen2019ownership}
Allen, F., Cai, J., Gu, X., Qian, J., Zhao, L., and Zhu, W. (2019).
\newblock Ownership network and firm growth: What do five million companies
  tell about chinese economy.
\newblock {\em SSRN 3465126}.

\bibitem[Banerjee et~al., 2013]{banerjee2013diffusion}
Banerjee, A., Chandrasekhar, A.~G., Duflo, E., and Jackson, M.~O. (2013).
\newblock The diffusion of microfinance.
\newblock {\em Science}, 341(6144).

\bibitem[Banerjee et~al., 2019]{banerjee2019using}
Banerjee, A., Chandrasekhar, A.~G., Duflo, E., and Jackson, M.~O. (2019).
\newblock Using gossips to spread information: Theory and evidence from two
  randomized controlled trials.
\newblock {\em The Review of Economic Studies}, 86(6):2453--2490.

\bibitem[Banerjee and Ma, 2017]{banerjee2017optimal}
Banerjee, D. and Ma, Z. (2017).
\newblock Optimal hypothesis testing for stochastic block models with growing
  degrees.
\newblock {\em arXiv preprint arXiv:1705.05305}.

\bibitem[Battaglini et~al., 2021]{battaglini2021endogenous}
Battaglini, M., Patacchini, E., and Rainone, E. (2021).
\newblock Endogenous social interactions with unobserved networks.
\newblock {\em The Review of Economic Studies}, 89(4):1694--1747.

\bibitem[Bickel and Sarkar, 2016]{bickel2016hypothesis}
Bickel, P.~J. and Sarkar, P. (2016).
\newblock Hypothesis testing for automated community detection in networks.
\newblock {\em Journal of the Royal Statistical Society Series B: Statistical
  Methodology}, 78(1):253--273.

\bibitem[Binkiewicz et~al., 2017]{binkiewicz2017covariate}
Binkiewicz, N., Vogelstein, J.~T., and Rohe, K. (2017).
\newblock Covariate-assisted spectral clustering.
\newblock {\em Biometrika}, 104(2):361--377.

\bibitem[Bonacich and Lloyd, 2004]{bonacichCalculatingStatusNegative2004}
Bonacich, P. and Lloyd, P. (2004).
\newblock Calculating status with negative relations.
\newblock {\em Social Networks}, 26(4):331--338.

\bibitem[Borgatti et~al., 2006]{borgatti2006robustness}
Borgatti, S.~P., Carley, K.~M., and Krackhardt, D. (2006).
\newblock On the robustness of centrality measures under conditions of
  imperfect data.
\newblock {\em Social Networks}, 28(2):124--136.

\bibitem[Bramoull{\'e} et~al., 2009]{bramoulle2009identification}
Bramoull{\'e}, Y., Djebbari, H., and Fortin, B. (2009).
\newblock Identification of peer effects through social networks.
\newblock {\em Journal of Econometrics}, 150(1):41--55.

\bibitem[Breza and Chandrasekhar, 2019]{breza2019social}
Breza, E. and Chandrasekhar, A.~G. (2019).
\newblock Social networks, reputation, and commitment: evidence from a savings
  monitors experiment.
\newblock {\em Econometrica}, 87(1):175--216.

\bibitem[Breza et~al., 2020]{breza2020using}
Breza, E., Chandrasekhar, A.~G., McCormick, T.~H., and Pan, M. (2020).
\newblock Using aggregated relational data to feasibly identify network
  structure without network data.
\newblock {\em American Economic Review}, 110(8):2454--84.

\bibitem[Brune et~al., 2024]{brune2024goodness}
Brune, B., Flossdorf, J., and Jentsch, C. (2024).
\newblock Goodness-of-fit testing based on graph functionals for homogeneous
  erd{\"o}s--r{\'e}nyi graphs.
\newblock {\em Scandinavian Journal of Statistics}.

\bibitem[Bubeck et~al., 2016]{bubeck2016testing}
Bubeck, S., Ding, J., Eldan, R., and R{\'a}cz, M.~Z. (2016).
\newblock Testing for high-dimensional geometry in random graphs.
\newblock {\em Random Structures \& Algorithms}, 49(3):503--532.

\bibitem[Cai and Zhang, 2018]{cai2018rate1}
Cai, T.~T. and Zhang, A. (2018).
\newblock Rate-optimal perturbation bounds for singular subspaces with
  applications to high-dimensional statistics.
\newblock {\em The Annals of Statistics}, 46(1):60--89.

\bibitem[Candelaria and Ura, 2022]{candelaria2022identification}
Candelaria, L.~E. and Ura, T. (2022).
\newblock Identification and inference of network formation games with
  misclassified links.
\newblock {\em Journal of Econometrics}.

\bibitem[Candes and Plan, 2010]{candes2010matrix}
Candes, E.~J. and Plan, Y. (2010).
\newblock Matrix completion with noise.
\newblock {\em Proceedings of the IEEE}, 98(6):925--936.

\bibitem[Cartwright and Harary, 1956]{cartwright1956structural}
Cartwright, D. and Harary, F. (1956).
\newblock Structural balance: a generalization of heider's theory.
\newblock {\em Psychological review}, 63(5):277.

\bibitem[Chiang et~al., 2014]{chiang2014prediction}
Chiang, K.-Y., Hsieh, C.-J., Natarajan, N., Dhillon, I.~S., and Tewari, A.
  (2014).
\newblock Prediction and clustering in signed networks: a local to global
  perspective.
\newblock {\em The Journal of Machine Learning Research}, 15(1):1177--1213.

\bibitem[De~Paula, 2017]{de2017econometrics}
De~Paula, A. (2017).
\newblock Econometrics of network models.
\newblock In {\em Advances in Economics and Econometrics: Theory and
  Applications: Eleventh World Congress}, volume~1, pages 268--323. Cambridge
  University Press, Cambridge.

\bibitem[De~Paula et~al., 2019]{de2019identifying}
De~Paula, {\'A}., Rasul, I., and Souza, P. (2019).
\newblock Identifying network ties from panel data: theory and an application
  to tax competition.
\newblock {\em arXiv preprint arXiv:1910.07452}.

\bibitem[Everett and Borgatti, 2014]{everettNetworksContainingNegative2014}
Everett, M.~G. and Borgatti, S.~P. (2014).
\newblock Networks containing negative ties.
\newblock {\em Social Networks}, 38:111--120.

\bibitem[Fan et~al., 2016]{fan2016projected}
Fan, J., Liao, Y., and Wang, W. (2016).
\newblock Projected principal component analysis in factor models.
\newblock {\em The Annals of Statistics}, 44(1):219--254.

\bibitem[Fosdick and Hoff, 2015]{fosdick2015testing}
Fosdick, B.~K. and Hoff, P.~D. (2015).
\newblock Testing and modeling dependencies between a network and nodal
  attributes.
\newblock {\em Journal of the American Statistical Association},
  110(511):1047--1056.

\bibitem[Frantz et~al., 2009]{frantz2009robustness}
Frantz, T.~L., Cataldo, M., and Carley, K.~M. (2009).
\newblock Robustness of centrality measures under uncertainty: Examining the
  role of network topology.
\newblock {\em Computational and Mathematical Organization Theory},
  15(4):303--328.

\bibitem[Gao and Lafferty, 2017]{gao2017testing}
Gao, C. and Lafferty, J. (2017).
\newblock Testing network structure using relations between small subgraph
  probabilities.
\newblock {\em arXiv preprint arXiv:1704.06742}.

\bibitem[Gao and Ma, 2021]{gao2021minimax}
Gao, C. and Ma, Z. (2021).
\newblock Minimax rates in network analysis: Graphon estimation, community
  detection and hypothesis testing.
\newblock {\em Statistical Science}, 36(1):16--33.

\bibitem[Gofman, 2017]{gofman2017efficiency}
Gofman, M. (2017).
\newblock Efficiency and stability of a financial architecture with
  too-interconnected-to-fail institutions.
\newblock {\em Journal of Financial Economics}, 124(1):113--146.

\bibitem[Gromov, 2025]{gromov2025social}
Gromov, D. (2025).
\newblock Social balance-based centrality measure for directed signed networks.
\newblock {\em Social Networks}, 80:1--9.

\bibitem[Han et~al., 2023]{han2023universal}
Han, X., Yang, Q., and Fan, Y. (2023).
\newblock Universal rank inference via residual subsampling with application to
  large networks.
\newblock {\em The Annals of Statistics}, 51(3):1109--1133.

\bibitem[Handcock and Gile, 2010]{handcock2010modeling}
Handcock, M.~S. and Gile, K.~J. (2010).
\newblock Modeling social networks from sampled data.
\newblock {\em The Annals of Applied Statistics}, 4(1):5.

\bibitem[Harary, 1953]{harary1953notion}
Harary, F. (1953).
\newblock On the notion of balance of a signed graph.
\newblock {\em Michigan Mathematical Journal}, 2(2):143--146.

\bibitem[Hochberg et~al., 2007]{hochberg2007whom}
Hochberg, Y.~V., Ljungqvist, A., and Lu, Y. (2007).
\newblock Whom you know matters: Venture capital networks and investment
  performance.
\newblock {\em The Journal of Finance}, 62(1):251--301.

\bibitem[Hoff, 2009]{hoff2009multiplicative}
Hoff, P.~D. (2009).
\newblock Multiplicative latent factor models for description and prediction of
  social networks.
\newblock {\em Computational and mathematical organization theory},
  15(4):261--272.

\bibitem[Hsieh and Lee, 2016]{hsieh2016social}
Hsieh, C.-S. and Lee, L.~F. (2016).
\newblock A social interactions model with endogenous friendship formation and
  selectivity.
\newblock {\em Journal of Applied Econometrics}, 31(2):301--319.

\bibitem[Hu et~al., 2021]{hu2021using}
Hu, J., Zhang, J., Qin, H., Yan, T., and Zhu, J. (2021).
\newblock Using maximum entry-wise deviation to test the goodness of fit for
  stochastic block models.
\newblock {\em Journal of the American Statistical Association},
  116(535):1373--1382.

\bibitem[Jackson, 2010]{jackson2010social}
Jackson, M.~O. (2010).
\newblock {\em Social and Economic Networks}.
\newblock Princeton University Press.

\bibitem[Jackson et~al., 2017]{jackson2017economic}
Jackson, M.~O., Rogers, B.~W., and Zenou, Y. (2017).
\newblock The economic consequences of social-network structure.
\newblock {\em Journal of Economic Literature}, 55(1):49--95.

\bibitem[Kleinberg, 1999]{kleinberg1999authoritative}
Kleinberg, J.~M. (1999).
\newblock Authoritative sources in a hyperlinked environment.
\newblock {\em Journal of the ACM (JACM)}, 46(5):604--632.

\bibitem[K{\"o}nig et~al., 2017]{konig2017networks}
K{\"o}nig, M.~D., Rohner, D., Thoenig, M., and Zilibotti, F. (2017).
\newblock Networks in conflict: Theory and evidence from the great war of
  africa.
\newblock {\em Econometrica}, 85(4):1093--1132.

\bibitem[Kornienko and Granger, 2018]{kornienko2018peer}
Kornienko, O. and Granger, D.~A. (2018).
\newblock Peer networks, psychobiology of stress response, and adolescent
  development.
\newblock {\em Oxford handbook of evolution, biology, and society}, pages
  327--348.

\bibitem[Labianca and Brass, 2006]{labiancaExploringSocialLedger2006}
Labianca, G. and Brass, D.~J. (2006).
\newblock Exploring the {{Social Ledger}}: {{Negative Relationships}} and
  {{Negative Asymmetry}} in {{Social Networks}} in {{Organizations}}.
\newblock {\em The Academy of Management Review}, 31(3):596--614.

\bibitem[Lakhina et~al., 2003]{lakhina2003sampling}
Lakhina, A., Byers, J.~W., Crovella, M., and Xie, P. (2003).
\newblock Sampling biases in {IP} topology measurements.
\newblock In {\em IEEE INFOCOM 2003. Twenty-second Annual Joint Conference of
  the IEEE Computer and Communications Societies}, volume~1, pages 332--341.
  IEEE.

\bibitem[Landefeld et~al., 2008]{landefeld2008taking}
Landefeld, J.~S., Seskin, E.~P., and Fraumeni, B.~M. (2008).
\newblock Taking the pulse of the economy: Measuring gdp.
\newblock {\em Journal of Economic Perspectives}, 22(2):193--216.

\bibitem[Le et~al., 2018]{le2018estimating}
Le, C.~M., Levin, K., and Levina, E. (2018).
\newblock Estimating a network from multiple noisy realizations.
\newblock {\em Electronic Journal of Statistics}, 12(2):4697--4740.

\bibitem[Le et~al., 2016]{le2016optimization}
Le, C.~M., Levina, E., and Vershynin, R. (2016).
\newblock Optimization via low-rank approximation for community detection in
  networks.
\newblock {\em The Annals of Statistics}, 44(1):373--400.

\bibitem[Le and Li, 2020]{le2020linear}
Le, C.~M. and Li, T. (2020).
\newblock Linear regression and its inference on noisy network-linked data.
\newblock {\em arXiv preprint arXiv:2007.00803}.

\bibitem[Lee, 2007]{lee2007identification}
Lee, L.-F. (2007).
\newblock Identification and estimation of econometric models with group
  interactions, contextual factors and fixed effects.
\newblock {\em Journal of Econometrics}, 140(2):333--374.

\bibitem[Lee et~al., 2010]{lee2010specification}
Lee, L.-f., Liu, X., and Lin, X. (2010).
\newblock Specification and estimation of social interaction models with
  network structures.
\newblock {\em The Econometrics Journal}, 13(2):145--176.

\bibitem[Lei, 2016]{lei2016goodness}
Lei, J. (2016).
\newblock A goodness-of-fit test for stochastic block models.

\bibitem[Lei and Rinaldo, 2015]{lei2015consistency}
Lei, J. and Rinaldo, A. (2015).
\newblock Consistency of spectral clustering in stochastic block models.
\newblock {\em The Annals of Statistics}, 43(1):215--237.

\bibitem[Li et~al., 2016]{li2016supervised}
Li, G., Yang, D., Nobel, A.~B., and Shen, H. (2016).
\newblock Supervised singular value decomposition and its asymptotic
  properties.
\newblock {\em Journal of Multivariate Analysis}, 146:7--17.

\bibitem[Li et~al., 2019]{li2019prediction}
Li, T., Levina, E., Zhu, J., et~al. (2019).
\newblock Prediction models for network-linked data.
\newblock {\em The Annals of Applied Statistics}, 13(1):132--164.

\bibitem[Lipsey, 2009]{lipsey20091}
Lipsey, R.~E. (2009).
\newblock {\em Measuring International Trade in Services}.
\newblock University of Chicago Press.

\bibitem[Liu, 2019]{liu2019industrial}
Liu, E. (2019).
\newblock Industrial policies in production networks.
\newblock {\em The Quarterly Journal of Economics}, 134(4):1883--1948.

\bibitem[Liu and Tsyvinski, 2020]{liu2020dynamical}
Liu, E. and Tsyvinski, A. (2020).
\newblock Dynamical structure and spectral properties of input-output networks.
\newblock Technical report, National Bureau of Economic Research.

\bibitem[Ma et~al., 2019]{ma2019clusters}
Ma, Y., Zhu, X., and Yu, Q. (2019).
\newblock Clusters detection based leading eigenvector in signed networks.
\newblock {\em Physica A: Statistical Mechanics and its Applications},
  523:1263--1275.

\bibitem[Ma et~al., 2020]{ma2020universal}
Ma, Z., Ma, Z., and Yuan, H. (2020).
\newblock Universal latent space model fitting for large networks with edge
  covariates.
\newblock {\em Journal of Machine Learning Research}, 21:4--1.

\bibitem[Manski, 1993]{manski1993identification}
Manski, C.~F. (1993).
\newblock Identification of endogenous social effects: The reflection problem.
\newblock {\em The Review of Economic Studies}, 60(3):531--542.

\bibitem[Martin and Niemeyer, 2019]{martin2019influence}
Martin, C. and Niemeyer, P. (2019).
\newblock Influence of measurement errors on networks: Estimating the
  robustness of centrality measures.
\newblock {\em Network Science}, 7(2):180--195.

\bibitem[Mojzisch et~al., 2021]{mojzisch2021interactive}
Mojzisch, A., Frisch, J.~U., Doehne, M., Reder, M., and H{\"a}usser, J.~A.
  (2021).
\newblock Interactive effects of social network centrality and social
  identification on stress.
\newblock {\em British Journal of Psychology}, 112(1):144--162.

\bibitem[Ouadah et~al., 2020]{ouadah2020degree}
Ouadah, S., Robin, S., and Latouche, P. (2020).
\newblock Degree-based goodness-of-fit tests for heterogeneous random graph
  models: Independent and exchangeable cases.
\newblock {\em Scandinavian Journal of Statistics}, 47(1):156--181.

\bibitem[Ozsoylev et~al., 2014]{ozsoylev2014investor}
Ozsoylev, H.~N., Walden, J., Yavuz, M.~D., and Bildik, R. (2014).
\newblock Investor networks in the stock market.
\newblock {\em The Review of Financial Studies}, 27(5):1323--1366.

\bibitem[Richmond, 2019]{richmond2019trade1}
Richmond, R.~J. (2019).
\newblock Trade network centrality and currency risk premia.
\newblock {\em The Journal of Finance}, 74(3):1315--1361.

\bibitem[Rohe, 2019]{rohe2019critical}
Rohe, K. (2019).
\newblock A critical threshold for design effects in network sampling.
\newblock {\em The Annals of Statistics}, 47(1):556--582.

\bibitem[Rohe et~al., 2011]{rohe2011spectral}
Rohe, K., Chatterjee, S., and Yu, B. (2011).
\newblock Spectral clustering and the high-dimensional stochastic blockmodel.
\newblock {\em The Annals of Statistics}, 39(4):1878--1915.

\bibitem[Rossi et~al., 2018]{rossi2018network}
Rossi, A.~G., Blake, D., Timmermann, A., Tonks, I., and Wermers, R. (2018).
\newblock Network centrality and delegated investment performance.
\newblock {\em Journal of Financial Economics}, 128(1):183--206.

\bibitem[Shabalin and Nobel, 2013]{shabalin2013reconstruction1}
Shabalin, A.~A. and Nobel, A.~B. (2013).
\newblock Reconstruction of a low-rank matrix in the presence of gaussian
  noise.
\newblock {\em Journal of Multivariate Analysis}, 118:67--76.

\bibitem[Sharpe, 1994]{sharpe1994sharpe}
Sharpe, W.~F. (1994).
\newblock The sharpe ratio.
\newblock {\em Journal of portfolio management}, 21(1):49--58.

\bibitem[Singh, 2019]{singh2019eigenvector}
Singh, R. (2019).
\newblock On eigenvector structure of weakly balanced networks.
\newblock {\em Physica A: Statistical Mechanics and its Applications},
  527:121093.

\bibitem[Soufiani and Airoldi, 2012]{soufiani2012graphlet}
Soufiani, H.~A. and Airoldi, E. (2012).
\newblock Graphlet decomposition of a weighted network.
\newblock In {\em Artificial Intelligence and Statistics}, pages 54--63. PMLR.

\bibitem[Van~Loan and Golub, 1996]{van1996matrix}
Van~Loan, C.~F. and Golub, G. (1996).
\newblock {\em Matrix Computations (Johns Hopkins Studies in Mathematical
  Sciences)}.
\newblock The Johns Hopkins University Press.

\bibitem[Wang et~al., 2012]{wang2012measurement}
Wang, D.~J., Shi, X., McFarland, D.~A., and Leskovec, J. (2012).
\newblock Measurement error in network data: A re-classification.
\newblock {\em Social Networks}, 34(4):396--409.

\bibitem[Yan et~al., 2019]{yan2019statistical}
Yan, T., Jiang, B., Fienberg, S.~E., and Leng, C. (2019).
\newblock Statistical inference in a directed network model with covariates.
\newblock {\em Journal of the American Statistical Association},
  114(526):857--868.

\bibitem[Yang et~al., 2016]{yang2016rate}
Yang, D., Ma, Z., and Buja, A. (2016).
\newblock Rate optimal denoising of simultaneously sparse and low rank
  matrices.
\newblock {\em The Journal of Machine Learning Research}, 17(1):3163--3189.

\bibitem[Yang and Zhu, 2020]{zhu2020networks1}
Yang, Y. and Zhu, W. (2020).
\newblock Networks and business cycles.
\newblock {\em Available at SSRN}.

\bibitem[Zhang and Chen, 2017]{zhang2017hypothesis}
Zhang, J. and Chen, Y. (2017).
\newblock A hypothesis testing framework for modularity based network community
  detection.
\newblock {\em Statistica Sinica}, pages 437--456.

\bibitem[Zhang et~al., 2016]{zhang2016community}
Zhang, Y., Levina, E., and Zhu, J. (2016).
\newblock Community detection in networks with node features.
\newblock {\em Electronic Journal of Statistics}, 10(2):3153--3178.

\bibitem[Zhao et~al., 2012]{zhao2012consistency}
Zhao, Y., Levina, E., and Zhu, J. (2012).
\newblock Consistency of community detection in networks under degree-corrected
  stochastic block models.
\newblock {\em The Annals of Statistics}, 40(4):2266--2292.

\bibitem[Zhu et~al., 2017]{zhu2017network}
Zhu, X., Pan, R., Li, G., Liu, Y., and Wang, H. (2017).
\newblock Network vector autoregression.
\newblock {\em The Annals of Statistics}, 45(3):1096--1123.

\end{thebibliography}

}

\newpage

{
\linespread{1.2}\selectfont{}

\appendix 

\renewcommand{\thesection}{\arabic{section}}
\let\thesectionWithoutS\thesection
\renewcommand\thesection{S\thesectionWithoutS}
\let\theequationWithoutS\theequation 
\renewcommand\theequation{S\theequationWithoutS}
\let\thefigureWithoutS\thefigure 
\renewcommand\thefigure{S\thefigureWithoutS}
\let\thetableWithoutS\thetable
\renewcommand\thetable{S\thetableWithoutS}
\renewcommand{\thetheorem}{S\arabic{theorem}}
\renewcommand{\thecor}{S\arabic{cor}}
\renewcommand{\thermk}{S\arabic{rmk}}
\renewcommand{\thealgocf}{S\arabic{algocf}}

\newtheorem{remark}{Remark}
\let\theremarkWithoutS\theremark
\renewcommand\theremark{S\theremarkWithoutS}

\let\thepropositionWithoutS\theproposition
\renewcommand\theproposition{S\thepropositionWithoutS}

\let\thelemmaWithoutS\thelemma
\renewcommand\thelemma{S\thelemmaWithoutS}

\begin{center}
{\Large\bf{Supplement to ``Network Regression and Supervised Centrality Estimation''}}
\end{center}

This supplementary material contains more details and proofs that are deferred from the main text and is organized as follows.
Section \ref{app:centrality-toy-exmaples} demonstrates the intuition behind the hub and authority centralities using a toy example.
Section \ref{app:net-exmaples} presents the spectral properties of four real networks as empirical evidence to support our network model.
Section \ref{app:algo-supercent} provides the detailed SuperCENT algorithm and a modified version for 
undirected networks with the eigenvector centrality along with their derivations and the proof of algorithmic convergence; we further justify and show the prediction algorithm.
Section \ref{app:theory-ts} explicates the theoretical properties of the two-stage and
provide the explicit mathematical expressions of the asymptotic covariances of the two-stage estimators.
Section \ref{app:more-sim} 
shows additional results for the simulation in Section \ref{sec:sim}
 and a phase-transition experiment to demonstrate the behaviors of SuperCENT and the two-stage estimators under different network signal-to-noise ratios.
Section \ref{app:case} provides details on data construction, additional results, and information for the case study.
Finally, the proofs of the theoretical results are in Section \ref{sec:proof}.

\section{A toy example for the hub and authority centralities}
\label{app:centrality-toy-exmaples}

In this section, we use a toy example to further illuminate the intuition behind  
{the hub and authority centralities.}
Consider a citation network where each paper is a node and 
an edge from Paper A to Paper B indicates Paper A cites Paper B.
Figure \ref{fig:intro-adj} shows an example of the adjacency matrix of such network.
{Figures \ref{fig:intro-hub} and \ref{fig:intro-aut} show the same network with different node sizes: the node sizes in Figure \ref{fig:intro-hub} are proportional to the hub centralities while {those in} Figure \ref{fig:intro-aut} {are proportional} to the authority centralities.

To understand the hub centrality, note that Papers 1 and 4 are the major citors: they both cite three papers with Paper 2 being the common one. Except for the common one, Paper 4 cites Papers 5 and 6, which are only cited by Paper 4, and Paper 1 cites Papers 4 and 3, among which Paper 3 
{is}
cited twice. Therefore, compared with Paper 4, Paper 1 cites the same number of papers with one being cited more than the others. This makes the hub centrality of Paper 1 larger than that of Paper 4. One can think of Paper 1 as a better survey paper than Paper 4. Paper 7 cites only one paper, which makes its hub centrality smaller than Papers 1 and 4. The rest of the papers have small hub centrality since they do not cite other papers.

As for the authority centrality, attention should be given to citees. Papers 2 and 3 both have two citations, but Paper 2 
{is}
cited by Papers 1 and 4 while Paper 3 
{is}
cited by Papers 1 and 7. Observe that Paper 4 as {a} hub is more influential than Paper 7. So the authority centrality for Paper 2 is the highest, followed by Paper 3. For the same reason, Paper 4 has higher authority centrality than Papers 5 and 6, since Paper 4 is cited by Paper 1 while Papers 5 and 6 are cited by Paper 4.}

\begin{figure}
   \centering
   \begin{subfigure}[b]{0.3\textwidth}
     \[
\begin{pmatrix} \cdot & 1 & 1 & 1 & \cdot & \cdot & \cdot \\
 \cdot & \cdot & \cdot & \cdot & \cdot & \cdot & \cdot \\
 \cdot & \cdot & \cdot & \cdot & \cdot & \cdot & \cdot \\
 \cdot & 1 & \cdot & \cdot & 1 & 1 & \cdot \\
 \cdot & \cdot & \cdot & \cdot & \cdot & \cdot & \cdot \\
 \cdot & \cdot & \cdot & \cdot & \cdot & \cdot & \cdot \\
 \cdot & \cdot & 1 & \cdot & \cdot & \cdot & \cdot \end{pmatrix}
\]
     \caption{The adjacency matrix of the citation network. For readability, we denote 0 as $\cdot$.}
     \label{fig:intro-adj}
   \end{subfigure}
   \hfill
   \begin{subfigure}[b]{0.3\textwidth}
     \centering
     \includegraphics[width=\textwidth, trim={6cm 3cm 5.5cm 0},clip]{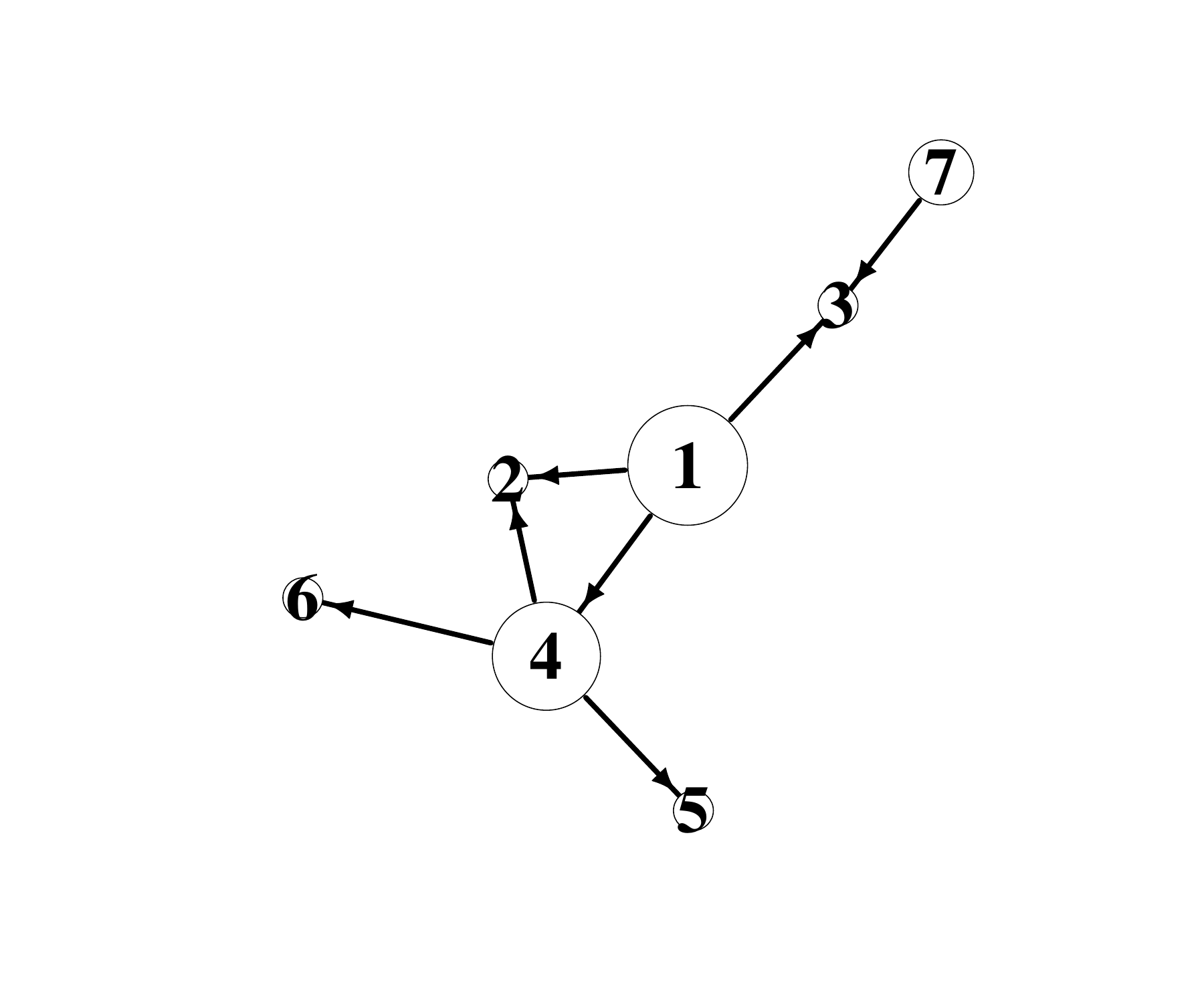}
     \caption{Node size by hub centralities.\newline }
     \label{fig:intro-hub}
   \end{subfigure}
   \hfill
   \begin{subfigure}[b]{0.3\textwidth}
     \centering
     \includegraphics[width=\textwidth, trim={6cm 3cm 5.5cm 0},clip]{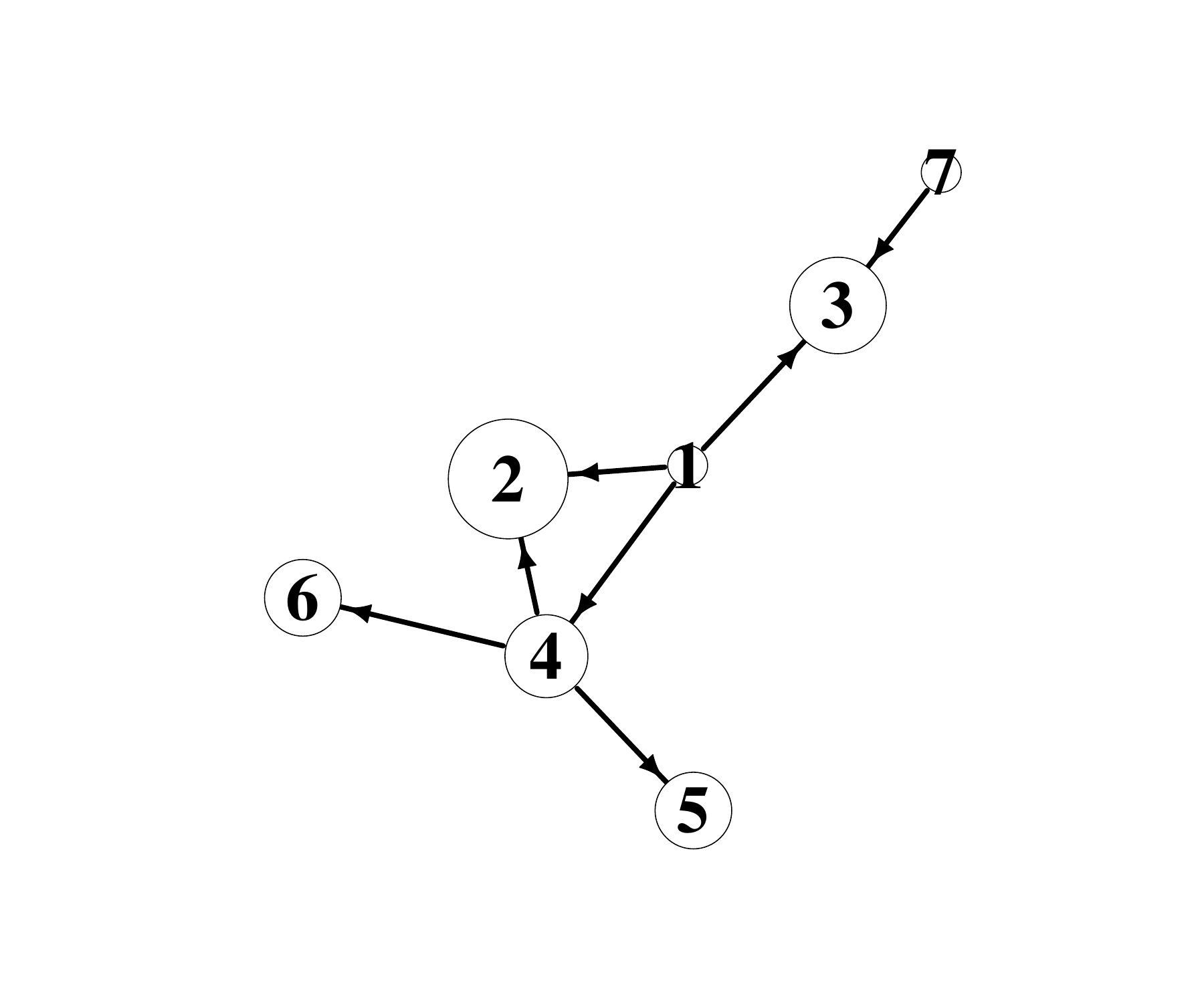}
     \caption{Node size by authority centralities.\newline}
     \label{fig:intro-aut}
   \end{subfigure}
    \caption{A toy network to illustrate the hub and authority centrality.}
    \label{fig:intro-cent}
\end{figure}


\section{Spectral properties of four empirical networks in Section \ref{sec:intro}}
\label{app:net-exmaples}


\begin{figure}
    \centering
    \includegraphics[width=\textwidth]{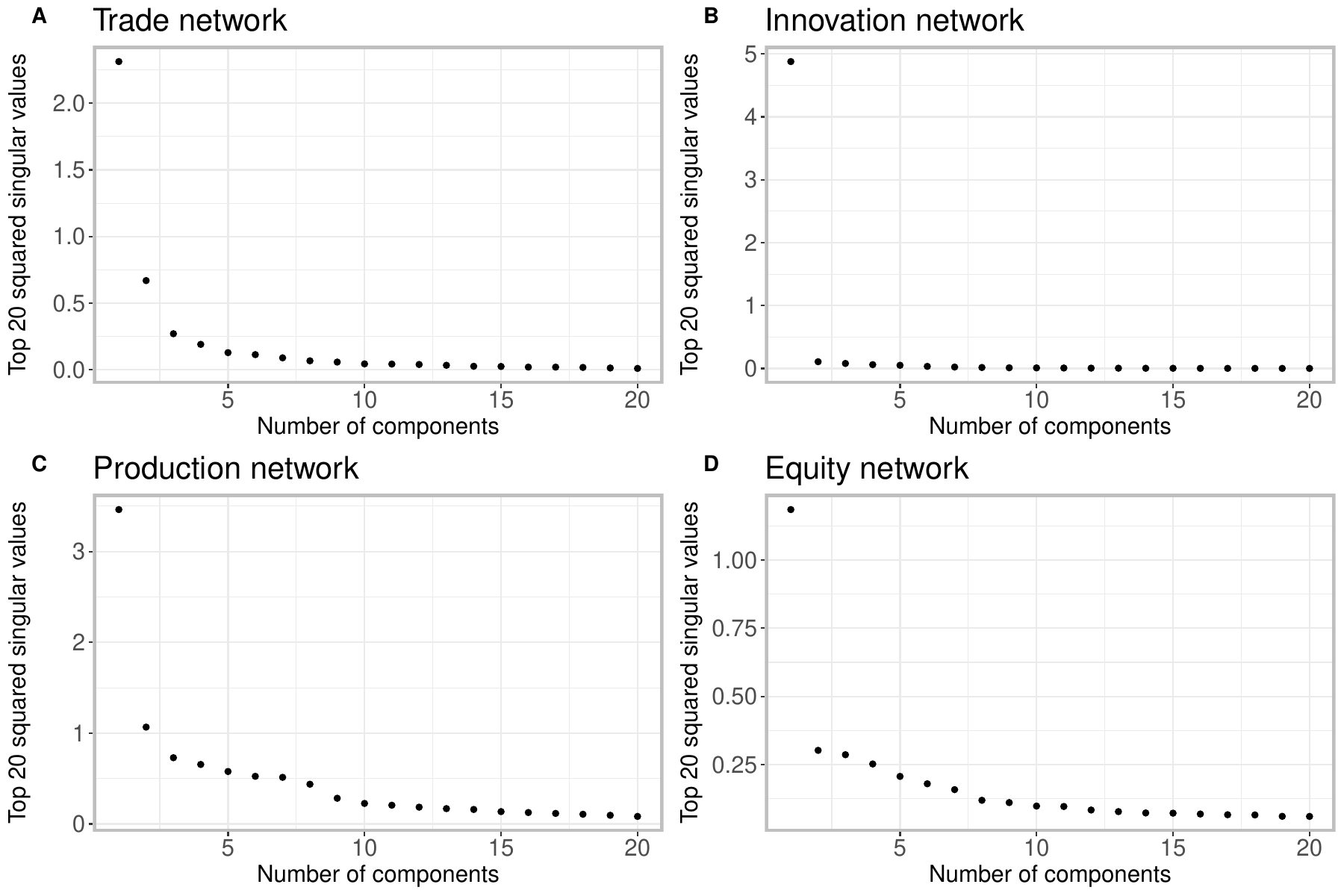}
    \caption{The leading 20 squared singular values of \ref{item:trade} global trade network,
\ref{item:inno} innovation network,
\ref{item:prod} production network, and 
\ref{item:equity} equity network in 2013.
All networks have a low-rank structure with the leading singular value dominates the non-leading ones.
}
    \label{fig:four-examples}
\end{figure}

In this section, we justify the low-rank assumption of our network model \eqref{eq:model-supercent1} by demonstrating that many (economic) networks are, in fact, low-rank
and each of their leading singular value dominates the non-leading ones.
Specifically, we examine the spectral properties of four empirical networks:
\ref{item:trade} global trade network, 
\ref{item:inno} innovation network,
\ref{item:prod} production network, and 
\ref{item:equity} equity network.
We will first describe each network and show the top 20 singular values of each network in 2013. 

\begin{enumerate}[(A)]
    \item \textbf{Trade network.} The global trade network is the country-level trade network as described in Section \ref{sec:case}. We include all 35 countries\footnote{In addition to the 24 countries that have exchange rates, we further include Austria, Belgium, Germany, Spain, Finland, France, Greece, Ireland, Italy, Netherlands, and Portuguese.} here in the bilateral trade data from the correlates of war project (COW). 
    The trade network is defined as the proportion of imports from other countries, i.e., the proportion of imports from country $i$ among all the imports of country $j$.
    {
    We can also define the trade network as the import-export amount in US dollars or as the proportion of exports to other countries. 
    The results are similar.
    }
    \label{item:trade}
    \item \textbf{Innovation network.} The innovation network is an industry-level network of knowledge flow based on patent citations in the US.
    The industry is classified by the 3-digit code of the North American Industry Classification System (NAICS) and we cover 87 industries in 2013.
    Please refer to the Data Appendix of \cite{zhu2020networks} for a detailed description of the construction of the innovation network. \label{item:inno}
    \item \textbf{Production network.} The production network is a network of the input-output flow at the industry level in the US. Particularly, the input-output flow from industry $i$ to $j$ is defined as the proportion of input from industry $i$ among all the inputs of industry $j$. Same as the innovation network, we use the 3-digit NAICS code and cover 87 industries in 2013. Please refer to the Data Appendix of  \cite{zhu2020networks} for a detailed description of the construction of the production network. \label{item:prod}
    \item \textbf{Equity network.} The equity network is a network of firm-level investor-investee shareholding relationships in China, i.e., the percentage of shares of firm $j$ owned by firm $i$. In 2013, we include 3.6 million firms that have at least one firm as their shareholders\footnote{Shareholders can be firms or individuals and we only consider firm-to-firm equity holding.} and we focus on the largest connected component\footnote{A connected component is defined as a network that is weakly connected, i.e., each pair of nodes has at least one path regardless of edge directions.} which includes 1.3 million firms. Please refer to \cite{cai2021microscopic} for a detailed description of the construction of the equity network. \label{item:equity}
\end{enumerate}

Figure \ref{fig:four-examples} shows the top 20 {\emph{squared} singular values of each network, because the proportion of the squared singular values among their sum measures the percentage of the variance explained by each singular component}. As a common pattern across four networks, the leading singular value dominates the non-leading ones. 
Since the leading singular vectors correspond to the hub and authority centralities, 
the centralities are, therefore, able to capture most information of the networks.
This consistent pattern across these empirical networks, along with their corresponding theoretical or empirical studies, motivates our network model and lays the foundation of our formulation.
In addition, as mentioned in Section \ref{sec:supercent-model}, there exist numerous empirical studies that demonstrate the importance of centralities and there is a rising amount of literature that develops information theory or economic theory based on centralities in recent years.

Another feature that shares across the four networks is low-rankness.
Each network has a low-rank structure since the first few leading singular values ``explain'' most of the variability and thus only the first few leading singular vectors are needed to represent the whole network. 
In particular, if we use the ``elbow'' rule to determine the rank of a network, 
the innovation network appears to be rank-one while the ranks of the rest are less than 10.

{
To further confirm the low-rankness of the networks, we adopted the rank inference via residual subsampling (RIRS), a
 universal approach for testing rank proposed by \cite{han2023universal}. 
 We follow the transformations in their real data analysis to handle asymmetric networks:
 \be
 \text{Method 1: } \bA + \bA^\top; \quad \quad 
 \text{Method 2: } 
 \begin{pmatrix}
     0 & \bA \\
     \bA^\top & 0
 \end{pmatrix}.
 \ee
Table \ref{tbl:network-rank} shows the test statistics and $p$-values based on the two transformations for different null hypotheses of
  \ref{item:trade} global trade network,
\ref{item:inno} innovation network, and
\ref{item:prod} production network\footnote{The equity network \ref{item:equity} is too large to run RIRS.}. 
Using $\alpha = 0.05$ as the significance level, RIRS estimates the number of rank as 1 for all three networks based on Method 2. While based on Method 1, the only difference is that the rank of the innovation network is estimated as 2.

\begin{table}[ht]
\centering
\begin{tabular}{c|c|c|c|c|c}
  \hline \multicolumn{1}{c}{} & \multicolumn{1}{c}{} & \multicolumn{2}{|c|}{Method 1} & \multicolumn{2}{c}{Method 2} \\
 \hline
Network & Null hypothesis & Test stats & p-value & Test stats & $p$-value \\ 
  \hline
Trade & $r=1$ & 0.56 & 0.29 & 0.70 & 0.24 \\ 
   \hline 
 Innovation & $r=1$ & -4.51 & 0.00 & -0.46 & 0.32 \\ 
   & $r=2$ & -1.12 & 0.13 & 0.21 & 0.42 \\ 
   \hline 
 Production & $r=1$ & -1.37 & 0.09 & 1.51 & 0.06 \\ 
   \hline
\end{tabular}
\caption{Test statistics and p-value from RIRS.} 
\label{tbl:network-rank}
\end{table}

}

In short, real networks often have a low-rank structure and
the gap between the leading singular value and the rest is large. 
These phenomena support our low rank assumption on the network model and demonstrates the importance of the hub and authority centralities.

\section{SuperCENT algorithms for different models and their derivations}
\label{app:algo-supercent}

In this section, we first derive the SuperCENT algorithm and prove the algorithmic convergence of SuperCENT algorithm in Section \ref{app:supercent-rankone}.
We then derive the SuperCENT algorithm for undirected networks in Section \ref{app:sym-A}.
We further provide the detailed algorithm for prediction in Section \ref{app:prediction-algo}.
Finally, we propose a cross-validation algorithm for selecting the tuning parameter $\lambda$ in Section \ref{sec:cv}.

\subsection{\jcM{Derivation of SuperCENT Algorithm \ref{algo:supercent} for directed network Model \eqref{eq:model-supercent}}}
\label{app:supercent-rankone}

{
In Section \ref{sec:method}, 
the proposed SuperCENT obtains estimates by optimizing the objective function \eqref{eq:supercent-obj}
which combines the network model and the network regression model in the unified framework \eqref{eq:model-supercent}. 
To solve \eqref{eq:supercent-obj}, we use a block descent algorithm by updating
$(\hu,\hv, \hbbeta, \hd)$ where $\hbbeta = (\hbetax^\top,\hbetau,\hbetav)^\top$.
The detailed algorithm is shown in Algorithm \ref{algo:supercent}.

The derivation of each step in each iteration of Algorithm \ref{algo:supercent} is described below. 
Denote {$\bW = (\bX, \bu, \bv)$, $\bbeta = (\bbeta_x, \betau, \betav)$}, and $\mathcal{L}(\bu,\bv,\bbeta, d) := \frac{1}{n}\|\by - \bX\bbeta_x - \bu\beta_u - \bv\beta_v\|_2^2 +
   \frac{\lambda}{n^2} \|\bA - d\bu\bv^\top\|_F^2$.
Given $\lambda$, we minimize the objective function \eqref{eq:supercent-obj} by setting the partial derivatives of all the parameters as zeros. The partial derivatives are as follows. }
{
\be
    \partial_\bbeta \mathcal{L}&=&-\frac{2}{n} \bW^\top (\by-\bW\bbeta), \label{eq:supercent-deriv-beta} \\
	\partial_d \mathcal{L} &=& \frac{2}{n^2} \lambda d \|\bu\|_2^2 \|\bv\|_2^2 - \frac{2}{n^2} \lambda \bu^\top \A \bv, \\
	\partial_\bu \mathcal{L}	&=& -\frac{2}{n}\beta_u(\by-\bX\betax-\bu\betau-\bv\betav) + \frac{2}{n^2} \lambda d^2 \bu \|\bv\|_2^2 - \frac{2}{n^2} \lambda d \A \bv,  \\
	\partial_\bv \mathcal{L}	&=& -\frac{2}{n}\beta_v(\by-\bX\betax-\bu\betau-\bv\betav) + \frac{2}{n^2}\lambda d^2 \bv \|\bu\|_2^2- \frac{2}{n^2}\lambda d\A^\top \bu.
\ee
Setting the partial derivatives above as zeros yields the estimates
\be
\small
\hu &=& \left(\hbetau^2 + \frac{1}{n} \lambda \hd^2 \|\hv\|_2^2 \right)^{-1}
\left[\hbetau(\y - \X\hbetax - \hv\hbetav)+ \frac{1}{n}\lambda \hd\A\hv\right],  \label{eq:algo-hu}\\
\hv &=& \left(\hbetav^2 + \frac{1}{n} \lambda \hd^2 \|\hu\|_2^2 \right)^{-1}
\left[\hbetav(\y - \X\hbetax - \hu\hbetau)+ \frac{1}{n}\lambda \hd\A^\top\hu\right]. \label{eq:algo-hv}\\
\hbbeta &=& ({\widehat\bW}^\top \widehat\bW)^{-1} \widehat\bW^\top \by \label{eq:algo-hbeta},\\
\hd &=& \frac{\hu^\top\bA\hv}{\|\hu\|_2^2 \|\hv\|_2^2},\label{eq:algo-hd}  
\ee 
}
with constraints
\be
\hu^\top\hu=n \quad \mbox{and}\quad
\hv^\top\hv=n \label{eq:algo-hvhv},
\ee
where $\widehat\bW = (\bX, \hu, \hv)$.
Denote $(\hu^{(t)}, \hv^{(t)}, \hbbeta^{(t)}, \hd^{(t)})$ as the estimations from the $t$-th iteration.
Combining \eqref{eq:algo-hu}-\eqref{eq:algo-hvhv}
and substituting the corresponding estimates from the previous updates,
we obtain each update step in each iteration.

To further obtain the estimates of $\bu_l$ and $\bv_l$ with their singular values $d_l$ for $l = 2,\ldots, r$, 
one may perform SVD on $\bA - \hd \hu \hv^\top$ {and scale appropriately}.

\subsubsection{Proof of algorithmic convergence of SuperCENT Algorithm \ref{algo:supercent}}
\label{app:supercent-convergence}

In Remark \ref{rmk:supercent-convergence}, we claim that SuperCENT will converge to stationary points that have smaller objective values than the objective function value of our initial point. Precisely, 
we have $\lim_{t\to \infty} \| \partial \mathcal L (\bu^{(t)}, \bv^{(t)} ,\bbeta^{(t)}, d^{(t)} ) \|_2 =0$ and $ \sup_{t\ge 1}\mathcal L(\bu^{(t)}, \bv^{(t)}, \bbeta^{(t)}, d^{(t)} ) \le \mathcal L(\bu^{(0)}, \bv^{(0)}, \bbeta^{(0)}, d^{(0)} )$.

\begin{proof}


\jcM{First note that with given $\bu$ and $\bv$, the $\bbeta$ and $d$ that minimize the objective function have explicit expression, provided in \eqref{eq:algo-hbeta} and \eqref{eq:algo-hd} with the hats removed. Therefore, minimization of $\mathcal L(\bu,\bv,\bbeta,d)$ with respect to $(\bu,\bv,\bbeta,d)$ reduces to minimization of another appropriate function $\tilde{\mathcal L}(\bu,\bv)$ with respect to $(\bu,\bv)$. The detailed expression of $\tilde{\mathcal L}$ can be obtained by plugging \eqref{eq:algo-hbeta} and \eqref{eq:algo-hd} without the hats into the the expression of $\mathcal L(\bu,\bv,\bbeta,d)$. 
Furthermore, due to the existence of $\bbeta$ and $d$, the scaling of $\bu$ and $\bv$ does not affect the optimal value of the objective function. In other words, $\tilde{\mathcal L}(\bu,\bv) = \tilde{\mathcal L}(c_u\bu,c_v\bv)$, where $c_u, c_v$ are two non-zero constants. 

In the $(t+1)$-th iteration, Steps \ref{alg-u} and \ref{alg-v}, before the normalization in Step \ref{alg-norm}, decrease the value of the objective function
\be
     \mathcal L( \bu^{(t)}, \bv^{(t)}, \bbeta^{(t)}, d^{(t)} ) \ge  
     \mathcal L ( \bu^{(t+1,inter)}, \bv^{(t+1,inter)}, \bbeta^{(t)}, d^{(t)} ),
\ee
where $\bu^{(t+1,inter)}$ and $\bv^{(t+1,inter)}$ are the intermediate outputs from Steps \ref{alg-u} and \ref{alg-v} in iteration $t+1$. The superscript $^{inter}$ indicates the intermediate output in the algorithm.  

Because of the scaling properties of $d\bu\bv^\top, \betau\bu$, and $\betav\bv$, there exist $\bbeta^{(t,inter)}$ and $d^{(t,inter)}$ such that
\be 
\mathcal L ( \bu^{(t+1,inter)}, \bv^{(t+1,inter)}, \bbeta^{(t)}, d^{(t)} )
=
\mathcal L ( \bu^{(t+1)}, \bv^{(t+1)}, \bbeta^{(t,inter)}, d^{(t,inter)} ),
\ee
where $\bu^{(t+1)}$ and $\bv^{(t+1)}$ are from Step \ref{alg-norm}, which are root-$n$ normalization of $\bu^{(t+1,inter)}$ and $\bv^{(t+1,inter)}$. Since Steps \ref{alg-w}, \ref{alg-beta}, \ref{alg-d} optimize the objective function with respect to $\bbeta$ and $d$, we have
\be 
\mathcal L ( \bu^{(t+1)}, \bv^{(t+1)}, \bbeta^{(t,inter)}, d^{(t,inter)} )
\ge
\mathcal L( \bu^{(t+1)}, \bv^{(t+1)}, \bbeta^{(t+1)}, d^{(t+1)} ).
\ee

Combining the three displays above, we have proved that 
\be
     \mathcal L( \bu^{(t)}, \bv^{(t)}, \bbeta^{(t)}, d^{(t)}
     ) 
     &=&\tilde{\mathcal L}( \bu^{(t)}, \bv^{(t)}) \\ 
     &\ge&  
     \mathcal L( \bu^{(t+1)}, \bv^{(t+1)}, \bbeta^{(t+1)}, d^{(t+1)} )
     =\tilde{\mathcal L}( \bu^{(t+1)}, \bv^{(t+1)}). 
\ee
If  $\partial_{\bu} \mathcal L( \bu^{(t)}, \bv^{(t)}, \bbeta^{(t)}, d^{(t)}) \neq \mathbf{0}_n$ or $\partial_{\bv} \mathcal L( \bu^{(t)}, \bv^{(t)}, \bbeta^{(t)}, d^{(t)} ) \neq \mathbf{0}_n$, the inequality is strict, meaning that the algorithm will never revisit point $( \bu^{(t)}, \bv^{(t)}, \bbeta^{(t)}, d^{(t)})$ again. 

Recall that the scaling of $\bu^{(t)},\bv^{(t)}$ does not affect $\tilde{\mathcal L}( \bu^{(t)}, \bv^{(t)})$, let us focus on the unit sphere $\mathcal S$. 
Note that since $\bu^{(t)}/\|\bu^{(t)}\|_2$ and $\bv^{(t)}/\|\bv^{(t)}\|_2$ are on the unit sphere $\mathcal S$, there are limiting points for the pair. Clearly, the limiting points are stationary points (otherwise, for any point within a sufficiently small neighborhood, by continuity and continuous differentiability of $\mathcal L$ together with the descending nature of the algorithm at the limiting point, one iteration will lead to a point that has a smaller objective function value than any point in the neighborhood). Next, we use common techniques in mathematical analysis to show that $\lim_{t\to \infty} \| \partial \mathcal L (\bu^{(t)}, \bv^{(t)} ,\bbeta^{(t)}, d^{(t)} ) \|_2 =0$. 

For simplicity of notation, we denote $$\bbeta(\bu,\bv), d(\bu,\bv) = \argmin_{\bbeta,  d } \mathcal L(\bu, \bv, \bbeta,d).$$
Clearly, the partial derivative of objective function $\mathcal L$ w.r.t. $\bbeta$ and $d$ are zeros:
$$ \partial_{ \bbeta, d} \mathcal L (\bu, \bv, \bbeta, d ) |_{\bbeta=\bbeta(\bu,\bv), d=d(\bu,\bv)} = \mathbf{0}. $$
For any small $\eta >0$, we construct the following covering of $\mathcal S \times \mathcal S $. For any $\mathbf{z_u},\mathbf{z_v} \in \mathcal S $, such that 
$\partial_{\bu,\bv,\bbeta,d} \mathcal{L}(\bu,\bv,\bbeta,d)|_{\bu=\mathbf{z_u},\bv=\mathbf{z_v} , \bbeta=\bbeta(\mathbf{z_u},\mathbf{z_v}), d=d(\mathbf{z_u},\mathbf{z_v}) } \neq \mathbf{0}$,
define $C(\mathbf{z_u},\mathbf{z_v})$ to be the largest ball centered by $(\mathbf{z_u},\mathbf{z_v})$ such that 
\begin{multline}
\inf_{(\mathbf{z}_u',\mathbf{z}_v') \in C(\mathbf{z_u},\mathbf{z_v})} \| \partial_{\bu,\bv,\bbeta,d} \mathcal{L}(\bu,\bv,\bbeta,d)|_{ \bu=\mathbf{z}_u',
\bv = \mathbf{z}_v' , 
\bbeta=\bbeta(\mathbf{z}_u',\mathbf{z}_v'), 
d= d(\mathbf{z}_u',\mathbf{z}_v') } \|_2 \\
>  \frac{1}{2} \| \partial_{\bu,\bv,\bbeta,d} \mathcal{L}(\bu,\bv,\bbeta,d)|_{ \bu=\mathbf{z}_u,
\bv = \mathbf{z}_v , 
\bbeta=\bbeta(\mathbf{z}_u,\mathbf{z}_v), 
d= d(\mathbf{z}_u,\mathbf{z}_v) }   \|_2 .
\end{multline}
For any $\mathbf{z_u},\mathbf{z_v} \in \mathcal S $ such that $\partial_{\bu,\bv,\bbeta,d} \mathcal{L}(\bu,\bv,\bbeta,d)|_{\bu=\mathbf{z_u},\bv=\mathbf{z_v} , \bbeta=\bbeta(\mathbf{z_u},\mathbf{z_v}), d=d(\mathbf{z_u},\mathbf{z_v}) } = \mathbf{0}$,
define $C(\mathbf{z_u},\mathbf{z_v})$ to be the largest ball centered by $(\mathbf{z_u},\mathbf{z_v})$ such that 
\be
\sup_{(\mathbf{z}_u',\mathbf{z}_v') \in C(\mathbf{z_u},\mathbf{z_v})} \| \partial_{\bu,\bv,\bbeta,d} \mathcal{L}(\bu,\bv,\bbeta,d)|_{ \bu=\mathbf{z}_u',
\bv = \mathbf{z}_v' , 
\bbeta=\bbeta(\mathbf{z}_u',\mathbf{z}_v'), 
d= d(\mathbf{z}_u',\mathbf{z}_v') } \|_2< \eta.
\ee
Clearly, $ \mathcal S \times \mathcal S \subset \bigcup_{ (\mathbf{z_u},\mathbf{z_v}) \in \mathcal S \times \mathcal S} C( \mathbf{z_u},\mathbf{z_v}) $. By the compactness of $ \mathcal S \times \mathcal S$, it can be covered by finite many sets, denoted as $C( \mathbf{z_u^1},\mathbf{z_v^1}), \dots, C( \mathbf{z_u^k},\mathbf{z_v^k})$. 

Since the limiting points are all stationary points, the sequence of pairs, $\bu^{(t)}/\|\bu^{(t)}\|_2$ and $\bv^{(t)}/\|\bv^{(t)}\|_2$, will never visit the balls centered around non-stationary points after a certain number of iterations. That is, there exists an integer $T_{\eta}>1$ such that after $T_{\eta}$ iterations, all the pairs, $\bu^{(t)}/\|\bu^{(t)}\|_2$ and $\bv^{(t)}/\|\bv^{(t)}\|_2$, generated by the algorithm are in one of the balls centered around some stationary points. In other words, 
$$\| \partial_{\bu,\bv,\bbeta,d} \mathcal{L}(\bu,\bv,\bbeta,d)|_{  \bu={ \bu^{(t)} \over \sqrt{n}}, \bv={ \bv^{(t)}  \over \sqrt{n} },\bbeta={ \sqrt{n}\bbeta^{(t)} }, d = n d^{(t)}  }\|_2 \le \eta , \mbox{ for } t> T_\eta,$$ 
which implies 
\begin{multline}
    \|  \partial_{\bu,\bv,\bbeta,d} \mathcal{L}(\bu,\bv,\bbeta,d)|_{\bu={ \bu^{(t)} }, \bv={ \bv^{(t)}   },\bbeta={ \bbeta^{(t)} },  d=d^{(t)} } \|_2 
    \\
    = \frac{1}{\sqrt{n}} \| \partial_{\bu,\bv,\bbeta,d} \mathcal{L}(\bu,\bv,\bbeta,d)|_{ \bu={ \bu^{(t)} \over \sqrt{n}}, \bv={ \bv^{(t)}  \over \sqrt{n} },
    \bbeta={ \sqrt{n}\bbeta^{(t)} }, d=n d^{(t)} } \|_2 \le \eta , \mbox{ for } t> T_\eta,
\end{multline}
which establishes the claim $\lim_{t\to \infty} \| \partial \mathcal L (\bu^{(t)}, \bv^{(t)} ,\bbeta^{(t)}, d^{(t)} ) \|_2 =0$.

}

\end{proof}

\subsection{SuperCENT algorithm for undirected networks}
\label{app:sym-A}

When the network is undirected with the eigenvector centrality,
it can be represented by a symmetric matrix $\bA$. 
Denote $\bu$ as the eigenvector centrality.
The objective function of SuperCENT estimation is a special case of \eqref{eq:supercent-obj}, i.e.,
\be
\label{eq:supercent-sym}
(\hbetax,\hbetau,\hd,\hu) :=
   \argmin_{\substack{\bbeta_x,\beta_u \\ d,\|\bu\|_2=\sqrt{n}}}
\frac{1}{n} \|\y -	\X\bbeta_x - \bu \beta_u  \|_2^2 +
\frac{\lambda}{n^2} \|\A - d\bu\bu^\top \|^2_F.
\ee



To solve \eqref{eq:supercent-sym}, we adopt a similar strategy as Algorithm \ref{algo:supercent} --
a partial block descent algorithm by updating $(\hbbeta, \hd, \hu)$ iteratively until convergence,
where $\hbbeta = (\hat{\bbeta}_x^\top,\hbetau)^\top$. 
To update $\hu$, we use gradient descent with backtracking line search instead. 
The initialization can be obtained from the eigen decomposition of $\bA$. 

Given a tuning parameter $\lambda$, Algorithm \ref{algo:l2-2} describes the algorithm for a symmetric matrix $\bA$. 
Similarly, the tuning parameters $\lambda$ can be chosen using cross-validation as described in Section \ref{sec:cv}.

\begin{algorithm}
\SetAlgoLined
\jcM{
\KwResult{$\hd$, $\hu$ and $\hbbeta$.} 
 \textbf{Input:} the observed network $\A\in\mathbb{R}^{n\times n}$, the design matrix $\X\in\mathbb{R}^{n\times p}$, the response vector $\y\in\mathbb{R}^{n}$, the tuning penalty parameter $\lambda$, the tolerance parameter $\rho>0$, the maximum number of iteration $T$ \;
Initiate $t=0$,\\
\quad \quad \quad $\bu^{(0)} = \argmin_{\|\bu\|_2 = \sqrt{n}} \|\bA- d\bu \bu^\top\|^2_F$,\\
\quad \quad \quad $\W^{(0)} =  (\X, \bu^{(0)})$,\\
\quad \quad \quad $\bbeta^{(0)} = (\W^{(0)\top} \W^{(0)})^{-1}\W^{(0)\top}\by$,\\
\quad \quad \quad $d^{(0)} = {\bu^{(0)}}^\top \A \bu^{(0)} / \|\bu^{(0)}\|_2^4$\;
     \While{$t \leq 1$ or $(\| \bP_{\bu^{(t-1)}} - \bP_{\bu^{(t)}}\|_2 > \rho$ and $t<T)$ }{
 	\begin{enumerate}[topsep=0pt,itemsep=-1ex,partopsep=1ex,parsep=1ex]
        \item $t \leftarrow t+1$\;
        \item $\bu^{(t)} \leftarrow BLS(\A, \X, \y, \lambda, \bu^{(t-1)}, \bbeta^{(t-1)}, d^{(t-1)})$  of Algorithm \ref{algo:gd-u}\;
		\item Normalize $\bu^{(t)}$ to have norm $\sqrt{n}$: $\bu^{(t)} = \sqrt{n} \bu^{(t)} / \|\bu^{(t)}\|_2 $\;
        \item $\W^{(t)} = (\X, \bu^{(t)})$\;
        \item $\bbeta^{(t)} = (\W^{(t)\top} \W^{(t)})^{-1}\W^{(t)\top}\by$\;
 		\item $d^{(t)} = {\bu^{(t)}}^\top \A \bu^{(t)} / \|\bu^{(t)}\|_2^4$\;
 	\end{enumerate}
    }
    $\hu = \bu^{(t)},  \hbbeta = \bbeta^{(t)}, \hd = d^{(t)}$.
}
 \caption{$\supercent(\A, \X, \y, \lambda)$ to solve \eqref{eq:supercent-sym} for a symmetric $\bA$. 
 }
 \label{algo:l2-2}
\end{algorithm}

\begin{algorithm}
\SetAlgoLined
\KwResult{$\bu^{(t)}$}
 \textbf{Input:} the observed network $\A\in\mathbb{R}^{n\times n}$, the design matrix $\X\in\mathbb{R}^{n\times p}$, the response vector $\y\in\mathbb{R}^{n}$, the tuning penalty parameter $\lambda$, the tolerance parameter $\rho_2>0$, the maximum number of iteration $T_2$, the step size shrinkage $\gamma\in(0,1)$, $\bu^{(t-1)}$, $\bbeta^{(t-1)}=(\bbeta_x^{(t-1)}, \beta_u^{(t-1)})$,  $d^{(t-1)}$, $\alpha = 1$\;
 
 Let $\mathcal{L}_{sym, t}(\bu) = \frac{1}{n} \|\by -	\X\bbeta_x^{(t-1)} - \bu \beta_u^{(t-1)}  \|_2^2 +
\frac{\lambda}{n^2} \|\A - d^{(t-1)} \bu \bu^\top \|^2_F$\;

 Initiate $\bu_0 = \bu^{(t-1)}, \tau=1$\;
 
    \While{$\tau\leq2$ or $(\| \bP_{\bu_{\tau-1}} - \bP_{\bu_{\tau-2}}\|_2 > \rho_2$ and $\tau<T_2)$ }{
 	\begin{enumerate}[topsep=0pt,itemsep=-1ex,partopsep=1ex,parsep=1ex]
        \item $\alpha_\tau = \alpha$\;
        \item $\nabla_\tau = -\frac{2}{n}\beta_u^{(t-1)}(\by-\bX\bbeta_x^{(t-1)}-\bu_{\tau-1}\beta_u^{(t-1)})$ \\ 
      \hspace{1in} $+ \frac{4}{n^2}\lambda (d^{(t-1)})^2 \|\bu_{\tau-1}\|^2 \bu_{\tau-1} - \frac{4}{n^2} \lambda d^{(t-1)} \A \bu_{\tau-1}$\;
        \item \textbf{while} {$\mathcal{L}_{sym, t} (\bu_{\tau-1} - \alpha_\tau \nabla_\tau) > \mathcal{L}_{sym, t}(\bu_{\tau-1}) - \frac{1}{2} \alpha_\tau \|\nabla_\tau\|^2$} \textbf{do} \\
                $\vert$ \quad  Update the step size: $\alpha_{\tau} = \gamma \alpha_{\tau}$; \\
                \textbf{end}
        \item $ \bu_{\tau} = \bu_{\tau-1} - \alpha_\tau \nabla_\tau$\;
        \item $\tau \leftarrow \tau +1$\;
    \end{enumerate}
 }
 $\bu^{(t)} = \bu_{\tau-1}.$
 \caption{$BLS(\A, \X, \y, \lambda, \bu^{(t-1)}, \bbeta^{(t-1)}, d^{(t-1)})$:  backtracking line search (BLS) to obtain $\bu^{(t)}$. 
 \label{algo:gd-u}
 }
\end{algorithm}
First, our results now apply to multi-rank networks, which require handling non-leading vectors. Second, we provide results not only for the solution of the objective function but also for the solution produced by the algorithm, which adds complexity to the analysis.

The derivation of Algorithm \ref{algo:l2-2} is similar to Algorithm \ref{algo:supercent}.
Denote {$\bW = (\bX, \bu)$, $\bbeta = (\bbeta_x, \betau)$}, and $\mathcal{L}_{sym} := \frac{1}{n} \|\y -	\X\bbeta_x - \bu \beta_u  \|_2^2 +
\frac{\lambda}{n^2} \|\A - d\bu\bu^\top \|^2_F$ where the subscript $sym$ denotes the objective function for a symmetric matrix $\bA$.
Given $\lambda$, we minimize the objective function \eqref{eq:supercent-obj} by setting the partial derivatives with respect to $\bbeta$ and $d$ as zero and then applying gradient descent with line backtracking search for $\bu$. The partial derivatives are as follows. 
\be
    \partial_\bbeta \mathcal{L}_{sym}  &=&-\frac{2}{n} \bW^\top (\by-\bW\bbeta), \\
	\partial_d \mathcal{L}_{sym} &=& \frac{2}{n^2} \lambda d \|\bu\|_2^4 - \frac{2}{n^2} \lambda \bu^\top \A \bu, \\
	\partial_\bu \mathcal{L}_{sym} &=& -\frac{2}{n}\beta_u(\by-\bX\betax-\bu\betau) + \frac{4}{n^2}\lambda d^2 \|\bu\|_2^2 \bu - \frac{4}{n^2} \lambda d \A \bu.
\ee
Setting the partial derivatives above as zero yields the estimates
\be
\small
\hbbeta &=& ({\widehat\bW}^\top \widehat\bW)^{-1} \widehat\bW^\top \by, \label{eq:algo-beta-sym} \\
\hd &=& \frac{ \hu^\top\bA\hu}{\|\hu\|_2^4} \label{eq:algo-hd-sym},
\ee
where $\widehat\bW= (\bX, \hu)$. 
Similarly, denote $(\hbbeta^{(t)}, \hd^{(t)}, \hu^{(t)})$ as the estimations from the $t$-th iteration.
Taking together \eqref{eq:algo-beta-sym}-\eqref{eq:algo-hd-sym} with Algorithm \ref{algo:gd-u}
and substituting corresponding estimates  from the previous update, 
we obtain each update step in each iteration.

\subsection{Prediction algorithm}
\label{app:prediction-algo}

\jcM{
In this section, we explain and demonstrate that under the unified framework, the fitted model can be used for prediction. 

Assume that there are $n_1$ training observations, $n_2$ testing observations, and in total $n=n_1+n_2$ observations. All of these $n$ observations form an adjacency matrix $\bA =
\begin{pmatrix}
\bA_{11} & \bA_{12} \\
\bA_{21} & \bA_{22}
\end{pmatrix}$, which has block structure. In what follows, we will assume $r=1$ for simplicity. At the end of this section, we will provide the algorithm for the multi-rank cases. 

Given $\bA, \bX,$ and $\by$, suppose the unified framework holds for all these $n$ training and testing observations altogether, i.e.,
\begin{subequations}
\label{eq:prediction-model-supercent}
  \begin{align}[left ={\empheqlbrace}]
	\bA &=  d\bu\bv^\top 
    + \bE, \label{eq:prediction-model-supercent1} \\
	\by &= \bX\bbeta_x + \bu\beta_u + \bv\beta_v + \bepsilon, \label{eq:prediction-model-supercent2}
  \end{align}
\end{subequations}
where we make assumptions on the length of the centralities that $\|\bu\|=\|\bv\|=\sqrt{n}$. 

Denote the sub-blocks of $\bu,\bv, \bE, \by, \bX, \bepsilon$ appropriately with subscripts $_1$ and $_2$ corresponding to the training and testing parts respectively. Then, the model for the training part of \eqref{eq:prediction-model-supercent} becomes
\begin{subequations}
\label{eq:prediction-model-supercent-training}
  \begin{align}[left ={\empheqlbrace}]
	\bA_{11} &=  d\bu_1\bv_1^\top 
    + \bE_{11}, \label{eq:prediction-model-supercent1-training} \\
	\by_1 &= \bX_1\bbeta_x + \bu_1\beta_u + \bv_1\beta_v + \bepsilon_1. \label{eq:prediction-model-supercent2-training}
  \end{align}
\end{subequations}
And the testing part of \eqref{eq:prediction-model-supercent} becomes
\begin{subequations}
\label{eq:prediction-model-supercent-testing}
  \begin{align}[left ={\empheqlbrace}]
	\bA_{22} &=  d\bu_2\bv_2^\top 
    + \bE_{22}, \label{eq:prediction-model-supercent1-testing} \\
	\by_2 &= \bX_2\bbeta_x + \bu_2\beta_u + \bv_2\beta_v + \bepsilon_2. \label{eq:prediction-model-supercent2-testing}
  \end{align}
\end{subequations}

Note that because of the joint model \eqref{eq:prediction-model-supercent} for both training and testing data, the regression coefficients $\bbeta_x,\betau,\betav$ in the training \eqref{eq:prediction-model-supercent2-training} and in the testing \eqref{eq:prediction-model-supercent2-testing} are the same. Therefore, if the estimations of the regression coefficients $\hbbeta_x,\hbetau,\hbetav$ are obtained from the training data $\bA_{11}, \bX_1, \by_1$ based on model \eqref{eq:prediction-model-supercent-training} by our SuperCENT algorithm, and $\bu_2,\bv_2$ are available, we can plug them into \eqref{eq:prediction-model-supercent2-testing} for prediction and obtain $\hat\by_2$.

Although such prediction seems straightforward, a few subtle points worth clarification. First, the model \eqref{eq:prediction-model-supercent-training} for the training data does not satisfy the norm constraints, $\|\bu_1\|=\|\bv_1\|=\sqrt{n_1}$. An equivalent model can be expressed which satisfies the norm constraints,
\begin{subequations}
\label{eq:prediction-model-supercent-training-norm}
  \begin{align}[left ={\empheqlbrace}]
	\bA_{11} &=  \left(\frac{d\|\bu_1\|\|\bv_1\|}{n_1} \right)
    \left(\frac{\sqrt{n_1}}{\|\bu_1\|} \bu_1\right)
    \left(\frac{\sqrt{n_1}}{\|\bv_1\|} \bv_1\right)^\top 
    + \bE_{11}, \label{eq:prediction-model-supercent1-training-norm} \\
	\by_1 &= \bX_1\bbeta_x + 
    \left(\frac{\sqrt{n_1}}{\|\bu_1\|} \bu_1\right)
    \left(\frac{\|\bu_1\|}{\sqrt{n_1}}\beta_u\right) + \left(\frac{\sqrt{n_1}}{\|\bv_1\|} \bv_1\right)
    \left(\frac{\|\bv_1\|}{\sqrt{n_1}}\beta_v\right)
    + \bepsilon_1. \label{eq:prediction-model-supercent2-training-norm}
  \end{align}
\end{subequations}
Therefore, implicitly, we assume the following model for the training data
\begin{subequations}
\label{eq:prediction-model-supercent-training-norm-tilde}
  \begin{align}[left ={\empheqlbrace}]
	\bA_{11} &=  \td\tbu_1\tbv_1^\top 
    + \bE_{11}, \label{eq:prediction-model-supercent1-training-norm-tilde} \\
	\by_1 &= \bX_1\bbeta_x + \tbu_1\tbetau + \tbv_1\tbetav + \bepsilon_1, \label{eq:prediction-model-supercent2-training-norm-tilde}
  \end{align}
\end{subequations}
where $\td = \frac{d\|\bu_1\|\|\bv_1\|}{n_1}, \tbu_1=\frac{\sqrt{n_1}}{\|\bu_1\|} \bu_1, \tbv_1 = \frac{\sqrt{n_1}}{\|\bv_1\|} \bv_1, \tbetau = \frac{\|\bu_1\|}{\sqrt{n_1}}\beta_u$, and $\tbetav = \frac{\|\bv_1\|}{\sqrt{n_1}}\beta_v$. Hence, applying SuperCENT algorithm \ref{algo:supercent} to the observed training data $\bA_{11}, \bX_1, \by_1$ will produce estimates $\htbu_1, \htbv_1, \hbbeta_x, \htbetau, \htbetav, \htd$ for $\tbu_1, \tbv_1, \bbeta_x, \tbetau, \tbetav, \td$ in model \eqref{eq:prediction-model-supercent-training-norm-tilde}, but not for $\bu_1, \bv_1, \bbeta_x, \betau, \betav, d$ in \eqref{eq:prediction-model-supercent-training}. Therefore, these estimates cannot be plugged into \eqref{eq:prediction-model-supercent2-testing} for prediction, because there is a mismatch of the scaling. 

The model for the testing data that depends on the tilde version of the coefficients is
\begin{subequations}
\label{eq:prediction-model-supercent-testing-norm}
  \begin{align}[left ={\empheqlbrace}]
	\bA_{22} &=  \left(\frac{d\|\bu_1\|\|\bv_1\|}{n_1} \right)
    \left(\frac{\sqrt{n_1}}{\|\bu_1\|} \bu_2\right)
    \left(\frac{\sqrt{n_1}}{\|\bv_1\|} \bv_2\right)^\top 
    + \bE_{22}, \label{eq:prediction-model-supercent1-testing-norm} \\
	\by_2 &= \bX_2\bbeta_x + 
    \left(\frac{\sqrt{n_1}}{\|\bu_1\|} \bu_2\right)
    \left(\frac{\|\bu_1\|}{\sqrt{n_1}}\beta_u\right) + \left(\frac{\sqrt{n_1}}{\|\bv_1\|} \bv_2\right)
    \left(\frac{\|\bv_1\|}{\sqrt{n_1}}\beta_v\right)
    + \bepsilon_1, \label{eq:prediction-model-supercent2-testing-norm}
  \end{align}
\end{subequations}
which is equivalent to 
\begin{subequations}
\label{eq:prediction-model-supercent-testing-norm-tilde}
  \begin{align}[left ={\empheqlbrace}]
	\bA_{22} &=  \td\tbu_2\tbv_2^\top 
    + \bE_{22}, \label{eq:prediction-model-supercent1-testing-norm-tilde} \\
	\by_2 &= \bX_2\bbeta_x + \tbu_2\tbetau + \tbv_2\tbetav + \bepsilon_2, \label{eq:prediction-model-supercent2-testing-norm-tilde}
  \end{align}
\end{subequations}
where $\td, \tbetau,\tbetav$ are defined above and $\tbu_2=\frac{\sqrt{n_1}}{\|\bu_1\|} \bu_2, \tbv_2 = \frac{\sqrt{n_1}}{\|\bv_1\|} \bv_2$. 

It is clear that to make prediction for $\by_2$, we need estimates $\hbbeta_x, \htbetau, \htbetav$ for model \eqref{eq:prediction-model-supercent-training-norm-tilde} from training data $\bA_{11}, \bX_1, \by_1$ via our Algorithm \ref{algo:supercent}, as well as estimates $\htbu_2, \htbv_2$. 

The discussion above \eqref{eq:prediction-model-supercent}-\eqref{eq:prediction-model-supercent-testing-norm-tilde}  first assumes the root-$n$ constraints for the centralities from all $n$ observations including training and testing data in model \eqref{eq:prediction-model-supercent}, and then adjusts the training and testing models accordingly to satisfy the root-$n_1$ constraints for the training data. Following this discussion, there are two conclusions: (i) the same $\tbetau,\tbetav$ can be used for both training and testing data, as long as the $\tbu_1,\tbu_2,\tbv_1,\tbv_2$ are scaled appropriately; (ii) the concatenation of the scaled centralities for both training data $\tbu_1$ and testing data $\tbu_2$ together is proportional to original leading singular vectors for all data, i.e.,  $(\tbu_1^\top,\tbu_2^\top)^\top \propto (\bu_1^\top,\bu_2^\top)^\top = \bu$, similarly for $\bv$.

The discussion above \eqref{eq:prediction-model-supercent}-\eqref{eq:prediction-model-supercent-testing-norm-tilde} still follows, if one starts with root-$n_1$ constraints for the centralities for the training data, and adjust the models for all data and testing data accordingly. Similar conclusions can be obtained. 
As we mentioned in Section \ref{sec:supercent-est},
the scales of $\bu_1, \bv_1, \betau, \betav$ are not identifiable (one can multiple the centralities by non-zero constant and divide the regression coefficients by the same non-zero constant), only the products $\bu_1\betau, \bv_1\betav$ are identifiable. Hence, it does not matter what norm-constraint is imposed on the centralities of the training data, that is $\|\hbu_1\|,\|\hbv_1\|$ can be anything, as long as $\hbetau,\hbetav$ are adjusted accordingly. Despite this flexibility, we still assume root-$n_1$ constraints above for clarity purpose. The main take-away is that, for any pair of estimates of $\hbu_1,\hbetau$ (respectively, $\hbv_1,\hbetav$), whose product is fully determined, it is not necessary for them to be $\htbu_1,\htbetau$, what matters for proper prediction is to ensure that $(\hbu_1^\top, \hbu_2^\top)^\top \propto  (\bu_1^\top,\bu_2^\top)^\top = \bu$ (respectively, $(\hbv_1^\top, \hbv_2^\top)^\top \propto  (\bv_1^\top,\bv_2^\top)^\top = \bv$).

The only remaining question is how to obtain estimate $\htbu_2$. The procedure for $\hat{\tilde{\bv}}_2$ is similar and will be omitted. Two approaches will be introduced. The first approach is relatively simple, which is based on the SVD of the large network matrix $\bA$ and is described in Algorithm S\ref{algo:pred-uv-svd}. Since $\tbu_2=\frac{\sqrt{n_1}}{\|\bu_1\|} \bu_2$, where $(\bu_1^\top,\bu_2^\top)^\top = \bu$ is the leading left singular vector of the signal in \eqref{eq:prediction-model-supercent1}. Suppose $\hu$ is the leading left singular vector of $\bA$. Assume that $\hu_{1:n_1}$ and the SuperCENT estimate $\htbu_1$ have an angle less than 90 degrees; otherwise, flip the sign of $\hu_{1:n_1}$. Naturally, we have 
\be
\htbu_2=\frac{\sqrt{n_1}}{\|\hu_{1:n_1}\|} \hu_{(n_1+1):n}.
\ee
}

\jcM{
\begin{algorithm}[H]
\label{algo:pred-uv-svd}
\SetAlgoLined
\KwResult{$\htbu_2$ and $\htbv_2$.}
 \textbf{Input:} 
 The augmented network $\bA\in\mathbb{R}^{n\times n}$, $\htbu_1$, and $\htbv_1$.
 \begin{enumerate}
 	\item $(\hu, \hv)$ are the leading left and right singular vectors of $\bA$\;
 	\item {$\hu = \text{sign}(\htbu_1^\top \hu_{1:n_1}) \hu$ and $\hv = \text{sign}(\htbu_1^\top \hu_{1:n_1}) \hv$\;}
 	\item Rescale $\htbu_2 = \frac{\sqrt{n_1}}{\|\hu_{1:n_1}\|_2} \hu_{(n_1+1):n}$ and $\htbv_2 = \frac{\sqrt{n_1}}{\|\hv_{1:n_1}\|_2} \hv_{(n_1+1):n} $.
 \end{enumerate}
 \caption{{Modified} SVD of $\bA$ to obtain $\htbu_2$ and $\htbv_2$.}
\end{algorithm}
}

\jcM{
The first approach only uses the adjacency matrix $\bA$ to estimate $\tbu_2$. The second approach leverages the better estimate from SuperCENT comparing to plain SVD and is described in Algorithm S\ref{algo:pred-uv-svd-2}. It is based on the following observation. Focusing on the bottom left block of the observed network matrix $\bA$ in \eqref{eq:prediction-model-supercent1}, we have 
$$
\bA_{21} = d\bu_2\bv_1^\top + \bE_{21} = \td \tbu_2\tbv_1^\top + \bE_{21},
$$
where $\td,\tbu_2,\tbv_1$ were defined the same as above in \eqref{eq:prediction-model-supercent-training-norm-tilde} and \eqref{eq:prediction-model-supercent-testing-norm-tilde}. Since $\htd,\htbv_1$ are provided by applying SuperCENT on the training data, we can estimate $\tbu_2$ by minimizing $\|\bA_{21}-\htd \tbu_2\htbv_1^\top\|_F^2$. This leads to an estimate 
\be 
\htbu_2 = \htd^{-1}(\htbv_1^\top \htbv_1)^{-1}\bA_{21}\htbv_1.
\ee
Plugging in the model of $\bA_{21}$ into the estimation, it is easily seen that $\htbu_2\approx \tbu_2$ plus a term related to noise $\bE_{21}$. 

Moving to the multi-rank case, the same reasoning follows and the only difference is the model for the network for all training and testing changes from \eqref{eq:prediction-model-supercent1} to
\be 
\bA =  \bU\bD\bV^\top  + \bE
    =  \tbU\tbD\tbV^\top  + \bE, \label{eq:prediction-model-rankr1}
\ee
where for the terms without tilde, $\bU = (\bu,\bu_2,\cdots, \bu_r)\in\R^{n\times r}, \bV = (\bv,\bv_2,\cdots, \bv_r)\in\R^{n\times r}, \bD=\mathrm{diag}(d,d_1,\ldots,d_r)\in\R^{r\times r}$, which satisfy $\bU^\top\bU=\bV^\top\bV=n\bI$; and for the terms with tilde, they are defined as follows
\begin{eqnarray}
\bU\bD\bV^\top 
&=& \begin{pmatrix}
\bU_1\bD\bV_1^\top &\; \bU_1\bD\bV_2^\top \\
\bU_2\bD\bV_1^\top &\; \bU_2\bD\bV_2^\top
\end{pmatrix}\\
&=& \begin{pmatrix}
\bU_1\Lambda_u\Lambda_u^{-1}\bD\Lambda_v^{-1}\Lambda_v\bV_1^\top &\; \bU_1\Lambda_u\Lambda_u^{-1}\bD\Lambda_v^{-1}\Lambda_v\bV_2^\top \\
\bU_2\Lambda_u\Lambda_u^{-1}\bD\Lambda_v^{-1}\Lambda_v\bV_1^\top &\; \bU_2\Lambda_u\Lambda_u^{-1}\bD\Lambda_v^{-1}\Lambda_v\bV_2^\top
\end{pmatrix}\\
&=&
\begin{pmatrix}
\tbU_1\tbD\tbV_1^\top &\; \tbU_1\tbD\tbV_2^\top \\
\tbU_2\tbD\tbV_1^\top &\; \tbU_2\tbD\tbV_2^\top
\end{pmatrix}\\
&=&  \tbU\tbD\tbV^\top,
\end{eqnarray}
where $\Lambda_u = \sqrt{n_1}(\mathrm{diag}(\bU_1^\top\bU_1))^{-1/2}, \Lambda_v = \sqrt{n_1}(\mathrm{diag}(\bV_1^\top\bV_1))^{-1/2}, \tbU_1 = \bU_1\Lambda_u,\tbU_2 = \bU_2\Lambda_u,\tbV_1 = \bV_1\Lambda_v,\tbV_2 = \bV_2\Lambda_v, \tbD = \Lambda_u^{-1}\bD\Lambda_v^{-1}, \tbU = (\tbU_1^\top, \tbU_2^\top)^\top, \tbV = (\tbV_1^\top, \tbV_2^\top)^\top$. The coefficients $\tbetau,\tbetav$ are the same as in the rank one case. 

}

\jcM{
Again, for the training data part, estimations $\htbU_1,\htbV_1,\htbD$ can be obtained via SuperCENT. 
To estimate $\tbU_2$ and $\tbV_2$,
first note that $\bA_{21}=\tbU_2\tbD\tbV_1^\top+\bE_{21}$
and 
$\bA_{12}=\tbU_1\tbD\tbV_2^\top+\bE_{12}$.
Given $\htbU_1, \htbV_1,\htbD$, minimizing the Frobenius norm of the approximation errors of $\bA_{21}$ and $\bA_{12}$ yields the estimates $\htbU_2$ and $\htbV_2$ respectively, i.e., 
\be 
\htbU_2 = \argmin_{\bU_2} \|\bA_{21}-\bU_2 \htbD \htbV_1^\top\|_F^2
\quad \mbox{and} \quad
\htbV_2  = \argmin_{\bV_2} \|\bA_{12} - \htbU_1 \htbD \bV_2^\top\|_F^2.
\ee 
Take the first columns of $\htbU_2$ and $\htbV_2$ as $\htbu_2$ and $\htbv_2$, which are the estimated centralities of the testing data.

Finally, with the testing covariates $\bX_2$, the estimated coefficient $\hbbeta_x,\htbetau,\htbetav$ from SuperCENT on training data, we predict 
$\hy_2 = \bX_2 \hbbeta_x + \htbu_2 \htbetau + \htbv_2 \htbetav$.

\begin{algorithm}[H]
\label{algo:pred-uv-svd-2}
\SetAlgoLined
\KwResult{$\htbu_2$ and $\htbv_2$.}
 \textbf{Input:} 
 The augmented network $\bA\in\mathbb{R}^{n\times n}$, $\htbD$, $\htbU_1$, and $\htbV_1$.
 \begin{enumerate}
    \item $\htbU_2 = \left((\htbV_1^\top \htbD^2 \htbV_1)^{-1}\htbD \htbV_1^\top \bA_{21}^\top\right)^\top, 
\htbV_2 = \left((\htbU_1^\top \htbD^2 \htbU_1)^{-1} \htbD \htbU_1^\top \bA_{12}\right)^\top$\;
    \item Take the first columns of $\htbU_2$ and $\htbV_2$ as ${\htbu_2}$ and ${\htbv_2}$ respectively.
 \end{enumerate}
 \caption{An improved method using SuperCENT estimate of $\bU_1$ and $\bV_1$ to obtain $\htbu_2$ and $\htbv_2$.}
\end{algorithm}

}

\subsection{Selection of the tuning parameter $\lambda$}
\label{sec:cv}

The tuning parameter $\lambda$ can be selected using {the} $K$-fold cross-validation.
Given the prediction procedure in Section \ref{app:prediction-algo}, the cross-validation procedure can be easily carried out as follows.
For each fold of validation data, we first fit the model using the remaining $K-1$ folds with the corresponding induced subnetwork and obtain the estimates for the regression coefficients by implementing Algorithm \ref{algo:supercent}; we then obtain the estimates of the centralities for the validation data by applying Algorithm S\ref{algo:pred-uv-svd}; we last obtain the total prediction error for the validation data by combining the outcomes from the first two steps. The best tuning parameter $\lambda$ is set to be the minimizer of the total cross-validation error that sums over all folds.
Algorithm \ref{algo:supercent-cv} outlines this procedure in more detail.

\begin{algorithm}
\SetAlgoLined
\KwResult{$\lambda_{min}$
.}
\textbf{Input:} $\A,\X,\y$.\\
 \For{$\lambda$ on a exponentially regular grid 
 }{
 \For{ each fold }{
 \begin{enumerate}
	 	\itemsep0em
  \item[0.] Split the covariates and response into training $\X_{fold, train}, ~\y_{fold, train}$ and \\ validation $\X_{fold, val}, ~\y_{fold, val}$ and denote the 
  {the induced sub-network} \\ corresponding to the training data $\A_{fold, train}$\;
  \item[1.] $(\hbbeta_{fold, \lambda}, ~\hu_{fold, \lambda}, \hv_{fold, \lambda}) \leftarrow$ \\ 
   \quad SuperCENT$(\A_{fold, train}, ~\X_{fold, train}, ~\y_{fold, train}, ~\lambda)$ by Algorithm \ref{algo:supercent}\;
  \item[2.] $\hu_{fold, val}, \hv_{fold, val} \leftarrow$ SVD$(\bA)$ and re-scale by Algorithm S\ref{algo:pred-uv-svd}\;
  \item[3.] $SSE_{fold, \lambda} = \|\y_{fold, val} - (\X_{fold, val}, \hu_{fold, val}, \hv_{fold, val})~ \hbbeta_{fold, \lambda} \|^2_2$\;
 \end{enumerate}


	}
 }

 $\lambda_{min} = \min_{\lambda} \sum_{fold} SSE_{fold, \lambda}$


 \caption{The cross-validation algorithm for {SuperCENT} to choose $\lambda$.}
 \label{algo:supercent-cv}
\end{algorithm}


{As another strategy,}
Remark \ref{rmk:supercent-plugin-lambda} offers an alternative way to choose the tuning parameter based on the theoretical analysis of SuperCENT, which is less time-consuming than cross-validation.
{However, we recommend using the cross-validation strategy for the best performance based on the simulation results. }

\begin{remark}
\label{rmk:cv}
The cross-validation procedure will not change the underlying network structure under our unified framework {if all the entries of the network noise $\bE$ are assumed to be i.i.d., given that the centralities $\bU$ and $\bV$ are fixed parameters. 
This is because the generative distribution of a subset of a matrix of low-rank mean with i.i.d. noise remains the same.
For cross-validation under other model assumptions}, one may consider the edge sampling procedure proposed by \cite{li2020network}
or an ``out-of-sample'' procedure based on embeddings \citep{levin2021limit}.
\end{remark}

\section{Theoretical properties of the two-stage procedure}
\label{app:theory-ts}


In this section, we present the theoretical properties for the two-stage estimator.
{Under Assumptions \ref{assump:normal}-\ref{assump:consistent},
the two-stage estimators are consistent, with their asymptotic distributions given in Theorem \ref{thm:two-stage-normality} and their convergence rates given in Propositions \ref{prop:two-stage-rate1} and \ref{prop:two-stage-rate2}.
The proof is deferred to Section \ref{sec:proof-ts}.

We first introduce the following notations. Let $\bK$ be the $n^2\times n^2$ commutation matrix  such that $\vec1(\bE^\top) = \bK \vec1(\bE)$ and 
$\otimes$ denote the Kronecker product.
Define $\tu = (\bI-\bP_{\bX})\bu$,
$\tv = (\bI-\bP_{\bX})\bv$, which are the centralities projected onto the orthogonal space of $\bX$.
Denote $c=\tu^\top\tu\tv^\top\tv - (\tu^\top\tv)^2$ and $
\bCuv =
 \left(\tu,~ \tv\right)^\top
 \left(\tu,~ \tv\right)
$, which will show up in the asymptotic expressions.



{
Recall that the two-stage estimates are denoted as 
$\hdts$, $\huts$, $\hvts$ and $\hbetats = ((\hbetaxts)^\top, \hbetauts,\hbetavts)^\top$. }

\begin{theorem}
\label{thm:two-stage-normality}

Under 
{the unified framework}
\eqref{eq:model-supercent} and Assumptions \ref{assump:normal}-\ref{assump:consistent}, the two-stage estimates have the following asymptotic distribution,
\begin{enumerate}[nolistsep]
\item Centralities: 
\be
\huts - \bu = \etauts + o(\etauts)
\quad\mbox{and}\quad
\hvts - \bv = \etavts + o(\etavts),
\ee
\item Network effect:
\be
\hbbetats - \bbeta = \etabetats + o(\etabetats)
= \left( (\eta_\betax^{ts})^\top, \etabetauts, \etabetavts\right)^\top + o\left(\left( (\eta_\betax^{ts})^\top, \etabetauts, \etabetavts \right)^\top \right),
\ee
\end{enumerate}
where $\left(
 \begin{array}{c}
  \etauts \\
  \etavts \\
  \etabetats
 \end{array}
\right) \sim 
N\Big(
\zero_{(2n+2+p)\times 1}, 
\bCts\left(
              \begin{array}{cc}
                \sigma_y^2\bI_n & \zero_{n\times n^2} \\
                \zero_{n^2\times n} & \sigma_a^2 \bI_{n^2} \\
              \end{array}
              \right){\bCts}^\top
\Big)$,
$\frac{\|o\left(\etauts\right)\|}{\|\etauts\|} \overset{P}{\longrightarrow} 0$,
$\frac{\|o\left(\etavts\right)\|}{\|\etavts\|} \overset{P}{\longrightarrow} 0$,
$\frac{\left\|o\left(\etabetaxts\right)\right\|}{\left\|\etabetaxts\right\|} \overset{P}{\longrightarrow} 0$,
$\frac{|o\left(\etabetauts\right)|}{|\etabetauts|} \overset{P}{\longrightarrow} 0$,
$\frac{|o\left(\etabetavts\right)|}{|\etabetavts|} \overset{P}{\longrightarrow} 0$,
and 
$\bCts = \left(
  \begin{array}{cc}
    \bCts_{11} & \bCts_{12} \\
    \bCts_{21} & \bCts_{22} \\
    \bCts_{31} & \bCts_{32} \\
    \bCts_{41} & \bCts_{42} \\
    \bCts_{51} & \bCts_{52} 
  \end{array}
\right)$ whose specific forms are as follows.

The matrices related to $\huts$ and $\hvts$ are
\be
\left(
 \begin{array}{c}
  \bCts_{12} \\
  \bCts_{22} \\
 \end{array}
\right)
=
(d n)^{-1}
\left(
 \begin{array}{c}
	 \bv^\top \otimes (\bI-\pu) \\
   \left(\bu^\top \otimes (\bI-\pv)\right)\bK \\
 \end{array}
\right),
\ee
the matrices related to $\hbetauts$ and $\hbetavts$ are
\be
\left(
 \begin{array}{cc}
  \bCts_{41} & \bCts_{42} \\
  \bCts_{51} & \bCts_{52} \\
 \end{array}
\right)
 =
 \bCuvi
\left(
 \begin{array}{c}
  \tu^\top \\
  \tv^\top \\
 \end{array}
\right)
\left(-\betau\bI_n~~ -\betav\bI_n ~~ \bI_n\right)
\left(
 \begin{array}{cc}
  \zero_{n\times n} & \bCts_{12} \\
  \zero_{n\times n} & \bCts_{22} \\
  \bI_n & \zero_{n\times n^2}
 \end{array}
\right),
\ee
and the matrices related to $\hbetaxts$ are
\be
 \left(
 \begin{array}{cc}
  \bCts_{31} & \bCts_{32} \\
 \end{array}
\right) =
(\bX^\top\bX)^{-1}\bX^\top
 \left(
 -\betau\bI_n ~~ -\betav\bI_n ~~ -\bu ~~ -\bv ~~ \bI_n
\right)
\left(
 \begin{array}{cc}
\zero_{n\times n} & \bCts_{12} \\
\zero_{n\times n} & \bCts_{22} \\
  \bCts_{31} & \bCts_{32} \\
  \bCts_{41} & \bCts_{42} \\
  \bI_n & \zero_{n\times n^2}
 \end{array}
\right).
\ee
\end{theorem}


Recall the two-stage procedure first estimates the centralities $\bu$ and $\bv$
and then plugs the estimated centralities into the regression model. 
Therefore, the asymptotic distributions for $\huts$  and $\hvts$ only depend on the noise $\bE$ from the network model, not the regression noise $\bepsilon$. 
This can also be seen in the definitions of $\bsigmats_\bu$ and $\bsigmats_\bv$, all of which involve only $\sigma_a^2$, but not $\sigma_y^2$.
On the other hand, the asymptotic variance of $\hbetats$, $\bsigmats_\bbeta$, involves both $\sigma_y^2$ and $\sigma_a^2$, {not just $\sigma_y^2$}.
We will highlight this fact 
in Remarks \ref{rmk:random-uv} below.

\vspace{1em}
\begin{remark}
\label{rmk:random-uv}
(Non-classical covariance of $\hbetats$)
One important fact to emphasize is that the covariance of 
{$\hbetats$}
is \textit{not} $\sigma_y^2 (\hW^\top\hW)^{-1}$ where $\hW = (\bX, \huts,\hvts)$, which {\emph{would be}}
{the classical results of regression}
{\emph{if}} $\bX, \huts,\hvts$ {\emph{were}} considered {given and} fixed. This makes sense, as in our model, the observed network contains noise, which makes the estimated centralities $\huts,\hvts$ from the first stage random quantities and invalidates the {classical} result.
\end{remark}

\vspace{1em}

{Following Theorem \ref{thm:two-stage-normality}, we present the convergence rates of
the estimated centralities $\huts$  and $\hvts$ in Proposition \ref{prop:two-stage-rate1}.}
{To measure the difference between any estimate $\hu$ and the true $\bu$, one typically uses} the loss function $\| \bP_{\hu} - \bP_{\bu}\|_2^2$, which equals the squared sine of the angle between $\hu$ and $\bu$, $\sin^2 \angle(\hu,\bu)$. However, the exact form of this loss function is 
{not clean mathematically}. Instead, we use the scaled Euclidean distance $\|\hu-\mathrm{sign}(\hu^\top\bu)\bu\|_2^2/n = 2- 2\cos^2\angle(\hu,\bu)$, which has a cleaner expression and is connected to the squared sine through $\| \bP_{\hu} - \bP_{\bu}\|_2^2 = (\|\hu-\mathrm{sign}(\hu^\top\bu)\bu\|_2^2/n ) [1-(\|\hu-\mathrm{sign}(\hu^\top\bu)\bu\|_2^2/n)/4]$.
These two losses are approximately equivalent when the estimator is consistent, i.e., the loss goes to zero. 
In the following, the theorems are for $\argmax_{\mathbf{h}\in \{\huts,-\huts\}} \mathrm{sign}(\mathbf{h}^\top\bu) $ and $\argmax_{\mathbf{g}\in \{\hvts,-\hvts\}} \mathrm{sign}(\mathbf{g}^\top\bv) $, and we still use $\huts,\hvts$ to denote them. While these notions are a bit of an abuse of notation, it is reasonable since both the objective function and algorithm are sign-invariant (i.e., proper flipping of signs gives the same value or another valid iteration sequence). The same notation applies in the simulations as well.

\begin{proposition}
\label{prop:two-stage-rate1} (Convergence rates of $\huts$  and $\hvts$)
	Under the unified framework \eqref{eq:model-supercent} assume $\bA_0$ to be rank-one and Assumptions \ref{assump:normal}-\ref{assump:consistent},
	the two-stage estimators satisfy the following,
\be
\frac{1}{n}\E \|\huts - \bu\|^2_2 &=&
\frac{1}{n}\E \|\hvts - \bv\|^2_2 =
\frac{\sigma_a^2(n-1)}{d^2 n^2}(1+o(1)) \\
&=& \frac{\sigma_a^2}{d^2 n} (1+o(1))  = \netsnr(1+o(1)).
\label{eq:two-stage-mse-uv-exact}
\ee
\end{proposition}
The convergence rate of the two-stage depends on the interplay of three parameters:  $d,\sigmaa, n$.
When the signal strength $d$ and the noise level $\sigmaa$ are of constant order while $n$ diverges, the two-stage converges at a fast rate.
However, it is highly possible that the two-stage may converge slowly for the following reasons: 1) it is understandable that the observed network might become noisier with more nodes, i.e., large $\sigmaa$, because it may become exponentially costly to collect data for a larger network of the same level of noise; 2) the signal strength $d$ decays as more nodes are included into the network, since the network edge density might decay with more nodes.

We next present the convergence rates of the regression coefficients $\hbetauts,\hbetavts,\hbetaxts$ 
in Proposition \ref{prop:two-stage-rate2}.}

\begin{proposition}
\label{prop:two-stage-rate2}
({Asymptotic property of $\hbetats$})
	Under 
	{the unified framework}
	\eqref{eq:model-supercent} and Assumptions \ref{assump:normal}-\ref{assump:consistent}, the two-stage estimators satisfy
{
\scriptsize

\be
\E (\hbetauts - \betau )^2
&=& \left(\frac{\sigma_y^2 }{c} \tv^\top\tv \right.\label{eq:two-stage-betau-exact-term1}\\
&& \left.+ \frac{\sigma_a^2}{c^2} \frac{1}{d^2n}
\left[
\beta_v^2 \tv^\top\tv\tu^\top\opv\tu\tv^\top\tv
+ \beta_u^2 \tu^\top\tv\tv^\top \opu \tv\tu^\top\tv
\right]\right)(1+o(1))
\label{eq:two-stage-betau-exact}\\
&=&O\left(\frac{\sigma_y^2}{n} +
\frac{\sigma_a^2(\beta_u^2 + \beta_v^2) }{d^2n^2} \right),
\label{eq:two-stage-mse-betau-rate}\\
\E (\hbetavts - \betav )^2
&=& \left(\frac{\sigma_y^2 }{c} \tu^\top\tu\right.\label{eq:two-stage-betav-exact-term1}\\
&& \left.+ \frac{\sigma_a^2}{c^2} \frac{1}{d^2n}
\left[
\beta_u^2 \tu^\top\tu\tv^\top\opu\tv\tu^\top\tu
+ \beta_v^2 \tv^\top\tu\tu^\top \opv \tu\tv^\top\tu
\right]\right)(1+o(1)) \label{eq:two-stage-betav-exact}\\
&=&O\left(\frac{\sigma_y^2}{n} +
\frac{\sigma_a^2(\beta_u^2 + \beta_v^2) }{d^2n^2} \right), 
\label{eq:two-stage-mse-betav-rate} \\
Cov\left(\hbetaxts - \Bbetax\right)
&=&
\sigma_y^2
\left[
(\bX^\top\bX)^{-1}
+ (\bX^\top\bX)^{-1}\bX^\top
\left(
\begin{array}{cc}
	\bu & \bv
\end{array}
\right)
\bCuvi
\left(
 \begin{array}{c}
  \bu^\top \\
  \bv^\top
 \end{array}
\right)
\bX(\bX^\top\bX)^{-1}
\right]
\label{eq:two-stage-betax-exact-part1}\\
\hspace{-1in}
&& 
+ \sigma_a^2
\frac{1}{d^2n}
(\bX^\top\bX)^{-1}\bX^\top
\Bigg[
\beta_u^2 \opu + \beta_v^2 \opv
\label{eq:two-stage-betax-exact-part2}\\
\hspace{-1in}
&&
+
\left(
\begin{array}{cc}
	\bu & \bv
\end{array}
\right)
\bCuvi
\left(
\begin{array}{cc}
	\beta_v^2\tu^\top\opv\tu & 0\\
	0 & \beta_u^2\tv^\top\opu\tv \\
\end{array}
\right)
\bCuvi
\left(
 \begin{array}{c}
  \bu^\top \\
  \bv^\top
 \end{array}
\right)
\Bigg] \\
&& \hspace{2.5in}
\bX(\bX^\top\bX)^{-1}(1+o(1)).
\label{eq:two-stage-betax-exact-part3}
\ee
}
\end{proposition}

\begin{remark}(Comments on \eqref{eq:two-stage-betau-exact-term1}-\eqref{eq:two-stage-mse-betau-rate} for $\hbetauts$)
\label{rmk:compare-two-betau}
For the variance of $\hbetauts$, the first term \eqref{eq:two-stage-betau-exact-term1} is
the same as the variance in the classical regression results with deterministic predictors by treating $\huts$ and $\hvts$ as given and fixed.
The additional term \eqref{eq:two-stage-betau-exact} is 
due to
the randomness nature of $\huts$ and $\hvts$. Note that the second term \eqref{eq:two-stage-betau-exact} is non-negative and it becomes zero if $\sigmaa=0$, or $\tu\perp\tv$. 
Therefore, if ad-hoc inference were made by only considering the first term \eqref{eq:two-stage-betau-exact-term1} as in Remark \ref{rmk:ts-inference}, it 
would be valid only if the network is noiseless or the hub and authority centralities are completely orthogonal after accounting for the 
{covariates $\bX$, which are rarely the case in real networks.}
Similar messages can be obtained for the variance of $\hbetavts$
and covariance of $\hbetaxts$.
\end{remark}

\section{More simulation results}
\label{app:more-sim}
In this section, we show more simulation results that are deferred from Section \ref{sec:sim}.
Section \ref{sec:more-sim-inconsistent} provides additional results for Section \ref{sec:sim-incons}.
We then show a phase-transition experiment in Section \ref{sec:more-sim-high-rank} to demonstrate the behaviors of SuperCENT and the two-stage estimators under different network signal-to-noise ratios.

To give an overview of the simulation results, 
Table \ref{tab:sum-ts-vs-supercent} summarizes the comparison of the two-stage and SuperCENT from the perspectives of both estimation and inference.
SuperCENT universally outperforms the two-stage in terms of centrality estimation, regression coefficients estimation, and inference.

\begin{table}
\centering
\caption{Comparison of the two-stage and SuperCENT when the network signal-to-noise ratio is low and ${\beta_u^2}/{\sigma_y^2}\gg \beta_v^2/\sigma_y^2$. 
If 
$\beta_v^2/\sigma_y^2\gg \beta_u^2/\sigma_y^2$, the results for $\bu$ and $\bv$, $\betau$ and $\betav$ will be switched. 
In the estimation panel, \cmark~indicates accurate estimation, \xmark~indicates inaccurate estimation.
In the inference panel, \cmark~indicates that the empirical coverage of confidence interval is no less than the nominal level, \xmark~indicates that the confidence interval fails to reach the nominal level.
{In each row, SuperCENT is underlined whenever it outperforms the two-stage.}}
\label{tab:sum-ts-vs-supercent}
\begin{tabular}{ l|c|cc } 
\toprule
\multicolumn{2}{l|}{} & Two-stage & SuperCENT \\
\midrule \midrule
\multicolumn{2}{l}{} & \multicolumn{2}{c}{Estimation} \\
\midrule
\multicolumn{2}{c|}{$\bu$} & \xmark & \cmark \\
\multicolumn{2}{c|}{$\bv$} & \xmark & \xmark \, ({\underline{Slightly Improved}})\\
\multicolumn{2}{c|}{$\bA_0$} & \xmark & \underline{\cmark} \\
\multicolumn{2}{c|}{$\betau$} & \xmark & \underline{\cmark} \\
\multicolumn{2}{c|}{$\betav$} & \xmark & \xmark \\
\midrule \midrule
\multicolumn{2}{l}{} & \multicolumn{2}{c}{Inference} \\
\midrule
\multicolumn{2}{c|}{$CI_\betau$} & \xmark & \underline{\cmark}  \\
\multicolumn{2}{c|}{{$CI_\betav$}} & \xmark & \xmark  \\
\multicolumn{2}{c|}{$CI_{a_{ij}}$} & \xmark & \underline{\cmark}  \\
\bottomrule
\end{tabular}

\end{table}

\subsection{Additional results for Section \ref{sec:sim-incons}}
\label{sec:more-sim-inconsistent}

Section \ref{sec:sim-incons} shows that SuperCENT greatly improves over two-stage in
 terms of estimation of $\bu$ and $\betau$ as well as the inference of $\betau$.
In this section, we demonstrate the behaviors of the SuperCENT-based and the two-stage-based estimators of $\bv$, $\bA_0$, and $\betav$ and their corresponding confidence intervals. 

\plotfig[.9]{\simpathtwo}{v}{The boxplot of $\logl{\hv}{\bv}$ for the four estimators across different $\sigmaa$, $\sigmay$ and $\betau$ with $d = 1$ and $\betav = 1$. 
where $l(\hv,{\bv}) = \| \bP_{\hv} - \bP_{\bv}\|^2_2$. The super-imposed red symbols show the theoretical rates of the two-stage in Proposition \ref{prop:two-stage-rate1} and SuperCENT in Proposition \ref{prop:supercent-rate1}.}{fig:v-2}

\plotfig[.9]{\simpathtwo}{A_norm}{The boxplot of $\log_{10}(l(\hA,\bA_0))$ for four estimators across different $\sigmaa$, $\sigmay$ and $\betau$ with $d = 1$ and $\betav = 1$
{where $l(\hA,{\bA_0}) = \|\hA - \bA_0\|_F^2/\|\bA_0\|_F^2$. 
The super-imposed red symbols show the theoretical rates of the two-stage in Proposition \ref{prop:two-stage-rate1} and SuperCENT in Proposition \ref{prop:supercent-rate1}.}
}{fig:A-2}

\plotfig[.9]{\simpathtwo}{rate_betau}{The boxplot of $\log_{10}(l(\hbetau,\betau))$ for four estimators across different $\sigmaa$, $\sigmay$ and $\betau$ with $d = 1$ and $\betav = 1$ where $l(\hbetau,\betau) = (\hbetau - \betau)^2/\beta_u^2$.
The super-imposed red points show the median of $\log_{10}(l(\hbetau,\betau))$.
}{fig:betau-2}

 For the estimation of $\bv$ shown in Figure \ref{fig:v-2}, the improvement of SuperCENT over two-stage is not as large as that of the estimation of $\bu$ when $\betau \in 2^{2,4}$, because 
{$\frac{\beta_u^2}{\sigma_y^2}\gg \frac{{\beta_v^2}}{\sigma_y^2}$.}
But the improvement is still quite significant when $\betau=2^0\approx \betav=1$.
It is worth noting that the supervised effect to $\hv$ shrinks as $\betau$ increases, leading to a different trend comparing Figures \ref{fig:u-2} and \ref{fig:v-2}.
This phenomena aligns with Remark \ref{rmk:inconsistent}
where we discuss the estimation of $\bu$ and $\bv$ comparing SuperCENT and the two-stage.
Specifically, the roles of $\bu$ and $\bv$ are not exchangeable, 
because here we have $\betav \leq \betau$ by fixing $\betav=1$ and varying $\betau\in 2^{0,2,4}$.
On the other hand, when $\betav \gg \betau$ we should expect the improvement in estimating $\bv$ to increase.

{The conclusion for the estimation of $\bA_0$ is similar to that of $\bu$  as shown in Figure \ref{fig:A-2}.  }
With the improvement from estimating $\bu$ and $\bv$, $\supercentcv$ estimates $\bA_0$ more accurately across all the settings. As claimed in Remark \ref{rmk:inconsistent}, the convergence of $\hA$ in this regime only requires 
{$\frac{\beta_u^2}{\sigma_y^2}\rightarrow \infty$ or 
$\frac{\beta_v^2}{\sigma_y^2}\rightarrow \infty$.}
Therefore, with $\betav = 1$, $\betau > 1$, $\hA$ converges and $l(\hA,\bA_0) < l(\hAts,\bA_0)$. Comparing Figures \ref{fig:u-2}, \ref{fig:v-2} and \ref{fig:A-2} altogether, when $\betau=2^0$, 
{SuperCENT improves the estimation of both $\bu$ and $\bv$ significantly;}
when $\betau\in 2^{2,4}$, SuperCENT improves the estimation of $\bu$ a lot; therefore, SuperCENT improves the estimation of $\bA_0$ a lot for all {the} ranges of $\betau$.

Figure \ref{fig:betav-bias-2} shows $\hbetav - \beta_v$. 
We observe an over-estimation of $\betav$ when $\betau$ is large,
which is different from the under-estimation in $\hbetau$.
The inaccuracy is larger as $\betau$ increases. 
SuperCENT in this case can still improve the accuracy of the estimation of $\betav$
but the improvement is not as large as that of $\betau$
since the improvement in estimation of $\bv$ when $\betau$ is large is limited as shown in Figure \ref{fig:v-2},
thereby slightly improving over the two-stage from the perspective of $\betav$ estimation in Figure \ref{fig:betav-bias-2} 
as well as the squared error loss in Figure  \ref{fig:betav-2}.


\plotfig[0.9]{\simpathtwo}{bias_betav}{The boxplot of $\hbetav - \beta_v$ across different $\sigmaa$, $\sigmay$ and $\betau$ with fixed $d = 1$ and $\betav = 1$. The dashed lines show $\hbetav - \betav = 0$.
The super-imposed red points show the median of $\hbetav - \beta_v$.
}
{fig:betav-bias-2}

\plotfig[0.9]{\simpathtwo}{rate_betav}{The boxplot of $\log_{10}(l(\hbetav,\betav))$ across different $\sigmaa$, $\sigmay$ and $\betau$ with $d = 1$ and $\betav = 1$ where $l(\hbetav,{\betav}) = (\hbetav - \betav)^2/\beta_v^2$. 
The super-imposed red points show the median of $\log_{10}(l(\hbetav,\betav))$.
}{fig:betav-2}


{
Similar to $\betau$, the inaccurate estimation of $\betav$ further affects its confidence interval. 
Figures \ref{fig:confint-v-2} and \ref{fig:width-v-2} show the empirical coverage and $\log_{10}$ of the average width, respectively, of the 95\% confidence interval for $\betav$. 
For the empirical coverage, when $\sigmaa$ remains small (top panels), all methods remain close to the nominal level, 
again with  different reasons for different methods as discussed for the coverage of $CI_{\betau}$.
When $\sigmaa$ remains small, the inaccuracy in $\hbetav$ is relatively small, leaving a relatively small impact on the coverage. 
As $\betau$ increases, the oracles tend to under cover since the inaccuracy increases. 
Two-stage is still conservative because  $\sigma_y^2$ is overestimated, covering up the issue of inaccurate estimation.
Similar to the phenomena we observed in the consistent regime, 
the two-stage-adhoc is on par with the two-stage-oracle even with $\sigma_y^2$ being overestimated
and thus below the nominal level.
When $\sigmaa$ increases
and the inaccuracy is not too severe with $\betau = 2^2$ (the mid-bottom panel), 
the two-stage is still close to the nominal level while two-stage-oracle and two-stage-adhoc are no longer valid.
SuperCENT does not improve much from two-stage as the improvement of estimating $\betav$ is limited. 
The trade-off between $\betau$ and $\betav$ for SuperCENT is desirable 
--- SuperCENT provides valid and shorter intervals for both $\betau$ and $\betav$ if 
$\betau$ and $\betav$ are similar; 
if $\betau$ and $\betav$ differ a lot, $\supercent$~provides a valid and shorter interval for the larger effect which is more of one's interest.

As for the width of the CI for $\betav$, Figure \ref{fig:width-v-2} shows that when the SuperCENT methods reach the nominal level,
the widths are shorter than two-stage. 


}

\plotfig[0.9]{\confintpathtwo}{cov_v}{Empirical coverage of $CI_\betav$ across different $\sigmaa$, $\sigmay$ and $\betau$ with $d = 1$ and $\betav = 1$.
$\supercent$ variants are labelled as circles ($\circ\;\bullet$) and the two-stage variants are labelled as triangles ($\vartriangle\mathlarger{\mathlarger{\mathlarger{\blacktriangledown}}}\;\blacktriangle$). The hollow ones are for oracles and the solid ones are for non-oracles.
The dashed lines show the nominal confidence level 0.95.
}{fig:confint-v-2}

\plotfig[0.9]{\confintpathtwo}{width_v}{Width of $CI_{\beta_v}$ across different $\sigmaa$, $\sigmay$ and $\betau$ with $d = 1$ and $\betav = 1$.
$\supercent$ variants are labelled as circles ($\circ\;\bullet$) and the two-stage variants are labelled as triangles ($\vartriangle\mathlarger{\mathlarger{\mathlarger{\blacktriangledown}}}\;\blacktriangle$). The hollow ones are for oracles and the solid ones are for non-oracles.
}{fig:width-v-2}

Finally, we investigate the average coverage and the average width of confidence intervals for all the entries $a_{ij}$ of $\bA_0$ respectively. 
The average coverage probability of all the methods, $\text{Average}_{ij} \big( \text{CP}(\text{CI}{a_{ij}}) \big)$,
achieves the nominal level of 95\%.
The coverage tends to be slightly below the nominal coverage as $\sigmaa$ increases,  
because the estimation becomes worse and the theorem only holds up to $1+o(1)$. 
$\supercentcv$ is the closest to the nominal coverage in all the settings compared to the others.
Figure \ref{fig:width-A-2} shows the $\log_{10}$ of the average width of the CIs, $\text{Average}_{ij} \big( \text{Width}(\text{CI}{a_{ij}}) \big)$. $\supercentoracle$-oracle provides the shortest width among the four methods, followed by $\supercentcv$. The widths of the confidence intervals of both SuperCENT-based methods are shorter than those of the two-stage methods. Again, the improvement of SuperCENT over the two-stage increases as $\sigmaa$ and $\betau$ increase or $\sigmay$ decreases.

\plotfig[1]{\confintpathtwoold}{cov_A}{The average empirical coverage  $\text{Average}_{ij} ( \text{CP}(\text{CI}_{a_{ij}} ) )$
across different $\sigmaa$, $\sigmay$ and $\betau$ with $d = 1$ and $\betav = 1$.
$\supercent$ variants are labelled as circles ($\circ\;\bullet$) and the two-stage variants are labelled as triangles ($\vartriangle\mathlarger{\mathlarger{\mathlarger{\blacktriangledown}}}\;\blacktriangle$). The hollow ones are for oracles and the solid ones are for non-oracles.
The dashed lines show the nominal confidence level 0.95.
}{fig:confint-A-2}
\plotfig[1]{\confintpathtwoold}{width_A}{$log_{10}$ of the average width of $CI_{a_{ij}}$ across different $\sigmaa$, $\sigmay$ and $\betau$ with $d = 1$ and $\betav = 1$.
$\supercent$ variants are labelled as circles ($\circ\;\bullet$) and the two-stage variants are labelled as triangles ($\vartriangle\mathlarger{\mathlarger{\mathlarger{\blacktriangledown}}}\;\blacktriangle$). The hollow ones are for oracles and the solid ones are for non-oracles.
}{fig:width-A-2}

\subsection{Results of a phase-transition experiment}
\label{sec:more-sim-high-rank}

To further study the behaviors of SuperCENT and the two-stage and understand the advantages of SuperCENT under different network signal-to-noise ratios (SNRs), we perform a phase-transition experiment.

The simulation setup is similar to that in Section \ref{sec:sim-setup},
except that we only vary $\sigma_a \in 2^{1,1.25,\ldots,5}$ as well as the gap between the leading singular values and the non-leading ones, while fixing all the other parameters to investigate the continuous impact of the network noise level or equivalently the network SNR.
Specifically, we generate a network with $n = 256 = 2^8$ following the network model \eqref{eq:model-supercent1}, 
i.e., $\bA = {\bA_0 = } \bU \bD \bV^\top + \bE$
where $\bU, \bV \in\mathbb{R}^{n\times r}$, $\bD$ is a diagonal matrix of dimension $r\times r$ with the singular values $d > d_2 \geq  \ldots \geq d_r \geq 0$ as the diagonal entries, and all the entries of $\bE$ follow $N(0,\sigma_a^2)$ independently.
All entries of $\bU$ and $\bV$ are first generated from i.i.d. $N(0,1)$, then applied Gram?Schmidt to ensure orthogonality between columns, and finally each column of $\bU$ and $\bV$ is rescaled to have length $\sqrt{n}$.
We consider the case of $r = 10$ where the leading singular value $d = 1$ and the non-leading ones as $d_2 = \ldots = d_r \in \{0, 2^{-1}, 2^{-0.5}\}$. Note that when $d_2 = \ldots = d_{10} = 0$, it is a rank-one setting. 
We include this setting so as to compare the single rank setting with the multiple ranks in the network model. 
For the regression model, $\by = \bX\bbeta_x + \bu \betau + \bv \betav + \bepsilon$, we  include the covariate matrix $\bX$ and the hub and authority centralities, namely the leading singular vectors $\bu=\bu_1$ and $\bv=\bv_1$ 
instead of the entire $\bU$ and $\bV$.

We compare the performance of SuperCENT and the two-stage in terms of \emph{estimation accuracy} and \emph{inference property}. 
For \textit{estimation accuracy}, we compare the estimation error of the hub centrality $\bu$ and the hub effect $\betau$ between the two-stage and $\supercentoracle$ (labeled as SuperCENT); 
for \textit{inference property}, we compare the coverage probability and the width of $CI_{\betau}$ between the two-stage-adhoc (labeled as two-stage) and $\supercentoracle$-oracle (labeled as SuperCENT).

\plotfig[1]{\local plot/1992036-1991851_}{epsy-2_small_exmaple_rank_10_d_hat}{Top 20 squared singular value estimates using SVD under settings with $d_1 = 1$, $\sigma_a = 2$ and different non-leading singular values, i.e., $d_2 = \ldots = d_{10} \in \{0,\, 2^{-1}=0.5,\, 2^{-0.5}\approx 0.7 \}  $.
}{fig:multi-rank-d-hat}

Before comparing the performance of the two-stage and SuperCENT, we first show the top 20 squared singular values 
of the observed network adjacency matrix $\bA$ with fixed leading singular value $d_1=1$ for the noiseless component {$\bA_0$}, fixed noise level $\sigma_a=2$ and various non-leading singular values for {$\bA_0$} in Figure \ref{fig:multi-rank-d-hat}.
The purpose of these plots are to make connections and comparisons to the spectral properties of the four real networks in Section \ref{app:net-exmaples}. 
Under the rank-one setting when the noiseless singular values of $\bA_0$ as
$d_1 = 1$ and $d_2 = \ldots = d_{10} = 0$ (the plot on the left), 
the leading singular value $\hd^{ts}_1$ {of the observed network $\bA$,} {obtained via simple SVD or two-stage,}  dominates the non-leading ones, 
and the non-leading ones slowly decay, 
which resembles the spectral structure of the innovation network in Figure \ref{fig:four-examples}B.
For the multi-rank settings when $d = 1$ and $d_2 = \ldots = d_{10} \in \{2^{-1},\, 2^{-0.5}\}$ (the two plots on the right), 
the leading singular value $\hd^{ts}$ still dominates the non-leading ones,
and the non-leading ones are separated into two groups: $\{ \hd^{ts}_2,\ldots, \hd^{ts}_{10} \}$ as the ``signal'' group and the rest as the ``noise'' group.
{As $d_2, \ldots, d_{10}$ get larger, the two-stage estimates of the signal group are also larger as expected, and the gap between the last signal singular value $\hd^{ts}_{10}$ and the first noise singular value $\hd^{ts}_{11}$ is larger. }
The multi-rank setting {with small noiseless non-leading singular values $d_2,\ldots, d_{10}$, especially the middle plot}, resembles the
 global trade network,
production network, and 
equity network in Figures \ref{fig:four-examples}A, C, and D.

\plotfig[1]{\local plot/2023120312-2023120313_}{epsy-2_small_exmaple_rank_10}{Comparison between the two-stage and SuperCENT in terms of the estimations of $\bu$ and $\betau$ as well as the coverage probability and the width of $CI_{\betau}$ varying the network noise level $\sigmaa$ and the gap between the leading singular value and the non-leading ones.
The performance of the two-stage is shown in the red and SuperCENT in the green.
The solid line corresponds to the rank-one setting with $d_1=1$ and $d_2 = \ldots = d_{10}=0$,
while the dashed and dotted lines correspond to the multi-rank setting with $d_1=1,  d_2 = \ldots = d_{10}=2^{-1}=0.5$ and $d_1=1,  d_2 = \ldots = d_{10}=2^{-0.5}\approx 0.7$ respectively.
Subfigure A shows the estimation error of the hub centrality $\sin(\angle(\hu, \bu))$, i.e., sine of the angle between the true hub centrality $\bu$ and the estimate $\hu$.
Subfigure B shows $\hbetau - \betau$.
Subfigures C and D show the coverage probability and the width of the 95\% confidence interval of the hub centrality coefficient, $CI_\betau$, respectively. 
}{fig:multi-rank}


Figure \ref{fig:multi-rank}A shows the estimation error of the hub centrality $\sin(\angle (\hu, \bu))$, i.e., sine of the angle between the true hub centrality $\bu$ and the estimate $\hu$.
The relationship between $\sin(\angle (\hu, \bu))$ and $\| \bP_{\hu} - \bP_{\bu}\|_2^2$ has been discussed in Section \ref{app:theory-ts}. We use $\sin(\angle (\hu, \bu))$ for the propose of demonstration. 
The red curves correspond to the two-stage and the green ones correspond to SuperCENT; while the solid curves correspond to the rank-one setting and the dashed and dotted curves correspond to the multi-rank setting.
For the two-stage, the estimation error increases as the network noise $\sigma_a$ increases and $\huts$ eventually becomes orthogonal to $\bu$, i.e., $\sin(\angle (\huts, \bu))$ approaching 1. 
Comparing settings with different non-leading singular values among the two-stage estimates (red), 
the estimation error of the rank-one setting (solid) is smaller than 
those of the multi-rank settings (dashed, dotted), and
the estimation error is larger as $d_2 = \ldots = d_{10}$ get closer to $d_1$. 
The SuperCENT estimates $\hu$, on the other hand, have much smaller estimation errors in all settings 
and the supervision effect persists regardless of the rank of the network.
Comparing settings with different non-leading singular values among the SuperCENT estimates (green), the estimation accuracy is quite similar when $\sigma_a$ is smaller.
As $\sigma_a$ gets larger, 
the SuperCENT estimates of the multi-rank settings (dashed, dotted) are a bit less accurate than that of the rank-one setting (solid).

Figure \ref{fig:multi-rank}B shows $\hbetau-\betau$. 
The hub effect estimate of two-stage $\hbetauts$ is inaccurate due to the inaccurate centrality estimation as shown in Figure \ref{fig:multi-rank}A.
The larger the network noise $\sigma_a$ or noise-to-signal ratio $\netsnr$, the larger the estimation inaccuracy. Comparing settings with different non-leading singular values, the inaccuracy is larger  as $d_2 = \ldots = d_{10}$ get closer to $d$. 
For SuperCENT, the estimate of the hub effect $\hu$ is accurate until around $\sigma_a = 2^{-3.75}$. 
The estimation inaccuracy are slightly larger under the multi-rank settings than that of the rank-one setting,
but still much smaller than that of the two-stage.

As of the inference property, Figures \ref{fig:multi-rank}C-D show the coverage probability and the width of the 95\% confidence interval of the hub centrality coefficient, $CI_\betau$, respectively. 
When $\sigma_a \in 2^{1,1.25, 1.5}$ or the network noise is relatively small,
the two-stage confidence intervals are conservative and wider than SuperCENT. As $\sigma_a$ increases, the coverage probability of the two-stage confidence intervals sharply drop to zero. 
The SuperCENT confidence intervals, on the contrary, remain valid until $\sigmaa$ becomes large and are narrower than the two-stage confidence interval.
In particular, the confidence intervals under the multi-rank settings are slightly more conservative than that under the rank-one setting and the coverage probability is closer to the nominal 95\% as the gaps between the leading singular value $d_1$ and the non-leading ones become larger.

In sum, SuperCENT outperforms the two-stage in terms of both estimation accuracy and inference property and thus SuperCENT should be preferred over the two-stage regardless under different network SNRs.

\section{Details on the case study in Section \ref{sec:case}}
\label{app:case}

In this section, we provide details on data construction in Section \ref{app:case-data}, additional results and interpretations on the authority centrality in Section \ref{app:case-aut},
with more information on the case study in Section \ref{app:case-info}.

\subsection{Data construction}
\label{app:case-data}

In the case study, we consider a triplet of $\{\bA, \bX, \by\}$, where $\bA$ is the country-level trade network, $\by$ is the currency risk premium, and $\bX$ is 
the share of world's GDP.
All these quantities are not directly available, and we compute them according to \citet{richmond2019trade} as follows.

To compute the currency risk premium, we obtain the interest rates and the exchange rates from DataStream. 
The currency risk premium can be calculated 
as follows.
For an investor going long in a country/region $i$, the log risk premium ``rx'' at time $t+1$ is 
$y_{i,t+1} := r_{it} - r_{t} - \Delta q_{i,t+1}$,
where $r_{it}$ is log interest rate of country/region $i$, $r_t$ is the log interest rate of the U.S. and $ \Delta q_{i,t+1}$ is the appreciation of U.S. dollar. 
Only 25 countries/regions have exchange rates available during the period of interest. We exclude the region of Europe as it is not comparable to the others in the trade network, resulting in 24 countries/regions.\footnote{The list of country abbreviations is provided in Section \ref{app:case-info}.}
We use a 5-year moving average of the currency risk premium:
when considering year $t$, average is taken from year $t-4$ to year $t$. 

\cite{richmond2019trade} defined the trade linkage as the trade amount normalized by the pair-wise total GDP, which represents the relative trade (export/import) intensity between two countries. Specifically, the trade linkage between two countries is computed as
$a_{ijt}
= \frac{S_{ijt}}{GDP_{it} + GDP_{jt} }$,
where $S_{ijt}$ is the dollar value of goods and commodities exported from country $i$ to $j$ at time $t$, and $GDP_{it}$ is the GDP of country $i$ at time $t$ in U.S. dollar.\footnote{
The bilateral trade data comes from the correlates of war project (COW) 
\citep{barbieri2009trading} and the International Monetary Fund (IMF) Direction of Trade Statistics: \url{https://data.imf.org/?sk=9D6028D4-F14A-464C-A2F2-59B2CD424B85}. 
Current U.S. dollar GDP (using 2015 as the base year) data are from the World Bank's World Development Indicators: \url{https://databank.worldbank.org/source/world-development-indicators}.}
Same as the currency risk premium, we use the 5-year moving average in the analysis.

\subsection{Additional results on the authority centrality}
\label{app:case-aut}

In Section \ref{sec:case}, we focus on the hub centrality $\bu$ and the its coefficient $\betau$. In the following, we show the corresponding results for the authority centrality $\bv$ and the its coefficient $\betav$.

Figure \ref{fig:trade-authority} shows the time series plots of the ranking of the authority centrality estimated by two-stage and SuperCENT for the 24 countries/regions, together with the ranking of the currency risk premium. 
{We rank the centrality in ascending order and the risk premium in descending order. 
Based on the negative relationship between centralities and risk premium established in \cite{richmond2019trade},
the closer the trends of rankings between centralities and risk premium are, the better the centralities capture the time variation in the risk premium.}
Similar to the hub centrality, the authority centrality estimated by the two-stage procedure is relatively more stable over time compared to SuperCENT, since SuperCENT incorporates information of both the GDP share and currency risk premium, which is more volatile than the trade network itself. 

For the coefficient of the authority centrality $\betav$ in Table \ref{tab:rx-summary}, the estimate from two-stage-adhoc and two-stage is $-0.0005$,
while the estimate from SuperCENT is $-0.0003$, which is consistent with the inaccuracy we observed in the simulation due to $|\hbetav^\lambdacv| \ll |\hbetau^\lambdacv|$.
$\supercent$ still improves its estimation and confidence interval, even though the improvement is not as large as $\betau$ 
due to $(\hbetav^\lambdacv)^2 \ll (\hbetau^\lambdacv)^2$ and the nonexchangeable roles of $\bu$ and $\bv$.
{For $\betax$, the estimates from two-stage and SuperCENT are comparable, but the widths of the confidence intervals from two-stage-adhoc and two-stage are much larger than that from SuperCENT, again as a result of the over-estimation of $\sigma_y^2$. Hence, $\betax$ is barely significant when using two-stage-adhoc and two-stage, while very significant by SuperCENT.}


\plotfig[1]{\trade}{plot/authority_gap5}{
{Time series of authority centrality ranking in ascending order from 2003 to 2012.
Similar to the hub centrality, if the trend of centralities is close to the trend of risk premium, 
then the centralities capture the time variation of risk premium,
based on the negative relationship between the two as claimed in \cite{richmond2019trade}.
}
The vertical dashed line indicates 2008, the year of the financial crisis. 
}{fig:trade-authority}

\subsection{Additional information for the case study in Section \ref{sec:case}}
\label{app:case-info}

Table \ref{tab:country-code} provides the country abbreviations and full names. 
Figure \ref{fig:trade-circular} shows the average trade volume from 2003 to 2012 among the 24 countries/regions. The arrows reflect the trade directions and the widths represent the volume. 

\begin{table}[h]
  \centering
\begin{tabular}[t]{ll}
\toprule
Code & Country\\
\midrule
AUS & Australia\\
CAN & Canada\\
CHE & Switzerland\\
CZE & Czech Republic\\
DNK & Denmark\\
\addlinespace
GBR & United Kingdom\\
HKG & Hong Kong \\
HUN & Hungary\\
IDN & Indonesia\\
IND & India\\
\bottomrule
\end{tabular}
\hspace{1em}
\begin{tabular}[t]{ll}
\toprule
Code & Country\\
\midrule
JPN & Japan\\
KOR & Korea\\
KWT & Kuwait\\
MEX & Mexico\\
MYS & Malaysia\\
\addlinespace
NOR & Norway\\
NZL & New Zealand\\
PHL & Philippines\\
POL & Poland\\
SAU & Saudi Arabia\\
\bottomrule
\end{tabular}
\hspace{1em}
\begin{tabular}[t]{ll}
\toprule
Code & Country\\
\midrule
SGP & Singapore\\
SWE & Sweden\\
THA & Thailand\\
ZAF & South Africa\\
\bottomrule
\end{tabular}
\caption{List of country abbreviations.}
  \label{tab:country-code}
\end{table}

\plotfig[.6]{\trade}{plot/circular}{The average trade volume from 2003 to 2012 among the 24 countries/regions. {Each country is in different color.} }{fig:trade-circular}

\section{Additional theoretical results and proof of all theoretical results}
\label{sec:proof}

\jcM{
Section \ref{sec:proof-sc} is devoted to theoretical results and proof of all theoretical results for SuperCENT and Section \ref{sec:proof-ts} for the two-stage.
}

We begin by providing some basic properties of the Kronecker product
 and the commutation matrix.
 The Kronecker product of $\bM = (m_{ij})\in\mathbb{R}^{m\times n}$ and $\bN = (n_{ij})\in\mathbb{R}^{p\times q}$,
 denoted by $\bM\otimes \bN$, is defined as
 \be
 \bM\otimes \bN &=&
 \left(
  \begin{array}{ccc}
   m_{11} \bN & \cdots  & m_{1n} \bN \\
   \vdots & \cdots & \vdots \\
   m_{m1} \bN & \cdots  & m_{mn} \bN \\
  \end{array}
\right) \in \mathbb{R}^{mp\times nq}.
 \ee
Denote  $\bK_{mn} \in \{0,1\}^{mn\times mn}$ as the commutation matrix such that
 \be
 \vec1(\bM^\top) = \bK_{mn}\vec1(\bM).
 \ee
We list the following facts about  the Kronecker product
 and the commutation matrix, which are used in the section without specific references.
Proofs of these facts can be found in \cite{	magnus1979commutation}.

Let $\bM \in\mathbb{R}^{m\times n}$, $\bN \in\mathbb{R}^{p\times q}$,
$\bP \in \mathbb{R}^{n\times t}$, $\bQ \in \mathbb{R}^{q\times s}$, 
and $\bZ \in\mathbb{R}^{n\times p}$.
\begin{enumerate}[(i)]
\item $(\bM \otimes \bN)^\top  = \bM^\top \otimes \bN^\top$.
\item $(\bM \otimes \bN) (\bP \otimes \bQ) = (\bM\bP) \otimes (\bN\bQ)$.
\item $\vec1(\bM\bZ\bN) = \left( \bN^\top\otimes \bM \right)\vec1(\bZ)$.
\item  $\vec1(\bM\bP) = (\bI \otimes \bM)\vec1(\bP) = (\bP^\top \otimes \bI)\vec1(\bM)$.
\item $\bK^\top_{mn} = \bK_{nm}$.
\item $\bK^\top_{mn} \bK_{mn} = \bK_{mn}^\top \bK_{mn} = \bI$.
\item $\bK_{mp}  (\bM \otimes \bN) \bK_{qn}= (\bN \otimes \bM)$. Equivalently, $\bK_{mp}  (\bM \otimes \bN) = (\bN \otimes \bM)\bK_{qn}$.
\item $(\bM \otimes \bN) \bK_{nq} (\bP \otimes \bQ) = ((\bM\bP) \otimes (\bN\bQ))\bK_{st} = \bK_{mp} ((\bN\bQ) \otimes (\bM\bP))$.
\item $\tr\left( \bK_{mn}  (\bM \otimes \bN) \right)= \tr \left(\bM\bN \right) $ where $\bM, \bN \in \mathbb{R}^{m\times n} $.
\end{enumerate}

\small

\subsection{Additional theoretical results and proof of all theoretical results for SuperCENT}
\label{sec:proof-sc}

\jcM{
We start by introducing additional definitions and notations. 
Recall that Section \ref{sec:theory} is devoted for the theoretical results for the SuperCENT estimators, which are the optimizers of the objective function \eqref{eq:supercent-obj}. 
In particular, the main Theorem \ref{thm:supercent-normality} states the asymptotic distribution of the SuperCENT optimization estimators. 
In this section, we first present Theorem \ref{thm:supercent-normality-algo} in Section \ref{sec:thm-algo}, which is in parallel of Theorem \ref{thm:supercent-normality}, but for the SuperCENT estimators as the output of Algorithm \ref{algo:supercent}. 
The proofs for both Theorem \ref{thm:supercent-normality} and Theorem \ref{thm:supercent-normality-algo} are provided in Section \ref{app:proof-thm2}. 
Subsequently,
Section \ref{app:proof-supercent-rate1} provides the proof of Proposition \ref{prop:supercent-rate1}, i.e., the convergence rates of $\hu$ and $\hv$, and Section \ref{app:proof-supercent-rate2} provides the statement and proof of Proposition \ref{prop:supercent-rate2}, i.e., the convergence rate of $\hbbeta$.
Section \ref{sec:consistency} is dedicated to the consistency results of the SuperCENT estimators, which are needed to establish the asymptotic distribution in Theorems \ref{thm:supercent-normality} and \ref{thm:supercent-normality-algo}.
Specifically, we provide two versions of the consistency results: one for the solution of the objective function \eqref{eq:supercent-obj} and one for the solution by Algorithm \ref{algo:supercent}, which are later used to prove Theorem \ref{thm:supercent-normality} and Theorem \ref{thm:supercent-normality-algo}, respectively.
Section \ref{sec:prop1} is devoted to Proposition \ref{prop1}, which is essential to prove the consistency results in Section \ref{sec:consistency}, and its proof. 
Section \ref{sec:technical-lemmas} contains all the technical lemmas, which are needed for Proposition \ref{prop1}, along with their proofs.
} 

In the following, $d_1$ and $d$, $\bu_1$ and $\bu$, $\bv_1$ and $\bv$ are interchangeable, denoting the leading singular value and the leading singular vectors (hub and authority centralities). The relevant notations and definitions are given as follows.


\jcM{
\begin{defn} 
\label{defn0}
Let $\lambdanew = \frac{\sigma_a^2}{n \sigma_y^2}\lambda$.
\end{defn}
}

\begin{defn} 
\label{defn1}
Define function $f_{0}$ that takes in arguments $\bar{\bu} \in \mathbb{R}^{n}, \bar{\bv} \in \mathbb{R}^{n}, \bar{\boldsymbol{\beta}}_{x} \in \mathbb{R}^{p}, \bar{\beta}_{u} \in \mathbb{R}, \bar{\beta_{v}} \in \mathbb{R}, \bar{d} \in \mathbb{R}^+$,
\be
f_{0}: \left(\bar{\bu}, \bar{\bv}, \bar{\boldsymbol{\beta}}_{x}, \bar{\beta}_{u}, \bar{\beta}_{v}, \bar{d}\right) \longmapsto 
\frac{1}{\sigma_{y}^{2}}\left\|\by-\bX \bar{\boldsymbol{\beta}}_{x} - \bar{\bu} \bar{\beta}_{u} - \bar{\bv} \bar{\beta}_{v}\right\|^{2}+\frac{\lambdanew}{\sigma_{a}^{2}}\left\|\bar{d} \bar{\bu} \bar{\bv}^{\top}-\bA\right\|_{F}^{2} .
\ee
\end{defn}


\begin{defn}
\label{defn2}
Define function $\overrightarrow{\boldsymbol{\beta}}$ that takes in arguments $\bar{\bu} \in \mathbb{R}^{n}, \bar{\bv} \in \mathbb{R}^{n} $ such that $(\bX, \bar{\bu}, \bar{\bv})$ is full rank as follows

\be
\overrightarrow{\boldsymbol{\beta}}:(\bar{\bu}, \bar{\bv}) \longmapsto\left[(\bX, \bar{\bu}, \bar{\bv})^{\top}(\bX, \bar{\bu}, \bar{\bv})\right]^{-1}(\bX, \bar{\bu}, \bar{\bv})^{\top} \by .
\ee

Let $\overrightarrow{\boldsymbol{\beta}}_{x}, \overrightarrow{\beta}_{u}, \overrightarrow{\beta}_{v}$ be functions of $(\bar{\bu}, \bar{\bv})$ such that 

\be
& \overrightarrow{\boldsymbol{\beta}}_{x}(\bar{\bu}, \bar{\bv}) \in \mathbb{R}^{n}, \overrightarrow{\beta}_{u}(\bar{\bu}, \bar{\bv}) \in \mathbb{R}, \overrightarrow{\beta}_{v}(\bar{\bu}, \bar{\bv}) \in \mathbb{R},
\ee
and 
\be
& \left(\begin{array}{l}
\overrightarrow{\boldsymbol{\beta}}_{x}(\bar{\bu}, \bar{\bv}) \\
\overrightarrow{\beta}_{u}(\bar{\bu}, \bar{\bv}) \\
\overrightarrow{\beta}_{v}(\bar{\bu}, \bar{\bv})
\end{array}\right)=\overrightarrow{\beta}(\bar{\bu}, \bar{\bv}).
\ee
\end{defn}


\begin{defn}
\label{defn3}
Define function $\overrightarrow{d}$ that takes arguments $\bar{\bu} \in \mathbb{R}^{n}, \bar{\bv} \in \mathbb{R}^{n}$,
\be
\overrightarrow{d}:(\bar{\bu}, \bar{\bv}) \longmapsto \frac{\bar{\bu}^{\top} \bA \bar{\bv}}{\|\bar{\bu}\|^{2}\|\bar{\bv}\|^{2}}.
\ee

\end{defn}


\begin{defn}
\label{defn4}
Define function $f_{1}$ that takes in arguments $\bar{\bu} \in \mathbb{R}^{n}, \bar{\bv} \in \mathbb{R}^{n}$
\be
f_{1}:(\bar{\bu}, \bar{\bv}) \longmapsto f_{0}\left(\bar{\bu}, \bar{\bv}, \overrightarrow{\boldsymbol{\beta}}_{x}(\bar{\bu}, \bar{\bv}), \overrightarrow{\beta}_{u}(\bar{\bu}, \bar{\bv}), \overrightarrow{\beta}_{v}(\bar{\bu}, \bar{\bv}), \overrightarrow{d}(\bar{\bu}, \bar{\bv})\right).
\ee

\end{defn}

\jcM{
\begin{defn}
\label{defn-f2}
Define function $f_{2}$ that takes arguments $\bar{\bu} \in \mathbb{R}^n, \bar{\bv} \in \mathbb{R}^n$ as

{\small
\be
f_{2}:(\bar{\bu}, \bar{\bv}) \longmapsto \frac{1}{\sigma_{y}^{2}} \left\| P_{(\bX, \bar{\bu}, \bar{\bv})^{\perp}} \left( \Tilde{\varepsilon} + \Tilde{\bu}_{1} \beta_{u} + \Tilde{\bv}_{1} \beta_{v}\right)\right\|^{2} 
-\frac{\lambdanew}{n^2\sigma_{a}^{2}}\left(\bar{\bu}^{\top}\left(\sum_{i=1}^{r} d_{i} \bu_{i} \bv_{i}^{\top}+\bE\right) \bar{\bv}\right)^{2}.
\ee
}

\end{defn}
}

\begin{defn}
\label{defn8}
$\forall \bz \in \mathbb{R}^{n}$, let $\Tilde{\bz}=(\bI-\bP_{\bX}) \bz$.
\jcM{Specifically, $\Tilde{\bu} = (\bI-\bP_{\bX}) \bu$, 
$\Tilde{\bv} = (\bI-\bP_{\bX}) \bv$,
and $\Tilde{\varepsilon} = (\bI-\bP_{\bX}) \varepsilon$.}
\end{defn}

\jcM{
\subsubsection{Theorem \ref{thm:supercent-normality-algo}: Asymptotic {distribution} for SuperCENT estimator via Algorithm \ref{algo:supercent}}
\label{sec:thm-algo}

We present the asymptotic distribution for the SuperCENT estimator as the {output} produced by Algorithm \ref{algo:supercent}. 
Comparing to Theorem \ref{thm:supercent-normality}, which shows the asymptotic {distribution} of the minimizer of the objective function \eqref{eq:supercent-obj}, Theorem \ref{thm:supercent-normality-algo} shows that the solution of Algorithm \ref{algo:supercent} converges to the same asymptotic distribution under slightly different conditions. 

\begin{theorem}
\label{thm:supercent-normality-algo}
Under the unified framework
\eqref{eq:model-supercent} and Assumptions \ref{assump:normal}-\ref{assump:consistent},
suppose $(\bX, \hu, \hv)$ is full-rank, $\frac{\sigma_{y}}{\sqrt{\beta_{u}^{2}+\beta_{v}^{2}}} \sqrt{\frac{\log n}{n}} = o(1)$, 
$\frac{\sigma_{a}^{2}}{n\left(\done-d_{2}\right)^{2}} 
\frac{\sigma_{y}^{2}}{\beta_{u}^{2}+\beta_{v}^{2}} = o(1)$,
and $\left|\frac{\beta_{u}}{\beta_{v}}\right| \in[\underline{\alpha}, \bar{\alpha}]$
for positive constants $\bar{\alpha}>\underline{\alpha}>0$, 
then the SuperCENT estimators, defined as the solution of Algorithm \ref{algo:supercent}
with the two-stage estimators for initialization  and  a given tuning parameter $\lambda$ satisfying
$\frac{1}{\lambda} {\kappa} \frac{n\sigma_y^2}{\sigma_a^2} = o(1)$, 
$\frac{1}{\lambda} {\kappa} \frac{n\sigma_y^2}{\sigma_a^2} \frac{\sigma_y^2}{\beta_{u}^{2}+\beta_{v}^{2}} = o(1)$,
and
$ \frac{1}{\lambda} \frac{\kappa}{(d-d_2)^2} (\beta_{u}^{2}+\beta_{v}^{2}) = o(1)$,
converge to the following normal distributions asymptotically,

\begin{enumerate}[nolistsep]
\item Centralities: 
\be
\hu - \bu = \etau + o(\etau)
\quad\mbox{and}\quad
\hv - \bv = \etav + o(\etav),
\ee
\item Network effect: 
\be
\hbbeta - \bbeta = \etabeta + o(\etabeta)
= \left(\eta_\betax^\top, \etabetau, \etabetav\right)^\top + o\left(\left(\eta_\betax^\top, \etabetau, \etabetav \right)^\top \right),
\ee
\end{enumerate}
where $\left(
 \begin{array}{c}
  \etau \\
  \etav \\
  \etabeta
 \end{array}
\right) \sim 
N\Big(
\zero_{(2n+2+p)\times 1}, 
\bC\left(
              \begin{array}{cc}
                \sigma_y^2\bI_n & \zero_{n\times n^2} \\
                \zero_{n^2\times n} & \sigma_a^2 \bI_{n^2} \\
              \end{array}
              \right){\bC}^\top
\Big)$,
$\frac{\|o\left(\etau\right)\|}{\|\etau\|} \overset{P}{\longrightarrow} 0$,
$\frac{\|o\left(\etav\right)\|}{\|\etav\|} \overset{P}{\longrightarrow} 0$,
$\frac{\left\|o\left(\etabetax\right)\right\|}{\left\|\etabetax\right\|} \overset{P}{\longrightarrow} 0$,
$\frac{|o\left(\etabetau\right)|}{|\etabetau|} \overset{P}{\longrightarrow} 0$,
$\frac{|o\left(\etabetav\right)|}{|\etabetav|} \overset{P}{\longrightarrow} 0$,
and 
$\bC$ has the same form as in Theorem \ref{thm:supercent-normality}.
\end{theorem}


\begin{remark}
\label{rmk:full-rank-t}
    We would like to point out that, under our model assumptions, with high probability, the full-rank condition for $(\bX,\hat \bu, \hat \bv)$ is satisfied by all $(\bX,\hat \bu, \hat \bv)=(\bX,\bu^{(t)},\bv^{(t)})$, generated by Algorithm \ref{algo:supercent}, including $t=\infty$ which is the output of Algorithm \ref{algo:supercent}. Moreover, with high probability, the full-rank condition for $(\bX,\hat \bu, \hat \bv)$ is also satisfied by the optimizer of the objective function \eqref{eq:supercent-obj}. The reasoning is as follows. 
    Under Assumption \ref{assump:consistent}, the two-stage estimator is close to the truth $(\bu,\bv)$ with high probability. 
    By using the two-stage estimator as the initialization of our algorithm, the objective function value is already small and is smaller than that of any non-full-rank $(\bX,\hat\bu,\hat \bv)$.
    This is partly due to the assumption of the condition number, i.e., Assumption \ref{assump:x}, which implies a large gap between the minimum objective function value achievable by non-full-rank $(\bX,\hat \bu, \hat \bv)$ and that of the truth, $(\bX, \bu, \bv)$ with high-probability under reasonable noise level. Our algorithm is a descending algorithm, hence all the $(\bX,\bu^{(t)},\bv^{(t)})$ generated by the algorithm gives a smaller objective function value than that of the minimum objective function achievable by non-full-rank $(\bX,\hat \bu, \hat \bv)$, which implies that they are all full rank. 
    The minimizer of the objective function will have the smallest objective function value and thus it is impossible to be non-full-rank.
    The details of this proof are relatively tedious, so we have omitted them here.
    
    
\end{remark}

}

\subsubsection{Proof of Theorem \ref{thm:supercent-normality} and Theorem \ref{thm:supercent-normality-algo}}
\label{app:proof-thm2}

In this section, we provide the proofs of Theorem \ref{thm:supercent-normality} in Section \ref{app:proof-thm-supercent} and Theorem \ref{thm:supercent-normality-algo} in Section \ref{app:proof-thm-supercent-algo}, which state the asymptotic normality of the SuperCENT estimators as the minimizer of the objective function \ref{eq:supercent-obj} and as the solution produced by Algorithm \ref{algo:supercent}, respectively.

\paragraph{Proof of Theorem \ref{thm:supercent-normality}}
\label{app:proof-thm-supercent}

\begin{proof}

In Section \ref{app:algo-supercent}, we derive the estimates of $(\hd, \hu, \hv, \hbbeta)$
as \eqref{eq:algo-hbeta}-\eqref{eq:algo-hvhv}.
Recall that $\bAp = \bUp \bDp \bVp^\top$ where $\bUp = (\bu_2, \ldots, \bu_r)$, $\bVp = (\bv_2, \ldots, \bv_r)$ and 
$\bDp = diag(d_2, \ldots, d_r)$. Then, $\bAp^\top \bu = 0$ and $\bAp \bv = \bm 0$.
Let $\ttE = \bAp + \bE$ and $\Dy = \bu\dbetau+\du\betau+\bv\dbetav+\dv\betav+\du\dbetau+\dv\dbetav$.
Together they lead to the first order expansion:
\be
    &&(\bu+\du)^\top(\bu+\du)= n, \label{bu-du}\\
    &&(\bv+\dv)^\top(\bv+\dv)= n, \label{bv-dv}\\
    && \begin{multlined} 
    \betau + \dbetau =  \frac{1}{n}(\bu+\du)^\top(\bu\betau+\bepsilon -\bX\dbetax-\bv\dbetav-\dv\betav-\dv\dbetav)
    \end{multlined}, \label{dbu-1} \\
    &&\begin{multlined} 
    \betav + \dbetav= \frac{1}{n}(\bv+\dv)^\top(\bv\betav+\bepsilon
    -\bX\dbetax-\bu\dbetau-\du\betau -\du\dbetau)
    \end{multlined}, \label{dbv-1} \\	
    &&\betax +\dbetax = (\bX^\top\bX)^{-1}\bX^\top(\bX\betax+\bepsilon-\Dy), \label{dbx}\\
    &&d+\dd = (\bu+\du)^\top (d\bu\bv^\top+\ttE)(\bv+\dv)/n^2, \label{dd} \\
    &&\begin{multlined}[t]
    (\betau+\dbetau)(-\bepsilon+\bX\dbetax+\Dy) \\
      + \lambda (d+\dd)^2 (\bu+\du) -  \lambda (d+\dd) (d\bu\bv^\top+\ttE)(\bv+\dv)/n = 0, 
    \end{multlined} \label{du-1} \\
    &&\begin{multlined}[t]
    (\betav+\dbetav)(-\bepsilon+\bX\dbetax+\Dy) \\
      + \lambda (d+\dd)^2 (\bv+\dv) -  \lambda (d+\dd) (d\bv\bu^\top+\ttE^\top)(\bu+\du)/n = 0. \label{dv-1}
    \end{multlined}
\ee

Plugging \eqref{dbx} into \eqref{dbu-1}-\eqref{dbv-1} and using \eqref{bu-du}-\eqref{bv-dv} to eliminate $\dbetax$, we have
\be
&& \begin{multlined}
(\bu+\du)^\top(\bI-\bP_{\bX})\du\betau + (\bu+\du)^\top(\bI-\bP_{\bX})\dv\betav \\ 
+ (\bu+\du)^\top(\bI-\bP_{\bX})(\bu+\du)\dbetau + (\bu+\du)^\top(\bI-\bP_{\bX})(\bv+\dv)\dbetav \\
= (\bu+\du)^\top(\bI-\bP_{\bX})\bepsilon 
 \end{multlined}\label{dbu-2}\\ 
&& \begin{multlined}
(\bv+\dv)^\top(\bI-\bP_{\bX})\du\betau + (\bv+\dv)^\top(\bI-\bP_{\bX})\dv\betav \\ 
+ (\bv+\dv)^\top(\bI-\bP_{\bX})(\bu+\du)\dbetau + (\bv+\dv)^\top(\bI-\bP_{\bX})(\bv+\dv)\dbetav \\
= (\bv+\dv)^\top(\bI-\bP_{\bX})\bepsilon.
 \end{multlined}\label{dbv-2}
\ee

Let
$\tu = (\bI-\bP_{\bX})(\bu+\du)$,
$\tv = (\bI-\bP_{\bX})(\bv+\dv)$ and 
$\bCuv = \left(
\begin{array}{cc}
    \tu^\top\tu, & \tu^\top\tv \\
    \tu^\top\tv, & \tv^\top\tv
  \end{array}
\right)$. 
Then \eqref{dbu-2}-\eqref{dbv-2} can be written as
\be
\left(
 \begin{array}{c}
  \tu^\top \\
  \tv^\top \\
 \end{array}
\right)
\left(\betau\bI_n~~ \betav\bI_n \right)
\left(
  \begin{array}{c}
    \du \\
    \dv 
  \end{array}
\right)
+ \bCuv
\left(
  \begin{array}{c}
    \dbetau \\
    \dbetav \\
  \end{array}
\right)
&=& 
\left(
  \begin{array}{c}
    \tu^\top \\
    \tv^\top 
  \end{array}
\right)
\bepsilon.
\ee
Solving for $\dbetau,~\dbetav$ gives
\be
\left(
  \begin{array}{c}
    \dbetau \\
    \dbetav \\
  \end{array}
\right)
&=& 
\bCuvi
\left(
 \begin{array}{c}
  \tu^\top \\
  \tv^\top \\
 \end{array}
\right)
\left(-\betau\bI_n~~ -\betav\bI_n ~~ \bI_n\right)
\left(
  \begin{array}{c}
    \du \\
    \dv \\
        \bepsilon
  \end{array}
\right).
\label{sc:dbetau-dbeutav}
\ee


Plugging \eqref{dbx}, \eqref{dd} and \eqref{sc:dbetau-dbeutav} into \eqref{du-1} and \eqref{dv-1} using \eqref{bu-du}-\eqref{bv-dv},
we have
\be
\begin{split}
&\Bigg[
\frac{\lambda \hd}{n}
\left(
  \begin{array}{cc}
    d\bv^\top\hv  \left(\bI - \frac{1}{2n} \hu \du^\top \right), & -\left(\bI - \bP_{\hu}\right)\bAp \\
    (-\bAp)^\top \left(\bI - \bP_{\hv}\right), & 
    d\bu^\top\hu  \left(\bI - \frac{1}{2n} \hv \dv^\top \right) \\
  \end{array}
\right) 
+  \\
&\underbrace{
\hspace{8em}
\left( 
  \begin{array}{cc}
    \hhbetau\betau, & \hhbetau\betav \\
    \betau\hhbetav, & \hhbetav\betav \\
  \end{array}
\right)
\otimes \left(\bI - \bP_{\left(\bX\tu\tv\right)}\right)
\Bigg]
}_{\bM_1} 
\left(
  \begin{array}{c}
    \du \\
    \dv \\
  \end{array}
\right) \\
& \hspace{10em} 
= 
\underbrace{
\left(
  \begin{array}{cc}
    \hhbetau\left(\bI - \bP_{\left(\bX\tu\tv\right)}\right) & \frac{\lambda \hd}{n} \hhv^\top \otimes (\bI-\bP_{\hu}) \\
    \hhbetav\left(\bI - \bP_{\left(\bX\tu\tv\right)}\right) & \quad \frac{\lambda \hd}{n} \left(\hhu^\top \otimes (\bI-\bP_{\hv}) \right)\bK \\
  \end{array}
\right)
}_{\bM_2}
\left(
  \begin{array}{c}
    \bepsilon \\
    \vec1(\bE) \\
  \end{array}
\right).
\label{sc:du-dv-2}
\end{split}
\ee

Then
\be
\left(
  \begin{array}{c}
    \du \\
    \dv \\
  \end{array}
\right) 
= 
\bM_1^{-1} \bM_2
\left(
  \begin{array}{c}
    \bepsilon \\
    \vec1(\bE) \\
  \end{array}
\right).
\label{sc:du-dv-3}
\ee

For $\dbetau$ and $\dbetav$, 
plugging \eqref{sc:du-dv-3} into \eqref{sc:dbetau-dbeutav}
\be
\left(
  \begin{array}{c}
    \dbetau \\
    \dbetav \\
  \end{array}
\right)
&=& 
\bCuvi
\left(
 \begin{array}{c}
  \tu^\top \\
  \tv^\top \\
 \end{array}
\right)
\left(-\betau\bI_n~~ -\betav\bI_n ~~ \bI_n\right)
\left(
  \begin{array}{c}
    \du \\
    \dv \\
        \bepsilon
  \end{array}
\right) \\
&=&
\underbrace{
\bCuvi
\left(
 \begin{array}{c}
  \tu^\top \\
  \tv^\top \\
 \end{array}
\right)
\left(-\betau\bI_n~~ -\betav\bI_n ~~ \bI_n\right)
\left(
\begin{array}{c}
  \bM_1^{-1} \bM_2 \\
  \begin{array}{cc}
    \bI_n  \; , \; & \zero_{n\times n^2}
  \end{array}
\end{array}  
\right)
}_{\bM_3}
\left(
  \begin{array}{c}
    \bepsilon \\
    \vec1(\bE) \\
  \end{array}
\right).
\label{sc:dbetau-dbetav}
\ee

Lastly for $\dbetax$, we plug \eqref{sc:du-dv-3} and \eqref{sc:dbetau-dbetav} into \eqref{dbx} 
\be
\dbetax &=& (\bX^\top\bX)^{-1}\bX^\top(\bepsilon-\bu\dbetau-\du\betau -\bv\dbetav-\dv\betav)\\
&=&
(\bX^\top\bX)^{-1}\bX^\top
 \left(
 -\betau\bI_n ~~ -\betav\bI_n ~~ -\bu ~~ -\bv ~~ \bI_n
\right)
\left(
  \begin{array}{c}
    \du\\
    \dv\\
    \dbetau \\
    \dbetav \\
    \bepsilon
  \end{array}
\right)\\
&=&
\begin{multlined}[t]
(\bX^\top\bX)^{-1}\bX^\top
 \left(
 -\betau\bI_n ~~ -\betav\bI_n ~~ -\bu ~~ -\bv ~~ \bI_n
\right)
\left(
\begin{array}{c}
  \bM_1^{-1} \bM_2 \\
  \bM_3 \\
  \begin{array}{cc}
    \bI_n \;,\; & \zero_{n\times n^2}
  \end{array}
\end{array}
\right)
\left(
  \begin{array}{c}
    \bepsilon \\
    \vec1(\bE) \\
  \end{array}
\right).
\end{multlined}
\label{sc:dbetax}
\ee

Based on the consistency result in Corollary \ref{thm1-cor}, we have
$\min \left\{\frac{\left\|\hat{\bu}-\bu\right\|}{\sqrt{n}}, \frac{\left\|\hat{\bu}+\bu\right\|}{\sqrt{n}}\right\}\rightarrow 0$, \\
$\min \left\{\frac{\left\|\hat{\bv}+\bv\right\|}{\sqrt{n}}, \frac{\left\|\hat{\bv}-\bv\right\|}{\sqrt{n}}\right\}\rightarrow 0$,
$\left|\frac{\hat{d}-\done}{\done}\right| \rightarrow 0$,
$\left|\frac{\hat{\beta}_{u}-\beta_{u}}{\beta_{u}}\right|\rightarrow 0$,
and 
$\left|\frac{\hat{\beta}_{v}-\beta_{v}}{\beta_{v}}\right|\rightarrow 0$.
Then
\be
\left \{
\begin{aligned}
\du & = \etau + o\left(\etau\right), \\
\dv & = \etav + o\left(\etav\right), \\
\dbetau & = \etabetau + o\left(\etabetau\right), \\
\dbetav & = \etabetav + o\left(\etabetav\right), \\
\dbetax & = \etabetax + o\left(\etabetax\right), 
\end{aligned}
\right .
\label{eq:etas}
\ee
where
\be
\left(\begin{array}{c}
    \etau \\
    \etav \\
  \end{array}
  \right ) &=& 
\left[
\frac{\lambda d}{n}
\left(
  \begin{array}{cc}
    dn \bI & -\bAp \\
    (-\bAp)^\top & dn \bI \\
  \end{array}
\right)
+
\left(
  \begin{array}{cc}
    \beta_u^2 & \betau\betav \\
    \betau\betav & \beta_v^2 \\
  \end{array}
\right)
\otimes \opxuv
\right]^{-1}
\\
&&\left(
  \begin{array}{cc}
    \betau\opxuv & \lambda d \bv^\top \otimes (\bI-\pu)/n \\
    \betav\opxuv & \lambda d \left(\bu^\top \otimes (\bI-\pv)/n\right)\bK \\
  \end{array}
\right)
\left(
  \begin{array}{c}
    \bepsilon \\
    \vec1(\bE) \\
  \end{array}
\right) \\
&\stackrel{def}{=}&\left(
  \begin{array}{cc}
    \bC_{11} & \bC_{12} \\
    \bC_{21} & \bC_{22} \\
  \end{array}
\right)
\left(
  \begin{array}{c}
    \bepsilon \\
    \vec1(\bE) \\
  \end{array}
\right),
\label{sc:dudv-C}
\ee

\be
\left(
  \begin{array}{c}
    \etabetau \\
    \etabetav \\
  \end{array}
\right)
&=& 
\bCuvi
\left(
 \begin{array}{c}
  \tu^\top \\
  \tv^\top \\
 \end{array}
\right)
\left(-\betau\bI_n~~ -\betav\bI_n ~~ \bI_n\right)
\left(
  \begin{array}{cc}
    \bC_{11} & \bC_{12} \\
    \bC_{21} & \bC_{22} \\
    \bI_n & \zero_{n\times n^2}
  \end{array}
\right)
\left(
  \begin{array}{c}
    \bepsilon \\
    \vec1(\bE) \\
  \end{array}
\right) \\
&\stackrel{def}{=}&
\left(
  \begin{array}{cc}
    \bC_{41} & \bC_{42} \\
    \bC_{51} & \bC_{52} \\
  \end{array}
\right)
\left(
  \begin{array}{c}
    \bepsilon \\
    \vec1(\bE) \\
  \end{array}
\right),
\label{sc:dbetau-dbetav-C}
\ee

\be
\etabetax &=& 
(\bX^\top\bX)^{-1}\bX^\top
 \left(
 -\betau\bI_n ~~ -\betav\bI_n ~~ -\bu ~~ -\bv ~~ \bI_n
\right)
\left(
  \begin{array}{cc}
    \bC_{11} & \bC_{12} \\
    \bC_{21} & \bC_{22} \\
    \bC_{41} & \bC_{42} \\
    \bC_{51} & \bC_{52} 
  \end{array}
\right)
\left(
  \begin{array}{c}
    \bepsilon \\
    \vec1(\bE) \\
  \end{array}
\right)\\
&\stackrel{def}{=}&
\left(
  \begin{array}{cc}
    \bC_{31} & \bC_{32} 
  \end{array}
\right)
\left(
  \begin{array}{c}
    \bepsilon \\
    \vec1(\bE) \\
  \end{array}
\right),
\label{sc:dbetax-C}
\ee
and
$\frac{\|o\left(\etau\right)\|}{\|\etau\|} \overset{P}{\longrightarrow} 0$,
$\frac{\|o\left(\etav\right)\|}{\|\etav\|} \overset{P}{\longrightarrow} 0$,
$\frac{|o\left(\etabetau\right)|}{|\etabetau|} \overset{P}{\longrightarrow} 0$,
$\frac{|o\left(\etabetav\right)|}{|\etabetav|} \overset{P}{\longrightarrow} 0$,
$\frac{\left\|o\left(\etabetax\right)\right\|}{\left\|\etabetax\right\|} \overset{P}{\longrightarrow} 0$.

Finally, recall that we assume 
\be
\left(
  \begin{array}{c}
    \bepsilon \\
    \vec1(\bE) \\
  \end{array}
\right) \sim
N\left(\zero_{(n+n^2)\times 1}, \left(
                             \begin{array}{cc}
                               \sigma_y^2\bI_n & \zero_{n\times n^2} \\
                               \zero_{n^2\times n} & \sigma_a^2 \bI_{n^2} \\
                             \end{array}
                           \right)
\right).
\ee
By \eqref{eq:etas} and putting \eqref{sc:dudv-C},  \eqref{sc:dbetau-dbetav-C} and \eqref{sc:dbetax-C} together, we have
\be
\left(
 \begin{array}{c}
  \hu-\bu\\
  \hv-\bv\\
  \hbetax-\betax\\
  \hbetau-\betau \\
  \hbetav-\betav \\
 \end{array}
\right)
=
\left(
  \begin{array}{c}
    \etau + o\left(\etau\right) \\
    \etav + o\left(\etav\right) \\
    \etabetau + o\left(\etabetau\right) \\
    \etabetav + o\left(\etabetav\right) \\
    \etabetax + o\left(\etabetax\right)
  \end{array}
\right)
\ee
where
\be
\left(
  \begin{array}{c}
    \etau \\
    \etav \\
    \etabetau \\
    \etabetav \\
    \etabetax
  \end{array}
\right) &=&
\left(
  \begin{array}{cc}
    \bC_{11} & \bC_{12} \\
    \bC_{21} & \bC_{22} \\
    \bC_{31} & \bC_{32} \\
    \bC_{41} & \bC_{42} \\
    \bC_{51} & \bC_{52} 
  \end{array}
\right)
\left(
  \begin{array}{c}
    \bepsilon \\
    \vec1(\bE) \\
  \end{array}
\right) \\
&=&
\bC
\left(
  \begin{array}{c}
    \bepsilon \\
    \vec1(\bE) \\
  \end{array}
\right) \\
& {\sim} &
N\Big(
\zero_{(2n+2+p)\times 1}, 
\bC\left(
              \begin{array}{cc}
                \sigma_y^2\bI_n & \zero_{n\times n^2} \\
                \zero_{n^2\times n} & \sigma_a^2 \bI_{n^2} \\
              \end{array}
              \right){\bC}^\top
\Big).
\ee



\end{proof}

\jcM{
\paragraph{Proof of Theorem \ref{thm:supercent-normality-algo}}
\label{app:proof-thm-supercent-algo}

\begin{proof}
The proof of Theorem \ref{thm:supercent-normality-algo} largely parallels  
that of Theorem \ref{thm:supercent-normality}.
However, to derive \eqref{eq:etas} from \eqref{sc:du-dv-3}, \eqref{sc:dbetau-dbetav} and \eqref{sc:dbetax},
we need to invoke Corollary \ref{thm2-cor}, i.e., the consistency results for the SuperCENT estimators as the solution of Algorithm \ref{algo:supercent}, instead of Corollary \ref{thm1-cor}.
This change follows from the slightly different conditions in Theorem \ref{thm:supercent-normality-algo} that satisfy the conditions \eqref{eq:thm2-cor-cond1}-\eqref{eq:thm2-cor-cond3} stated in Corollary \ref{thm2-cor}.
\end{proof}
}

\subsubsection{Proof of Proposition \ref{prop:supercent-rate1}}
\label{app:proof-supercent-rate1}

\begin{proof}

We establish the error bound of $\hat{\bu}$ and $\hat{\bv}$ through the stronger statement given in the proof for establishing the limiting distribution in Theorem \ref{thm:supercent-normality} and the fact that $\|\hat{\bu}\|_2 = \sqrt{n}$ and $\|\hat{\bv}\|_2 = \sqrt{n}$.


Equations \eqref{eq:etas} and \eqref{sc:dudv-C} in the proof of Theorem \ref{thm:supercent-normality} give
\be
\left \{
\begin{aligned}
\du & = \etau + o\left(\etau\right), \\
\dv & = \etav + o\left(\etav\right),
\end{aligned}
\right .
\ee
where
\be
\left(
  \begin{array}{c}
    \etau \\
    \etav \\
  \end{array}
\right)
&=&
\left(
  \begin{array}{cc}
    \bC_{11} & \bC_{12} \\
    \bC_{21} & \bC_{22} \\
  \end{array}
\right)
\left(
  \begin{array}{c}
    \bepsilon \\
    \vec1(\bE) \\
  \end{array}
\right),
\ee 
$\frac{\|o\left(\etau\right)\|}{\|\etau\|} \overset{P}{\longrightarrow} 0$, and
$\frac{\|o\left(\etav\right)\|}{\|\etav\|} \overset{P}{\longrightarrow} 0$ with probability at least $1-6e^{-n}$ as in Theorem \ref{thm1}.
Therefore, 
\be
\frac{1}{n}\E \| \hu - \bu \|^2  \le  \frac{1}{n} \tr(\sigma_y^2 \bC_{11} \bC_{11}^\top + \sigma_a^2 \bC_{12} \bC_{12}^\top ) \left(1+o(1)\right) + 6e^{-n}.
\ee

Next, we compute $\tr(\sigma_y^2 \bC_{11} \bC_{11}^\top + \sigma_a^2 \bC_{12} \bC_{12}^\top )$.
Note that, when $\bA_0$ is rank-one, 
\be
\left(
  \begin{array}{c}
    \bC_{11} \\
    \bC_{21} \\
  \end{array}
\right)
&=& (\lambda d^2 + \beta_u^2 + \beta_v^2)^{-1}
\left(
  \begin{array}{c}
    \betau \\
    \betav \\
  \end{array}
\right)
\otimes \opxuv
\ee
and
{\footnotesize{
\begin{multline}
\left(
  \begin{array}{c}
    \bC_{12} \\
    \bC_{22} \\
  \end{array}
\right)
= \frac{1}{dn}
\left[
\left(
  \begin{array}{c}
    \bv^\top \otimes (\bI-\pu) \\
    \left(\bu^\top \otimes (\bI-\pv)\right)\bK \\
  \end{array}
\right)\right.\\
\left.-\frac{1}{(\lambda d^2 + \beta_u^2 + \beta_v^2)}
\left(
  \begin{array}{c}
    \beta_u^2  \bv^\top \otimes \opxuv + \betau\betav \left(\bu^\top \otimes \opxuv \right)\bK \\
    \betau\betav  \bv^\top \otimes \opxuv + \beta_v^2 \left(\bu^\top \otimes \opxuv \right)\bK \\
  \end{array}
\right)
\right].
\end{multline}
} }
Then, 
\be
&& \tr(\sigma_y^2 \bC_{11} \bC_{11}^\top + \sigma_a^2 \bC_{12} \bC_{12}^\top )  \\
&=& \frac{\sigma_y^2}{(\lambda d^2 + \beta_u^2 + \beta_v^2)^2}\tr\left( \beta_u^2 \opxuv \right) \\
&& + \frac{\sigma_a^2}{n^2 d^2} \tr\left( \bv^\top\bv\right) \tr\left( \bI-\pu \right)  \\
&& + \frac{\sigma_a^2}{n^2 d^2(\lambda d^2 + \beta_u^2 + \beta_v^2)^2} 
\tr\Bigg(\betau^4  \bv^\top\bv \opxuv + \\
&& \hspace{.3in} 2\beta_u^3\betav \bu \bv^\top \otimes \opxuv \bK^\top  + \beta_u^2 \beta_v^2\bu \bu^\top \otimes \opxuv  \Bigg)\\ 
&& - \frac{2\sigma_a^2}{nd^2(\lambda d^2 + \beta_u^2 + \beta_v^2)} 
\tr\Bigg(
\beta_u^2/n
\left(
\bv^\top \otimes (\bI-\pu)\right)
\left(  
\bv \otimes \opxuv\right) + \\
&& \hspace{.3in} \betau\betav/n
 \left(\bv^\top \otimes (\bI-\pu)\right)  \label{eq:u-rate-1}
 \left(\bK^\top \bu \otimes \opxuv \right) \Bigg)\\
&=& \frac{\sigma_y^2\beta_u^2(n-p-2)}{(\lambda d^2 + \beta_u^2 + \beta_v^2)^2} + 
	\left(\frac{\sigma_a^2}{d^2} - \frac{\sigma_a^2}{d^2n}\right) \\
&& + \frac{\sigma_a^2\beta_u^2 (n-p-2)}{n^2 d^2(\lambda d^2 + \beta_u^2 + \beta_v^2)^2 }  
\left(  \beta_u^2 n + 2\betau \betav \tr(\bv^\top\bu) + \beta_v^2 n \right) \\
&& - \frac{2\sigma_a^2 (n-p-2)}{n d^2(\lambda d^2 + \beta_u^2 + \beta_v^2)} \beta_u^2 \\
&=& \left(\frac{\sigma_a^2}{d^2} - \frac{\sigma_a^2}{d^2n}\right)   \\
&& - \frac{\beta_u^2(n-p-2)}{(\lambda d^2 + \beta_u^2 + \beta_v^2)^2}
\Bigg[
\frac{2\lambda d^2 + \beta_u^2 + \beta_v^2}{d^2n} \sigma_a^2
-\sigma_y^2 
\Bigg] \label{eq:u-rate-2}.
\ee

Therefore, 
\be
&& \frac{1}{n} \E \| \hu - \bu \|^2 \\
&=& 
\left(\frac{\sigma_a^2(n-1)}{d^2n^2} - \frac{\beta_u^2 (n-p-2)}{n(\lambda d^2 + \beta_u^2 + \beta_v^2)^2}
\Bigg[
\frac{2\lambda d^2 + \beta_u^2 + \beta_v^2}{d^2n} \sigma_a^2
-\sigma_y^2 
\Bigg]
\right)\left(1+o(1)\right)\\ 
&=&
O\left(\frac{\sigma_a^2}{d^2n} - \beta_u^2 \delta_{ts,sc}\right)
\ee
where 
$\delta_{ts,sc}=\frac{1}{(\lambda d^2 + \beta_u^2 + \beta_v^2)^2}
\Big[
\frac{2\lambda d^2 + \beta_u^2 + \beta_v^2}{d^2 n}
\sigma_a^2
- \sigma_y^2
\Big]$.

To get the optimal $\lambda$ in Remark \ref{rmk:opt-lambda}, 
we take the partial derivative of 
$\ell_u \overset{\Delta}{=} $ \eqref{eq:u-rate-2}
with respect to $\lambda$ yields
\be
\frac{\partial \ell_u}{\partial \lambda}
&=& \frac{\beta_u^2(n-p-2)}{(\lambda d^2 + \beta_u^2 + \beta_v^2)^3}
\Bigg[2d^2 \sigma_y^2 
+ \frac{\sigma_a^2}{n} 
\Big(
2(\lambda d^2 + \beta_u^2 + \beta_v^2)
- 4\lambda d^2 - 2\beta_u^2 - 2\beta_v^2
\Big)
\Bigg]\\
&=& \frac{\beta_u^2(n-p-2)}{(\lambda d^2 + \beta_u^2 + \beta_v^2)^3}
\left[2d^2 \sigma_y^2 
 - \frac{2d^2 }{n} 
\sigma_a^2 \lambda 
\right].
\ee

Setting $\frac{\partial \ell_u}{\partial \lambda}=0$ yields
\be
\lambda_0 = \frac{n\sigma_y^2}{\sigma_a^2}.
\ee

When $\lambda \in (0, \lambda_0]$, $\ell_u$ increases as $\lambda$ increases; $\lambda \in (\lambda_0, \infty)$, $\ell_u$ decreases and converges to 0 as $\lambda$ increases. The maximum of $\ell_u$ is then taken at $\lambda_0$.

Similarly, we derive the rate of $\hv$ as
\be
&& \frac{1}{n} \E \| \hv - \bv \|^2 \\
&\le&
\left(
\frac{\sigma_a^2(n-1)}{d^2n^2} - \frac{\beta_v^2 (n-p-2)}{n(\lambda d^2 + \beta_u^2 + \beta_v^2)^2}
\Bigg[
\frac{2\lambda d^2 + \beta_u^2 + \beta_v^2}{d^2n} \sigma_a^2
-\sigma_y^2 
\Bigg] 
\right) \left(1+o(1)\right)\\ 
&= &
O\left(\frac{\sigma_a^2}{d^2n} - \beta_v^2 \delta_{ts,sc}\right).
 \label{eq:v-rate-2}
\ee

\subsubsection{Proposition \ref{prop:supercent-rate2}}
\label{app:proof-supercent-rate2}

We now provide the convergence rates of $\hbbeta = (\hbetax^\top, \hbetau, \hbetav)^\top$.
\begin{proposition}
\label{prop:supercent-rate2}
Under the assumptions and conditions in Theorem \ref{thm:supercent-normality} and further assume
$\bA_0$ to be rank-one,
	the SuperCENT estimators satisfy the following,
\be
\E (\hbetau - \betau )^2
&=& \E (\hbetauts - \betau )^2 
= O\left(\frac{\sigma_y^2}{n} +
\frac{\sigma_a^2 (\beta_u^2 + \beta_v^2)}{d^2n^2} \right),\nonumber
\\
\E (\hbetav - \betav )^2
&=& \E (\hbetavts - \betav )^2
= O\left(\frac{\sigma_y^2}{n} +
\frac{\sigma_a^2 (\beta_u^2 + \beta_v^2)}{d^2n^2} \right),\nonumber
\\
Cov\left(\hbbeta_x - \bbeta_x \right)
&=&
Cov\left(\hbetaxts - \bbeta_x \right).\nonumber
\ee
\vspace{-1em}
\end{proposition}


PROOF:
\paragraph*{(1) Rate of $\hbetau$ and $\hbetav$. }

From Theorem \ref{thm:supercent-normality}, we have
\be
&& \quad \bepsilon - ( \du\betau+\dv\betav ) \\
&& =
(\bI - \betau \bC_{11} + \betav \bC_{12}) \bepsilon +
(\betau \bC_{12} + \betav \bC_{22}) \vec1(\bE) \\
&& = \left[ \bI - \frac{\beta_u^2 + \beta_v^2}{\lambda d^2 + \beta_u^2 + \beta_v^2} \opxuv \right] \bepsilon \\
&& \quad - \frac{1}{dn}\left[ \betau \bv^\top \otimes \opu + \betav \left(\bu^\top \otimes \opv\right)\bK \right] \vec1(\bE) \\
&& \quad + \frac{\beta_u^2 + \beta_v^2}{dn(\lambda d^2 + \beta_u^2 + \beta_v^2)} 
\Bigg[
\betau \bv^\top\otimes \opxuv + \\
&& \hspace{1.5in} \betav \left(\bu^\top \otimes \opxuv \bK \Bigg) 
\right] \vec1(\bE) 
\\
&&\stackrel{def}{=} 
\bA_1 \bepsilon + (\bC_1 + \bC_2) \vec1(\bE). \label{lr:dbudbv-0}
\ee

Plugging \eqref{lr:dbudbv-0} into \eqref{sc:dbetau-dbetav-C}, we obtain
\be
\left(
  \begin{array}{c}
    \etabetau \\
    \etabetav \\
  \end{array}
\right)
&=&
\left(\bCuvi
\left(
  \begin{array}{c}
    \tu^\top \\
    \tv^\top
  \end{array}
\right) 
\Big[
\bepsilon - (\du\betau+\dv\betav)
\Big] 
\right) \\
&=& 
\left (
\bCuvi
\left(
  \begin{array}{c}
    \tu^\top \\
    \tv^\top
  \end{array}
\right) 
\left[
\bA_1 \bepsilon + (\bC_1 + \bC_2 ) \vec1(\bE)
\right]
\right ) \\
&\stackrel{def}{=}&
\left(\bB_1 \bepsilon + \bB_2 \vec1(\bE)\right).
\label{lr:dbudbv-1}
\ee

To get the rate of $\hbetau$ and $\hbetav$, we next calculate $\bB_1\bB_1^\top$ and $\bB_2\bB_2^\top$.

(a) $\bB_1\bB_1^\top$. 

Since
\be
\bA_1 \bA_1^\top
&=& \bI - \frac{(\beta_u^2 + \beta_v^2)(2\lambda d^2 + \beta_u^2 + \beta_v^2)}{(\lambda d^2 + \beta_u^2 + \beta_v^2)^2} \opxuv
\label{sc:a1a1}
\ee
and
\be\opxuv \tu = \bm{0}, \label{sc:IPu}\ee

consequently
\be
\bB_1\bB_1^\top
&=&
\bCuvi
\left(
  \begin{array}{c}
    \tu^\top \\
    \tv^\top
  \end{array}
\right) 
\bA_1 \bA_1^\top
\left(
  \begin{array}{cc}
    \tu & \tv
  \end{array}
\right) 
\bCuvi = \bCuvi.
\label{lr:b1b1}
\ee

(b) $\bB_2\bB_2^\top$. 

Since
{\footnotesize
\be
&& \bC_1 \bC_1^\top \\
&=&\frac{1}{(dn)^2} 
\Big[
\beta_u^2(\bv^\top \otimes \opu)(\bv \otimes \opu)
+ \beta_v^2(\bu^\top \otimes \opv)(\bu \otimes \opv) \\
&&\vspace{1.5in} 
+ \betau\betav(\bv^\top \otimes \opu)\bK^\top(\bu \otimes \opv)
+ \betau\betav(\bu^\top \otimes \opv)\bK(\bv \otimes \opu) 
\Big] \\
&=& \frac{1}{d^2n} \left[ \beta_u^2 \opu + \beta_v^2 \opv \right],
\ee
}
\be
\bC_2 \bC_2^\top 
&=& n \left(\frac{\beta_u^2 + \beta_v^2}{dn(\lambda d^2 + \beta_u^2 + \beta_v^2)} \right)^2 \left[ \beta_u^2 \opxuv + \beta_v^2 \opxuv \right]
\ee
and
\be
\bC_1 \bC_2^\top = \bC_2 \bC_1^\top 
&=& - \frac{1}{d^2n}\frac{\beta_u^2 + \beta_v^2}{\lambda d^2 + \beta_u^2 + \beta_v^2} \left[ \beta_u^2 \opxuv + \beta_v^2 \opxuv \right],
\ee
consequently
\begin{multline}
(\bC_1 + \bC_2)(\bC_1 + \bC_2)^\top
= \frac{1}{d^2n} \Bigg[ \big(\beta_u^2 \opu + \beta_v^2 \opv \big) -\\
\frac{(\beta_u^2 + \beta_v^2)^2(2\lambda d^2 + \beta_u^2 + \beta_v^2)}{(\lambda d^2 + \beta_u^2 + \beta_v^2)^2}  \opxuv \Bigg].
\label{lr:c1c2}
\end{multline}

Therefore,
\be
\bB_2 \bB_2^\top 
&=&  
\bCuvi
\left(
  \begin{array}{c}
    \tu^\top \\
    \tv^\top
  \end{array}
\right) 
(\bC_1 + \bC_2)(\bC_1 + \bC_2)^\top
\left(
  \begin{array}{cc}
    \tu & \tv
  \end{array}
\right) 
\bCuvi \\
&=& 
\frac{1}{d^2n}
\bCuvi
\left(
  \begin{array}{c}
    \tu^\top \\
    \tv^\top
  \end{array}
\right) 
 \big(\beta_u^2 \opu + \beta_v^2 \opv \big)
\left(
  \begin{array}{cc}
    \tu & \tv
  \end{array}
\right) 
\bCuvi \\
&=&
\frac{1}{d^2n}
\bCuvi
\left(
\begin{array}{cc}
	\beta_v^2 \tu^\top \opv \tu & 0 \\
	0 & \beta_u^2 \tv^\top \opu \tv \\
\end{array}
\right)
\bCuvi
\label{lr:b2b2}
\ee
where the first equality is due to \eqref{sc:IPu}.

Combining \eqref{lr:b1b1} and \eqref{lr:b2b2} as well as \eqref{eq:etas}, 
\be
Cov\left(
  \begin{array}{c}
    \dbetau \\
    \dbetav \\
  \end{array}
\right) &\approx & 
 \sigma_y^2 \bB_1\bB_1^\top + \sigma_a^2 \bB_2 \bB_2^\top  \\
&=& 
 \sigma_y^2    \bCuvi \\
&&\vspace{1.5in}
\quad \quad + \sigma_a^2 \frac{1}{d^2n} \bCuvi 
 \left(
  \begin{array}{c}
    \tu^\top \\
    \tv^\top
  \end{array}
\right) 
\left[ \beta_u^2 \opu + \beta_v^2 \opv \right]
\left(
  \begin{array}{c}
    \tu \\
    \tv
  \end{array}
\right) 
\bCuvi  \\
&=& 
 \sigma_y^2    \bCuvi  
+ \sigma_a^2 \frac{1}{d^2n} \bCuvi 
\left(
\begin{array}{cc}
	\beta_v^2\tu^\top\opv\tu & 0\\
	0 & \beta_u^2\tv^\top\opu\tv \\
\end{array}
\right)
\bCuvi . \label{eq:supercent-betauv-exact}
\ee

Therefore,
\be
\E (\hbetau - \betau )^2
= \E (\hbetauts - \betau )^2 
 \quad \mbox{and} \quad
\E (\hbetav - \betav )^2
= \E (\hbetavts - \betav )^2.
\ee

\paragraph*{ (2) Rate of $\hbetax$.} Recall that
\be
\tG =
\left(
\begin{array}{cc}
	\bu & \bv
\end{array}
\right)
\bCuvi
\left(
  \begin{array}{c}
    \tu^\top \\
    \tv^\top
  \end{array}
\right) 
 \quad \mbox{and} \quad
\bG = 
\left(
\begin{array}{cc}
	\bu & \bv
\end{array}
\right)
\bCuvi
\left(
  \begin{array}{c}
    \bu^\top \\
    \bv^\top
  \end{array}
\right).
\ee

By plugging \eqref{sc:du-dv-3}, \eqref{sc:dbetau-dbetav}, and \eqref{lr:dbudbv-0} into \eqref{sc:dbetax-C}, we have
{\footnotesize
\be
\etabetax &=& (\bX^\top\bX)^{-1}\bX^\top(\bepsilon-\bu\dbetau-\du\betau -\bv\dbetav-\dv\betav) \\
&=&
\left(
(\bX^\top\bX)^{-1}\bX^\top
\left[
\bI - 
\left(
\begin{array}{cc}
	\bu & \bv
\end{array}
\right)
\bCuvi
\left(
  \begin{array}{c}
    \tu^\top \\
    \tv^\top
  \end{array}
\right) 
\right]
\left[
\bA_1 \bepsilon +  (\bC_1 + \bC_2 ) \vec1(\bE)
\right]
\right )  \\
&=&
\left(
(\bX^\top\bX)^{-1}\bX^\top
(\bI -\tG)
\left[
\bA_1 \bepsilon +  (\bC_1 + \bC_2 ) \vec1(\bE)
\right]
\right)  \\ \label{sc:dbetax-1}
&\stackrel{def}{=}&
\left( \bF_1 \bepsilon + \bF_2 \vec1(\bE) \right).
\ee
}

To get the rate of $\hbetax$, we next calculate $\bF_1\bF_1^\top$ and $\bF_2\bF_2^\top$.

(a) $\bF_1\bF_1^\top$.

Since
\be
(\bI -\tG)(\bI -\tG)^\top
= 
\bI - \tG - \tG^\top + \bG
\ee
and
\be
\opxuv \tu = \bm{0} 
 \quad \mbox{and} \quad
 \opxuv \bu = \bm{0} \label{sc:IPu2},
\ee
consequently
{\footnotesize
\be
(\bI -\tG) \bA_1 \bA_1^\top (\bI -\tG)^\top
&=& (\bI -\tG)(\bI -\tG)^\top 
- \frac{(\beta_u^2 + \beta_v^2)(2\lambda d^2 + \beta_u^2 + \beta_v^2)}{(\lambda d^2 + \beta_u^2 + \beta_v^2)^2} \opxuv.
\ee
}

Further because $\opxuv\bX = \bm{0}$ and $\tu^\top \bX = \bm{0}$,
\be
\bF_1\bF_1^\top = (\bX^\top\bX)^{-1}
+ (\bX^\top\bX)^{-1}\bX^\top 
\left(
\begin{array}{cc}
	\bu & \bv
\end{array}
\right)
\bCuvi
\left(
  \begin{array}{c}
    \bu^\top \\
    \bv^\top
  \end{array}
\right)
\bX(\bX^\top\bX)^{-1}.
\label{sc:f1f1}
\ee

(b) $\bF_2\bF_2^\top$.


Note that due to \eqref{sc:IPu2}
\be
(\bI -\tG)\opxuv (\bI -\tG)^\top = \opxuv.
\ee
Combining with \eqref{lr:c1c2}, we have
\be
&& (\bI -\tG) (\bC_1 + \bC_2)(\bC_1 + \bC_2)^\top (\bI -\tG)^\top \\
&=& \frac{1}{d^2n}
\Bigg[
\beta_u^2 \opu + \beta_v^2 \opv \\
&&\vspace{2in} 
- 
\left(
	\begin{array}{cc}
		\beta_v^2 \opv \tu & \beta_u^2 \opu \tv
	\end{array}
\right)
\bCuvi
\left(
  \begin{array}{c}
    \bu^\top \\
    \bv^\top
  \end{array}
\right)
- \left(
	\begin{array}{cc}
		\bu & \bv
	\end{array}
\right)
\bCuvi
\left(
\begin{array}{c}
	\beta_v^2 \opv \tu \\ \beta_u^2 \opu \tv\end{array}
\right) \\
&&\vspace{2in} 
+ 
\left(
\begin{array}{cc}
	\bu & \bv
\end{array}
\right)
\bCuvi 
\left(
\begin{array}{cc}
	\beta_v^2\tu^\top\opv\tu & 0\\
	0 & \beta_u^2\tv^\top\opu\tv \\
\end{array}
\right)
\bCuvi 
\left(
  \begin{array}{c}
    \bu^\top \\
    \bv^\top
  \end{array}
\right) \\
&&\vspace{2in} 
-
\frac{(\beta_u^2 + \beta_v^2)(2\lambda d^2 + \beta_u^2 + \beta_v^2)}{(\lambda d^2 + \beta_u^2 + \beta_v^2)^2}  \opxuv
\Bigg].
\ee

Because $\opxuv\bX = \bm{0}$, $\bX^\top \tu = \bm{0}$ and $\bX^\top \tv = \bm{0}$,
\be
&&\bF_2\bF_2^\top \\
&=& \frac{1}{d^2n}
(\bX^\top\bX)^{-1}\bX^\top 
\Bigg[
\beta_u^2 \opu + \beta_v^2 \opv \\
&&\vspace{2in} 
+ 
\left(
\begin{array}{cc}
	\bu & \bv
\end{array}
\right) 
\bCuvi 
\left(
\begin{array}{cc}
	\beta_v^2\tu^\top\opv\tu & 0\\
	0 & \beta_u^2\tv^\top\opu\tv \\
\end{array}
\right)
\bCuvi 
\left(
  \begin{array}{c}
    \bu^\top \\
    \bv^\top
  \end{array}
\right)
\Bigg]
\bX(\bX^\top\bX)^{-1}.
\label{sc:f2f2}
\ee

Together with \eqref{sc:f1f1} and \eqref{sc:f2f2}, 
we obtain the variance-covariance matrix of $\dbetax$ as follows.
{\footnotesize
\be
&& Cov\left(\dbetax\right) \\
&\approx&
 \sigma_y^2 
\left[
(\bX^\top\bX)^{-1}
+ (\bX^\top\bX)^{-1}\bX^\top 
\left(
\begin{array}{cc}
	\bu & \bv
\end{array}
\right)
\bCuvi
\left(
  \begin{array}{c}
    \bu^\top \\
    \bv^\top
  \end{array}
\right)
\bX(\bX^\top\bX)^{-1}
\right] \label{eq:supercent-betax-exact-term1} \\
&& + \vspace{2in}
\sigma_a^2 
\frac{1}{d^2n}
(\bX^\top\bX)^{-1}\bX^\top 
\Bigg[
\beta_u^2 \opu + \beta_v^2 \opv \\
&&\vspace{2in} 
+ 
\left(
\begin{array}{cc}
	\bu & \bv
\end{array}
\right) 
\bCuvi 
\left(
\begin{array}{cc}
	\beta_v^2\tu^\top\opv\tu & 0\\
	0 & \beta_u^2\tv^\top\opu\tv \\
\end{array}
\right)
\bCuvi 
\left(
  \begin{array}{c}
    \bu^\top \\
    \bv^\top
  \end{array}
\right)
\Bigg]
\bX(\bX^\top\bX)^{-1}.
\label{eq:two-stage-betax-exact-exact}
\ee
}

\end{proof}

\subsubsection{Consistency of the SuperCENT estimators}
\label{sec:consistency}


We show the non-asymptotic high probability bounds for the SuperCENT estimators $\hd, \hu, \hv$ and $\hbbeta$, along with the corresponding consistency results. 
We provide two versions of results: one for the solution of the objective function, presented in Section \ref{app:consistency-obj},
as well as one for the solution produced by the algorithm, presented in Section \ref{app:consistency-algo}.

\paragraph{Consistency results for the solution of the objective function.}
\label{app:consistency-obj}

We first establish the high probability bounds for the SuperCENT estimators as the minimizer of the objective function \eqref{eq:supercent-obj} in Theorem \ref{thm1}, followed by the corresponding consistency results in Corollary \ref{thm1-cor}. The proofs are provided subsequently.

\begin{theorem}
\label{thm1}
Under the unified framework
\eqref{eq:model-supercent} and Assumptions \ref{assump:normal}-\ref{assump:consistent},
suppose $(\bX, \hat{\bu}, \hat{\bv})$ is full rank where $(\bX, \hat{\bu}, \hat{\bv})$ is the minimizer of the objective function \eqref{eq:supercent-obj}, then with probability at least $1-6 e^{-n}$,

\begin{multline}
 \min \left\{\left\|\frac{\hat{\bu}-\bu}{\sqrt{n}}\right\|^{2},\left\|\frac{\hat{\bu}+\bu}{\sqrt{n}}\right\|^{2}\right\}+\min \left\{\left\|\frac{\hat{\bv}-\bv}{\sqrt{n}}\right\|^{2},\left\|\frac{\hat{\bv}+\bv}{\sqrt{n}}\right\|^{2}\right\}  \\
\leq  \frac{200 \sigma_{a}^{2}}{\lambdanew\left(\done-d_{2}\right) \done n}+\frac{2176 \sigma_{a}^{2}}{n\left(\done-d_{2}\right)^{2}},
\end{multline}

\be
\left|\frac{\hat{d}-\done}{\done}\right| \leq  \frac{4 \sigma_{a}}{\sqrt{n} \done}+
\frac{100 \sigma_{a}^{2}}{\lambdanew\left(\done-d_{2}\right) \done n}+\frac{1088 \sigma_{a}^{2}}{n\left(\done-d_{2}\right)^{2}}.
\ee

If we further have $\left(\bX, \bu, \bv\right)^{\top}\left(\bX, \bu, \bv\right)$ has condition number smaller or equal to $1/\tau^{2}$ for some positive constant $\tau$,
and when $\tau^{2} \geq 8 \frac{\sigma_{a}^{2}}{n\left(\done-d_{2}\right)^{2}}\left(2176+\frac{120}{\lambdanew}\right)$, we have with probability at least $1-e^{-3 n}-\frac{1}{n^{2}}$,
\be
\|\hat{\boldsymbol{\beta}}-\boldsymbol{\beta}\| &\leq & \left[ 8 \tau^{-4}\left(2+\frac{1}{2} \tau\right)\left(\frac{120 \sigma_{a}^{2}}{\lambdanew n \done\left(\done-d_{2}\right)}+\frac{2176 \sigma_{a}^{2}}{n\left(\done-d_{2}\right)^{2}}\right) \right.\\
&& \quad \left. + 4 \sqrt{2} \tau^{-2}\left(1+\frac{1}{2} \tau\right) \sqrt{\frac{120 \sigma_{a}^{2}}{\lambdanew n \done\left(\done-d_{2}\right)}+\frac{2176 \sigma_{a}^{2}}{n\left(\done-d_{2}\right)^{2}}} \right]  \sqrt{\beta_{u}^{2}+\beta_{v}^{2}} \\
&& \quad +\sqrt{\frac{120 \sigma_{a}^{2}}{\lambdanew n \done\left(\done-d_{2}\right)}+\frac{2176 \sigma_{a}^{2}}{n\left(\done-d_{2}\right)^{2}}} \cdot \sqrt{10} \cdot \tau^{-2}\left(1+\frac{8 \tau^{-2}+2 \tau_{1}}{\sqrt{n}}\right) \sigma_{y} \\
&& \quad +2 \sigma_{y} \tau^{2} \sqrt{\frac{1+2 \log n}{n}}.
\ee

\end{theorem}


\begin{cor}
\label{thm1-cor}
Under the assumptions and conditions in Theorem \ref{thm1}, if
\be
&&\frac{\sigma_{a}}{\sqrt{n}\left(\done-d_{2}\right)} \sqrt{1+\frac{1}{\lambdanew}}\left(1+\frac{\sigmay}{\sqrt{\beta_{u}^{2}+\beta_{v}^{2}}}\right) \rightarrow 0, \label{eq:thm1-cor-cond1} \\
&&\sigma_y \sqrt{\frac{\log{n}}{n}} \rightarrow 0, \label{eq:thm1-cor-cond2}
\ee
then we have
\be
\min \left\{\frac{\left\|\hat{\bu}-\bu\right\|}{\sqrt{n}}, \frac{\left\|\hat{\bu}+\bu\right\|}{\sqrt{n}}\right\}&\rightarrow& 0,\\
\min \left\{\frac{\left\|\hat{\bv}+\bv\right\|}{\sqrt{n}}, \frac{\left\|\hat{\bv}-\bv\right\|}{\sqrt{n}}\right\}&\rightarrow& 0, \\
\left|\frac{\hat{d}-\done}{\done}\right| &\rightarrow& 0, \\
\left\|\frac{\hat{\boldsymbol{\beta}}-\boldsymbol{\beta}}{\sqrt{\beta_{u}^{2}+\beta_{v}^{2}}}\right\| &\rightarrow& 0.
\ee

If we further assume
$\left|\frac{\beta_{u}}{\beta_{v}}\right| \in[\underline{\alpha}, \bar{\alpha}]$
for positive constants $\bar{\alpha}>\underline{\alpha}>0$, then with high probability
\be
\left|\frac{\hat{\beta}_{u}-\beta_{u}}{\beta_{u}}\right|\rightarrow 0 \quad \mbox{and} \quad \left|\frac{\hat{\beta}_{v}-\beta_{v}}{\beta_{v}}\right|\rightarrow 0.
\ee
\end{cor}


\begin{proof}
\label{prof:thm2}

We start to prove the bounds  for $\hu, \hv,$ and $\hd$, then proceed to $\hbbeta$.

\paragraph*{(1) Bounds for $\hu, \hv$.}
Clearly, $f_{1}(\hat{\bu}, \hat{\bv}) \leq f_{1}\left(\bu, \bv\right)$. Let $\delta=0$ in Proposition \ref{prop1} gives 
\be
\left(1-\frac{\left\langle\hat{\bu}, \bu\right\rangle^{2}}{2 n^{2}}-\frac{\langle\hat{\bv}, \bv\rangle^{2}}{2 n^{2}}\right) \leq \frac{30 \sigma_{a}^{2}}{\lambdanew\left(\done-d_{2}\right) \done n}+\frac{544 \sigma_{a}^{2}}{n\left(\done-d_{2}\right)^{2}}.
\ee

Note that 

\begin{multline}
1-\frac{\left\langle\hat{\bu}, \bu\right\rangle^{2}}{2 n^{2}}-\frac{\left\langle\hat{\bv}, \bv\right\rangle^{2}}{2 n^{2}} \geq  \\
\frac{1}{4}\left(\min \left\{\left\|\frac{\hat{\bu}-\bu}{\sqrt{n}}\right\|^{2},\left\|\frac{\hat{\bu}+\bu}{\sqrt{n}}\right\|^{2}\right\}  \right.
\left. + \min \left\{\left\|\frac{\hat{\bv}-\bv}{\sqrt{n}}\right\|^{2},\left\|\frac{\hat{\bv}+\bv}{\sqrt{n}}\right\|^{2}\right\}\right).
\end{multline}

Combine the two inequalities above, we obtain the statement for $\hat{\bu}$ and $\hat{\bv}$.

\paragraph*{(2) Bounds for $\hd$.}
Note that 
\be
\left|\hat{d}-\done\right| & =& \left|\frac{\hat{\bu}^{\top} \bA \hat{\bv}}{n^2}-\frac{\bu^{\top} \bA_{0} \bv}{n^2}\right| \\
& =&\left|\frac{\hat{\bu}^{\top} \bE\hat{\bv}}{n^2}\right|+\left|\frac{\left(\hat{\bu}-\bu^{\top}\right) \bA_{0} \hat{\bv}}{n^2}\right|+ \left|\frac{\bu^{\top} \bA_{0}\left(\hat{\bv}-\bv\right)}{n^2}\right| \\
& \leq& \frac{\|\bE\|_{op}}{n}+\done\left(\frac{\left\|\hat{\bu}-\bu\right\|^{2}+\left\|\hat{\bv}_{1}-\bv\right\|^{2}}{2n}\right)
\ee

Note that, under the high probability event with probability $1-3 e^{-n}$, the bounds for $\hat{\bu}$ and $\hat{\bv}$ hold in Proposition \ref{prop1}, we have

\be
\|\bE\|_{op} \leq 4 \sqrt{n} \sigma_{a}.
\ee

Therefore, under the event that the bounds for $\hat{\bu}, \hat{\bv}$ hold, we have

\be
\left|\hat{d}-\done\right| \leq \frac{4 \sigma_{a}}{\sqrt{n}}+\done\left(\frac{60 \sigma_{a}^{2}}{\lambdanew n \done\left(\done-d_{2}\right)}+\frac{1088 \sigma_{a}^{2}}{n\left(\done-d_{2}\right)^{2}}\right) .
\ee

\paragraph*{(3) Bounds for $\hbbeta$.}

Consider the high probability event that the bounds for $\hat{\bu}, \hat{\bv}$ in Proposition \ref{prop1} holds.
Let 
$g(\omega)=\left(\omega^{\top} \omega\right)^{-1} \omega^{\top}$,
then 
\be
\hat{\boldsymbol{\beta}}&=&g(\bX, \hat{\bu}, \hat{\bv})[(\bX, \bu, \bv) \boldsymbol{\beta}+\varepsilon_{y}] \\
&=&\left(\begin{array}{l}\boldsymbol{\beta}_{x} \\ 0  \\ 0 \end{array}\right)+g(\bX, \hat{\bu}, \hat{\bv})\left[(\bX, \bu, \bv)\left(\begin{array}{l} 0 \\ \beta_{u} \\ \beta_{v} \end{array}\right)+\varepsilon_{y}\right] .
\ee

Let 
\be
\left \{
\begin{aligned}
\omega &=(\bX, \bu, \bv), \\
\Delta &= \left(0, \hat{\bu}-\bu, \hat{\bv}-\bv\right), \\
\bar{\boldsymbol{\beta}} &= \left(\begin{array}{l}0 \\ \beta_{u} \\ \beta_{v} \end{array}\right).
\end{aligned}
\right .
\ee

Then
\be
\label{eq:thm1-proof}
\|\hat{\boldsymbol{\beta}}-\boldsymbol{\beta}\| & \leq& \|\left[g(\omega+\Delta)-g(\omega)\right] \omega \bar{\boldsymbol{\beta}}\|+\|g(\omega+\Delta) \varepsilon_{y}\| \\
& \leq& \| \left[g(\omega+\Delta)-g(\omega)\right] \omega \bar{\boldsymbol{\beta}}\|+\|g(\omega+\Delta)-g(\omega)\|_{op}\left\|\varepsilon_{y}\right\|_{2}+\left\|g(\omega) \varepsilon_{y}\right\| .
\ee

We start with boundary the first term in Inequality \eqref{eq:thm1-proof}. Note that 
\be
g(\omega+\Delta)-g(\omega)=\left(\omega^{\top} \omega\right)^{-1} \Delta^{\top}+\left[\left[(\omega+\Delta)^{\top}(\omega+\Delta)\right]^{-1}-\left[\omega^{\top} \omega\right]^{-1}\right] \omega^{\top}.
\ee

Then,
\be
&& [g(\omega+\Delta)-g(\omega)] \omega \bar{\boldsymbol{\beta}} \\
&= &
\begin{multlined}[t]
\left(\omega^{\top} \omega\right)^{-1} \Delta^{\top} \omega \bar{\boldsymbol{\beta}} +\left[(\omega+\Delta)^{\top}(\omega+\Delta)\right]^{-1}\\
\left[\bI-\left(\omega^{\top} \omega+\Delta^{\top} \omega+\omega^{\top} \Delta+\Delta^{\top} \Delta\right)\left(\omega^{\top} \omega\right)^{-1}\right] \omega^{\top} \omega \bar{\boldsymbol{\beta}} 
\end{multlined} \\
&= & \left(\omega^{\top} \omega\right)^{-1} \Delta^{\top} \omega \bar{\boldsymbol{\beta}}-\left((\omega+\Delta)^{\top}(\omega+\Delta)\right)^{-1}\left(\Delta^{\top} \omega \bar{\boldsymbol{\beta}}+\omega^{\top} \Delta \bar{\boldsymbol{\beta}}+\Delta^{\top} \Delta \bar{\boldsymbol{\beta}}\right) \\
&= & 
\begin{multlined}[t] 
\left[(\omega+\Delta)^{\top}(\omega+\Delta)\right]^{-1}\left(\Delta^{\top} \omega+\omega^{\top} \Delta+\Delta^{\top} \Delta\right)\left(\omega^{\top} \omega\right)^{-1} \Delta^{\top} \omega \bar{\boldsymbol{\beta}} \\
 -\left[(\omega+\Delta)^{\top}(\omega+\Delta)\right]^{-1} \omega^{\top} \Delta \bar{\boldsymbol{\beta}}-\left[(\omega+\Delta)^{\top}(\omega+\Delta)\right]^{-1} \Delta^{\top} \Delta \bar{\boldsymbol{\beta}} .
\end{multlined} 
\ee

Let's consider the smallest eigenvalue of $(\omega+\Delta)^{\top}(\omega+\Delta)$,
\be
\lambdanew_{n}\left((\omega+\Delta)^{\top}(\omega+\Delta)\right)=\min _{z \in \mathbb{R}^{n}}\|(\omega+\Delta) z\|^{2}.
\ee

Note that 
\be
\|\omega z\| \geq \sqrt{\lambdanew_{n}\left(\omega^{\top} \omega\right)} \geq \tau\|\omega\|_{op} \geqslant \tau \sqrt{n},
\ee
and
\be
\|\Delta z\| \leq \sqrt{2 n} \max \left\{\frac{\left\|\hat{\bu}-\bu\right\|}{\sqrt{n}}, \frac{\left\|\hat{\bv}-\bv\right\|}{\sqrt{n}}\right\} \leq \sqrt{2 n} \sqrt{\frac{120 \sigma_{a}^{2}}{\lambdanew n \done\left(\done-d_{2}\right)}+\frac{2176 \sigma_{a}^{2}}{n\left(\done-d_{2}\right)^{2}}}.
\ee

Then we have 
\be
\lambdanew_{n}\left((\omega+\Delta)^{\top}(\omega+\Delta)\right) \geq \frac{1}{4} \tau^{2}\|\omega\|_{o p}^{2}.
\ee

Therefore,
\be
&& \|(g(\omega+\Delta)-g(\omega)) \omega \bar{\boldsymbol{\beta}}\| \\
&\leq & 4 \tau^{-4}\|\omega\|_{o p}^{-4} \cdot\left(2\|\Delta\|_{o p} \cdot\|\omega\|_{o p}+\|\Delta\|_{o p}^{2}\right)\|\Delta\|_{o p}\|\omega\|_{o p}\|\bar{\boldsymbol{\beta}}\|_{2} \\
&& + 4 \tau^{-2}\|\omega\|_{o p}^{-2}\left[\|\omega\|_{o p}\|\Delta \bar{\boldsymbol{\beta}}\|_{2}+\|\Delta\|_{o p}\|\Delta \bar{\boldsymbol{\beta}}\|_{2}\right] \\
&\leq & 4 \tau^{-4}\|\omega\|_{o p}^{-4} \cdot\left(2+\frac{1}{2} \tau\right)\|\omega\|_{o p}^{2} \cdot\|\Delta\|_{o p}^{2}\|\bar{\boldsymbol{\beta}}\|_{2} \\
&& + 4 \tau^{-2}\|\omega\|_{o p}^{-2}\left(1+\frac{1}{2} \tau\right)\|\omega\|_{o p}\|\Delta\|_{o p}\|\bar{\boldsymbol{\beta}}\|_{2} \\
&\leq & 4 \tau^{-4}\left(2+\frac{1}{2} \tau\right) 2\left[\frac{120 \sigma_{a}^{2}}{\lambdanew n \done\left(\done-d_{2}\right)}+\frac{2176 \sigma_{a}^{2}}{n\left(\done-d_{2}\right)^{2}}\right]\|\bar{\boldsymbol{\beta}}\|_{2} \\
&& + 4 \tau^{-2}\left(1+\frac{1}{2} \tau\right) \sqrt{2} \sqrt{\frac{120 \sigma_{a}^{2}}{\lambdanew n \done\left(\done-d_{2}\right)}+\frac{2176 \sigma_{a}^{2}}{n\left(\done-d_{2}\right)^{2}}}\|\bar{\boldsymbol{\beta}}\|_{2}
\ee

Now we turn to the second term in Inequality \eqref{eq:thm1-proof}.
\be
&& \|g(\omega+\Delta)-g(\omega)\|_{o p}\left\|\varepsilon_{y}\right\| \\
&\leq & \left[\tau^{-2}\|\omega\|_{op}^{-2}\|\Delta\|_{op}+4 \tau^{-2}\|\omega\|_{op}^{-2} \tau^{-2}\|\omega\|_{o p}^{-2}\left(2\|\Delta\|_{o p}\|\omega\|_{o p}+\|\Delta\|_{o p}^{2}\right)\right]\|\omega\|_{o p}\left\|\varepsilon_{y}\right\|_{2} \\
&\leq & \tau^{-2}\|\omega\|_{op}^{-1}\|\Delta\|_{op}\left[1+4 \tau^{-2}\|\omega\|_{o p}^{-1}\left(2+\frac{\tau}{2}\right)\right]\|\varepsilon_{y}\|_{2} \\
&\leq & \sqrt{\frac{120 \gamma_{a}^{2}}{\lambdanew n \done\left(\done-d_{2}\right)}+\frac{2176 \sigma_{a}^{2}}{n\left(\done-d_{2}\right)^{2}}} \cdot \sqrt{2} \tau^{-2}\left\|\frac{\varepsilon_{y}}{\sqrt{n}}\right\|\left(1+\frac{8 \tau^{-2}+2 \tau^{-1}}{\sqrt{n}}\right) .
\ee

Under the high probability event that the bounds for $\hat{\bu}, \hat{\bv}$ in Proposition \ref{prop1} hold, we have

\be
&&\left\|g(\omega+\Delta)-g(\omega)\right\|_{op} \left\|\varepsilon_{y} \right\| \\
&\leq& \sqrt{\frac{120 \gamma_{a}^{2}}{\lambdanew n \done\left(\done-d_{2}\right)}+\frac{2176 \sigma_{a}^{2}}{n\left(\done-d_{2}\right)^{2}}} \sqrt{10} \cdot \tau^{-2}\left(1+\frac{8 \tau^{-2}+2 \tau^{-1}}{\sqrt{n}}\right) \sigma_{y}.
\ee

Now we turn to the third term in Inequality \eqref{eq:thm1-proof}.
First note that 
\be
g(\omega) \cdot \varepsilon_y \sim N\left(\overrightarrow{0}, \sigma_{y}^{2} \cdot\left(\omega^{\top} \omega\right)^{-1}\right).
\ee
Let $h=g(\omega) \cdot \varepsilon_{y}$, then for $\xi > 0$, 

\be
\mathbb{E} e^{\xi\|h\|^{2}} & =& \int e^{\xi\left(\varepsilon_{y}^{\top} \omega\left(\omega^{\top} \omega\right)^{-2} \omega^{\top} \varepsilon_{y}\right)-\frac{1}{2 \sigma_{y}^{2}}\|\varepsilon_{y}\|^{2}} \frac{1}{(\sqrt{2 \pi} \sigma_{y})^{p+2}} d \varepsilon_{y} \\
& \leq& \int e^{\left(\xi \tau^{4} \cdot\|\omega\|_{op}^{-2}-\frac{1}{2 \sigma_{y}^{2}}\right)\left\|\varepsilon_{y}\right\|^{2}} \frac{1}{\left(\sqrt{2 \pi} \sigma_{y}\right)^{p+2}} d \varepsilon_{y} \\
&=&\left(\frac{1}{\sqrt{1-2 \sigma_{y}^{2} \xi \cdot \tau^{4}\|\omega\|_{op}^{-2}}}\right)^{p+2} \\
&\leq& \left(\frac{1}{\sqrt{1-2 \sigma_{y}^{2} \tau^{4} \cdot \frac{1}{n} \xi}}\right)^{p+2};
\ee

and for $\xi<\frac{n}{2 \sigma_{y}^{2} \tau^{4}}$, 

\be
P\left(\|h\|^{2}>k\right) \leq \left(\frac{1}{\sqrt{1-2 \sigma_{y}^{2} \tau^{4} \cdot \frac{1}{n} \xi}}\right)^{p+2}  e^{-\xi k}.
\ee

Let 
$\xi=\frac{n}{4 \sigma_{y}^{2}  \tau^{4}}, \,
k=\frac{4 \sigma_{y}^{2}  \tau^{4}}{n}\left(1+2 \log n \right)$, 
we have
\be
P\left(\left\|g(\omega)  \varepsilon_{y}\right\|>\frac{2 \sigma_{y}  \tau^{2}}{\sqrt{n}} \sqrt{1+2 \log n}\right)<\frac{1}{n^{2}}.
\ee

Combining the above results for the three terms in \eqref{eq:thm1-proof} gives, with probability at least $1-e^{3 n}-\frac{1}{n^{2}}$,

\be
\|\hat{\boldsymbol{\beta}}-\boldsymbol{\beta}\| &\leq & \left[
8 \tau^{-4}\left(2+\frac{1}{2} \tau\right)\left(\frac{120 \sigma_{a}^{2}}{\lambdanew n \done\left(\done-d_{2}\right)}+\frac{2176 \sigma_{a}^{2}}{n\left(\done-d_{2}\right)^{2}}\right) \right. \\
&& \left. +4 \sqrt{2} \tau^{-2}\left(1+\frac{1}{2} \tau\right) \sqrt{\frac{120 \sigma_{a}^{2}}{\lambdanew n \done\left(\done-d_{2}\right)}+\frac{2176 \sigma_{a}^{2}}{n\left(\done-d_{2}\right)^{2}}} \right] \sqrt{\beta_{u}^{2}+\beta_{v}^{2}} \\
&& +\sqrt{\frac{120 \sigma_{a}^{2}}{\lambdanew n \done \left( \done-d_{2}\right)}+\frac{2176 \sigma_{a}^{2}}{n\left(\done-d_{2}\right)^{2}}} \sqrt{10} \tau^{-2}\left(1+\frac{8 \tau^{-2}+2 \tau^{-1}}{\sqrt{n}}\right) \sigma_{y} \\
&& +2 \sigma_{y} \tau^{2} \sqrt{\frac{1+2 \log n}{n}}.
\ee

Finally, Corollary \ref{thm1-cor}  directly follows from Theorem \ref{thm1} under the conditions \eqref{eq:thm1-cor-cond1}-\eqref{eq:thm1-cor-cond2}.

\end{proof}

\paragraph{Consistency results for the solution of Algorithm \ref{algo:supercent}.}
\label{app:consistency-algo}

We first establish the high probability bounds for the SuperCENT estimators as the solution of Algorithm \ref{algo:supercent} in Theorem \ref{thm2}, followed by the corresponding consistency results in Corollary \ref{thm2-cor}. The proofs are provided subsequently.

\begin{theorem}
\label{thm2}
Under the unified framework
\eqref{eq:model-supercent} and Assumptions \ref{assump:normal}-\ref{assump:consistent},
suppose $(\bX, \bu, \bv)$ is full rank and $(\hu, \hv)$ is produced by Algorithm \ref{algo:supercent}, then with probability at least $1-6 e^{-n}$,

\begin{multline}
 \min \left\{\left\|\frac{\hat{\bu}-\bu}{\sqrt{n}}\right\|^{2},\left\|\frac{\hat{\bu}+\bu}{\sqrt{n}}\right\|^{2}\right\}+\min \left\{\left\|\frac{\hat{\bv}-\bv}{\sqrt{n}}\right\|^{2},\left\|\frac{\hat{\bv}+\bv}{\sqrt{n}}\right\|^{2}\right\}  \\
\leq  \frac{200 \sigma_{a}^{2}}{\lambdanew\left(\done-d_{2}\right) \done n}+\frac{2176 \sigma_{a}^{2}}{n\left(\done-d_{2}\right)^{2}} + 4096 \frac{\beta_{u}^{2}+\beta_{v}^{2}}{\sigma_{y}^{2}} \frac{\sigma_{a}^{2}}{n\left(\done-d_{2}\right)^{2}} \frac{\sigma_{a}^{2}}{n^{2} \lambdanew \done\left(\done-d_{2}\right)},
\end{multline}

\begin{multline}
\left|\frac{\hat{d}-\done}{\done}\right| \leq  \frac{4 \sigma_{a}}{\sqrt{n} \done}+
\frac{100 \sigma_{a}^{2}}{\lambdanew\left(\done-d_{2}\right) \done n}+\frac{1088 \sigma_{a}^{2}}{n\left(\done-d_{2}\right)^{2}} \\
+ 2048 \frac{\beta_{u}^{2}+\beta_{v}^{2}}{\sigma_{y}^{2}} \frac{\sigma_{a}^{2}}{n\left(\done-d_{2}\right)^{2}} \frac{\sigma_{a}^{2}}{n^{2} \lambdanew \done\left(\done-d_{2}\right)} .
\end{multline}

If we further have $\left(\bX, \bu, \bv\right)^{\top}\left(\bX, \bu, \bv\right)$ has condition number smaller or equal to $1/\tau^{2}$ for some positive constant $\tau$, then with probability $1-6 e^{-n}-\frac{1}{n^{2}}$, the above holds and
\begin{multline}
\|\hat{\boldsymbol{\beta}}-\boldsymbol{\beta}\| \leq  \left(8 \tau^{-4}\left(2+\frac{1}{2} \tau\right) \eta+4 \sqrt{2} \tau^{-2}\left(1+\frac{1}{2} \tau\right) \sqrt{\eta}\right) \sqrt{\beta_{u}^{2}+\beta_{v}^{2}} \\
 +\sqrt{10 \eta} \sigma_{y} \tau^{-2}\left(1+\frac{8 \tau^{-2}+2 \tau^{-1}}{\sqrt{n}}\right)+ 2\sigma_{y} \tau^{2} \sqrt{\frac{1+p/2+2 \log n}{n}} .
\end{multline}

\end{theorem}

\begin{cor}
\label{thm2-cor}
Under the assumptions and conditions of Theorem \ref{thm2}, if
\be
 &&\left(1+\frac{1}{\lambdanew}\right) \frac{\sigma_{a}^{2}}{n\left(\done-d_{2}\right)^{2}}\left(1+\frac{\sigma_{y}^{2}}{\beta_{u}^{2}+\beta_{v}^{2}}\right) \rightarrow 0, \label{eq:thm2-cor-cond1} \\
&&\sqrt{\frac{\beta_{u}^{2}+\beta_{v}^{2}}{\sigma_{y}^{2}}} \cdot \sqrt{\frac{1}{\lambdanew}} \cdot \frac{\sigma_{a}^{2}}{n\left(\done-d_{2}\right)^{2}} \rightarrow 0, \label{eq:thm2-cor-cond2} \\
&&\frac{\sigma_{y}}{\sqrt{\beta_{u}^{2}+\beta_{v}^{2}}} \sqrt{\frac{\log n}{n}} \rightarrow 0, \label{eq:thm2-cor-cond3}
\ee
then
with high probability
\be
\min \left\{\frac{\left\|\hat{\bu}-\bu\right\|}{\sqrt{n}}, \frac{\left\|\hat{\bu}+\bu\right\|}{\sqrt{n}}\right\}&\rightarrow& 0,\\
\min \left\{\frac{\left\|\hat{\bv}+\bv\right\|}{\sqrt{n}}, \frac{\left\|\hat{\bv}-\bv\right\|}{\sqrt{n}}\right\}&\rightarrow& 0, \\
\left|\frac{\hat{d}-\done}{\done}\right| &\rightarrow& 0, \\
\left\|\frac{\hat{\boldsymbol{\beta}}-\boldsymbol{\beta}}{\sqrt{\beta_{u}^{2}+\beta_{v}^{2}}}\right\| &\rightarrow& 0.
\ee

If we further assume
$\left|\frac{\beta_{u}}{\beta_{v}}\right| \in[\underline{\alpha}, \bar{\alpha}]$
for positive constants $\bar{\alpha}>\underline{\alpha}>0$, then with high probability
\be
\left|\frac{\hat{\beta}_{u}-\beta_{u}}{\beta_{u}}\right|&\rightarrow& 0,\\
\left|\frac{\hat{\beta}_{v}-\beta_{v}}{\beta_{v}}\right|&\rightarrow& 0.
\ee

\end{cor}

\begin{proof}
\label{prof:thm2}
Denote the leading singular vector of network $\bA$ as $\bar{\bu}$ and $\bar{\bv}$ with their corresponding leading singular value as $\bar{d}$ for simplicity in notation.

\paragraph*{(1) Bounds for $\hu, \hv$.}

Plugging the unified model \eqref{eq:model-supercent} into $f_{1}(\bar{\bu}, \bar{\bv})-f_{1}\left(\bu, \bv\right)$,
we have
\be 
&& f_{1}(\bar{\bu}, \bar{\bv})-f_{1}\left(\bu, \bv\right) \\
&= & \frac{1}{\sigma_{y}^{2}}\left\|\bP_{(\bX, \bar{\bu}, \bar{\bv})^{\perp}}\left(\Tilde{\varepsilon}+\Tilde{\bu}_{1} \beta_{u}+\Tilde{\bv}_{1} \beta_{v}\right)\right\|^{2} -\frac{\lambdanew}{\sigma_{a}^{2}}\left(\bar{\bu}^{\top}\left(\sum_{i=1}^{r} d_{i} \bu_{i} \bv_{i}^{\top}+\bE\right) \bar{\bv} / n\right)^{2} \\
&& -\left( \frac{1}{\sigma_{y}^{2}} \left\|\bP_{(\bX, \bu, \bv)^{\perp}} \left( \Tilde{\varepsilon} + \Tilde{\bu}_{1} \beta_{u}+ \Tilde{\bv}_{1} \beta_{v} \right) \right\|^{2}  - \frac{\lambdanew}{\sigma_{a}^{2}}\left(\bu^{\top}\left(\sum_{i=1}^{r} d_{i} \bu_{i} \bv_{i}^{\top}+\bE\right) \bv / n\right)^{2} \right) \\
&\leq & \frac{1}{\sigma_{y}^{2}}\left[\left\|\bP_{(\bX, \bar{\bu}, \bar{\bv})^{\perp}}\Tilde{\varepsilon}\right\|^{2}-\left\|\bP_{\left(\bX, \bu, \bv\right)^{\perp}}\Tilde{\varepsilon}\right\|^{2} \right. \\
&& \left.+2\left\|\bP_{(\bX, \bar{\bu}, \bar{\bv})^{\perp}}\Tilde{\varepsilon}\right\|\left\|\bP_{(\bX, \bar{\bu}, \bar{\bv})^{\perp}} \left( \Tilde{\bu}_{1} \beta_{u}+ \Tilde{\bv}_{1} \beta_{v} \right)\right\|^{2}  +\left\|\bP_{(\bX, \bar{\bu}, \bar{\bv})^{\perp}}\left(\Tilde{\bu}_{1} \beta_{u}+ \Tilde{\bv}_{1} \beta_{v}\right)\right\|^{2}\right] \\
&\leq & 2 \frac{\|\Tilde{\varepsilon}\|^{2}}{ \sigma_y^{2}}+2\left\|\bP_{(\bX, \bar{\bu}, \bar{\bv})^{\perp}}\left(\Tilde{\bu}_{1} \frac{\beta_{u}}{\sigma_{y}}+ \Tilde{\bv}_{1} \frac{\beta_{v}}{\sigma_{y}}\right)\right\|^{2} \\
&\leq & 2\frac{\|\varepsilon\|^{2}}{ \sigma_{y}^{2}}+
4n\left(\frac{\beta_{u}}{\sigma_{y}}\right)^{2}\left[1-\left(\frac{\langle \bu, \bar{\bu}\rangle}{n}\right)^{2}\right]+  
4n\left(\frac{\beta_{v}}{\sigma_{y}}\right)^{2}\left[1-\left(\frac{\langle \bv, \bar{\bv}\rangle}{n}\right)^{2}\right].
\label{eq:thm2-proof-eq1} 
\ee

We next focus on bounding the following terms,
\be
1-\left(\frac{\langle \bu, \bar{\bu}\rangle}{n}\right)^{2}
\quad
\mbox{and}
\quad
1-\left(\frac{\langle \bv, \bar{\bv}\rangle}{n}\right)^{2}.
\ee

\jcM{Recall that $\bar{\bu}$ and $\bar{\bv}$ are the leading singular vectors of the observed network $\bA$, where $\bA = \sum_{i=1}^{r} d_{i} \bu_{i} \bv_{i}^{\top}+\bE$ as defined in our network model \eqref{eq:model-supercent1}}.
By definition of $\bar{\bu}, \bar{\bv}$, we have
\be
\bar{\bu}^{\top}\left(\sum_{i=1}^{r} d_{i} \bu_{i} \bv_{i}^{\top}+\bE\right) \bar{\bv} \geq \bu^{\top}\left(\sum_{i=1}^{r} d_{i} \bu_{i} \bv_{i}^{\top}+\bE\right) \bv.
\ee
Rearrange with some algebra,
\begin{multline}
    \operatorname{tr}\left[\bE\left(\bar{\bv}\bar{\bu}^{\top}-\bv \bu^{\top}\right)\right] \geq 
    n^{2} \done\left(1-\frac{\left\langle \bv, \bar{\bv}\right\rangle}{n} \frac{\left\langle \bu, \bar{\bu}\right\rangle}{n}\right) \\
    -n^{2} d_{2} \sqrt{1-\left(\frac{\left\langle \bv, \bar{\bv}\right\rangle}{n}\right)^{2}} \sqrt{1-\left(\frac{\left\langle \bu, \bar{\bu}\right\rangle}{n}\right)^{2}}.
\end{multline}
Then
\be
\sqrt{2}\|\bE\|_{op}\left\|\bar{\bv} {\bar{\bu}}^{\top}-\bv \bu^{\top}\right\|_{F} \geq n^{2}\left(\done-d_{2}\right)\left(1-\frac{\left\langle \bv, \bar{\bv}\right\rangle}{n} \frac{\left\langle \bu, \bar{\bu}\right\rangle}{n}\right).
\ee

Note that 
\be
\left\|\bar{\bv} \bar{\bu}^{\top}-\bv \bu^{\top}\right\|_{F}^{2} & \leq n^{2}\left(2-2 \frac{\left\langle\bar{\bv}_{1} \bv\right\rangle\left\langle \bu, \bar{\bu}\right\rangle}{n^{2}}\right)
\leq 4 n^{2}\left(1-\frac{\left\langle\bar{\bv}_{1} \bv\right\rangle\left\langle \bu, \bar{\bu}\right\rangle}{n}\right).
\ee

Therefore,
\be
\label{eq:thm2-proof-eq2} 
\left(1-\frac{\left\langle \bv, \bar{\bv}\right\rangle}{n} \frac{\left\langle \bu, \bar{\bu}\right\rangle}{n}\right) \leq \frac{32\|\bE\|_{op}^{2}}{\left(\done-d_{2}\right)^{2} n^{2}}.
\ee

Further,

\be
1-\frac{\left\langle \bv, \bar{\bv}\right\rangle}{n} \frac{\left\langle \bu, \bar{\bu}\right\rangle}{n} 
&\geq& \max \left\{1-\left|\frac{\left\langle \bv, \bar{\bv}\right\rangle}{n}\right|, 1-\left|\frac{\langle \bu, \bar{\bu}\rangle}{n}\right|\right\} \\
&\geq& \max \left\{\min \left(\left\|\frac{\bv-\bar{\bv}}{\sqrt{n}}\right\|^{2},\left\|\frac{\bv+\bar{\bv}}{\sqrt{n}}\right\|^{2}\right)  , \min \left(\left\|\frac{\bu-\bar{\bu}}{\sqrt{n}}\right\|^{2},\left\|\frac{\bu+\bar{\bu}}{\sqrt{n}}\right\|^{2}\right)\right\}.
\label{eq:thm2-proof-eq3} 
\ee

Plug \eqref{eq:thm2-proof-eq2} and \eqref{eq:thm2-proof-eq3} into \eqref{eq:thm2-proof-eq1} gives
\be
&& f_{1}(\bar{\bu}, \bar{\bv})-f_{1}\left(\bu, \bv\right) \leq  2\|\varepsilon\|^{2} / \sigma_{y}^{2}
+ 4n\left(\frac{\beta_{u}}{\sigma_{y}}\right)^{2} \frac{32\|\bE\|_{op}^{2}}{\left(\done-d_{2}\right)^{2} n^{2}} 
+ 4n\left(\frac{\beta_{v}}{\sigma_{y}}\right)^{2} \frac{32\|\bE\|_{op}^{2}}{\left(\done-d_{2}\right)^{2} n^{2}}.
\ee

Using Lemma \ref{lemma3} and Lemma \ref{lemma4}, we know that with probability at least $1-3 e^{-n}$,

\be
f_{1}(\bar{\bu}, \bar{\bv})-f_{1}\left(\bu, \bv\right) \leq 10 n+512 \frac{\beta_{u}^{2}+\beta_{v}^{2}}{\sigma_{y}^{2}} \frac{\sigma_{a}^{2}}{n\left(\done-d_{2}\right)^{2}} .
\ee

Note that the algorithm descends in $f_{1}$,
so 
\be
f_{1}(\hat{\bu}, \hat{\bv})-f_{1}\left(\bu, \bv\right) \leq 10 n+512 \frac{\beta_{u}^{2}+\beta_{v}^{2}}{\sigma_{y}^{2}} \frac{\sigma_{a}^{2}}{n\left(\done-d_{2}\right)^{2}}
\ee
with probability at least $1-3 e^{-n}$.

Invoking Proposition \ref{prop1}, we have

\begin{multline}
\label{eq:1-uv-leq}
1-\frac{\left\langle\hat{\bu}_{1} \bu\right\rangle^{2}}{2 n^{2}}-\frac{\left\langle\hat{\bv}_{1} \bv\right)^{2}}{2 n^{2}}
\leq \\  \frac{50 \sigma_{a}^{2}}{\lambdanew\left(\done-d_{2}\right) \done n}+\frac{544 \sigma_{a}^{2}}{n\left(\done-d_{2}\right)^{2}} 
 +1024 \frac{\beta_{u}^{2}+\beta_{v}^{2}}{\sigma_{y}^{2}} \frac{\sigma_{a}^{2}}{n\left(\done-d_{2}\right)^{2}} \frac{\sigma_{a}^{2}}{n^{2} \lambdanew \done\left(\done-d_{2}\right)}.
\end{multline}

Note that

\be
\label{eq:1-uv-geq}
1-\frac{\left\langle\hat{\bu}, \bu\right\rangle^{2}}{2 n^{2}}-\frac{\left\langle\hat{\bv}, \bv\right\rangle^{2}}{2 n^{2}} \geq \frac{1}{4} & \left(\min \left\{\left\|\frac{\hat{\bu}-\bu}{\sqrt{n}}\right\|^{2},\left\|\frac{\hat{\bu}+\bu}{\sqrt{n}}\right\|^{2}\right\}+
\min \left\{\left\|\frac{\hat{\bv}-\bv}{\sqrt{n}}\right\|^{2},\left\|\frac{\hat{\bv}+\bv}{\sqrt{n}}\right\|^{2}\right\}\right).
\ee
Combining \eqref{eq:1-uv-leq} and \eqref{eq:1-uv-geq}, we have
\begin{multline}
 \min \left\{\left\|\frac{\hat{\bu}-\bu}{\sqrt{n}}\right\|^{2},\left\|\frac{\hat{\bu}+\bu}{\sqrt{n}}\right\|^{2}\right\}+\min \left\{\left\|\frac{\hat{\bv}-\bv}{\sqrt{n}}\right\|^{2},\left\|\frac{\hat{\bv}+\bv}{\sqrt{n}}\right\|^{2}\right\}  \\
\leq  \frac{200 \sigma_{a}^{2}}{\lambdanew\left(\done-d_{2}\right) \done n}+\frac{2176 \sigma_{a}^{2}}{n\left(\done-d_{2}\right)^{2}} + 4096 \frac{\beta_{u}^{2}+\beta_{v}^{2}}{\sigma_{y}^{2}} \frac{\sigma_{a}^{2}}{n\left(\done-d_{2}\right)^{2}} \frac{\sigma_{a}^{2}}{n^{2} \lambdanew \done\left(\done-d_{2}\right)} .
\end{multline}

\paragraph*{(2) Bounds for $\hd$.}
Now we turn to $\hat{d}$. Under the high probability event that the bounds for $\hat{\bu}$ and $\hat{\bv}$ hold in Proposition \ref{prop1},

\be
\left|\hat{d}-\done\right| & = & \left|\frac{\hat{\bu}^{\top} \bA \hat{\bv}}{n \cdot n}-\frac{\bu^{\top} \bA_{0} \bv}{n \cdot n}\right| \\
&= & \left|\frac{\hat{\bu}^{\top} \bE \hat{\bv}}{n \cdot n}\right|+\left|\frac{\left(\hat{\bu}-\bu^{\top}\right) \bA_{0} \hat{\bv}}{n \cdot n}\right|+ \left|\frac{\bu^{\top} \bA_{0}\left(\hat{\bv}-\bv\right)}{n \cdot n}\right|\\
&\leq & \frac{\| \bE\|_{op}}{n}+\done\left(\frac{\left\|\hat{\bu}-\bu\right\|^{2}+\left\|\hat{\bv}-\bv\right\|^{2}}{2 n}\right) \\
&\leq & 
\begin{multlined}[t]
\frac{4 \sigma_{a}}{\sqrt{n}}+\done \Bigg(\frac{100 \sigma_{a}^{2}}{\lambdanew\left(\done-d_{2}\right) \done n}+\frac{1088 \sigma_{a}^{2}}{n\left(\done-d_{2}\right)^{2}} \\
+ 2048 \frac{\beta_{u}^{2}+\beta_{v}^{2}}{\sigma_{y}^{2}} \frac{\sigma_{a}^{2}}{n\left(\done-d_{2}\right)^{2}} \frac{\sigma_{a}^{2}}{n^{2} \lambdanew \done\left(\done-d_{2}\right)} \Bigg).
\end{multlined}
\ee

\paragraph*{(3) Bounds for $\hbbeta$.}

Denote
\be
\eta &=&\frac{200 \sigma_{a}^{2}}{\lambdanew\left(\done-d_{2}\right) \done n}+\frac{2176 \sigma_{a}^{2}}{n\left(\done-d_{2}\right)^{2}} 
+ 4096 \frac{\beta_{u}^{2}+\beta_{v}^{2}}{\sigma_{y}^{2}} \frac{\sigma_{a}^{2}}{n\left(\done-d_{2}\right)^{2}} \frac{\sigma_{a}^{2}}{n^{2} \lambdanew \done\left(\done-d_{2}\right)}.
\ee

Define
$g(\omega)=\left(\omega^{\top} \omega\right)^{-1} \omega^{\top}$,
then 
\be
\hat{\boldsymbol{\beta}}&=&g(\bX, \hat{\bu}, \hat{\bv})[(\bX, \bu, \bv) \boldsymbol{\beta}+\varepsilon_{y}] \\
&=&\left(\begin{array}{l}\boldsymbol{\beta}_{x} \\ 0  \\ 0 \end{array}\right)+g(\bX, \hat{\bu}, \hat{\bv})\left[(\bX, \bu, \bv)\left(\begin{array}{l} 0 \\ \beta_{u} \\ \beta_{v} \end{array}\right)+\varepsilon_{y}\right] .
\ee

Let 
\be
\left \{
\begin{aligned}
\omega &= (\bX, \bu, \bv), \\
\Delta &= \left(0, \hat{\bu}-\bu, \hat{\bv}-\bv\right), \\
\bar{\boldsymbol{\beta}} &= \left(\begin{array}{l}0 \\ \beta_{u} \\ \beta_{v} \end{array}\right), 
\end{aligned}
\right .
\ee
then
\be
\hat{\boldsymbol{\beta}}-\boldsymbol{\beta}=\left[g(\omega+\Delta)-g(\omega)\right] \omega \bar{\boldsymbol{\beta}}+ g(\omega+\Delta)\varepsilon_{y} .
\ee

Note that
\be
\|\hat{\boldsymbol{\beta}}-\boldsymbol{\beta}\| & \leq& \|\left[g(\omega+\Delta)-g(\omega)\right] \omega \bar{\boldsymbol{\beta}}\|+\|g(\omega+\Delta) \varepsilon_{y}\| \\
& \leq& \| \left[g(\omega+\Delta)-g(\omega)\right] \omega \bar{\boldsymbol{\beta}}\|+\|g(\omega+\Delta)-g(\omega)\|_{op}\left\|\varepsilon_{y}\right\|_{2}+\left\|g(\omega) \varepsilon_{y}\right\| \label{eq:thm2-proof}.
\ee

We start with boundary the first term in Inequality \eqref{eq:thm2-proof}. Note that

\be
g(\omega+\Delta)-g(\omega)=\left(\omega^{\top} \omega\right)^{-1} \Delta^{\top}+\left[\left[(\omega+\Delta)^{\top}(\omega+\Delta)\right]^{-1}-\left[\omega^{\top} \omega\right]^{-1}\right] \omega^{\top}.
\ee

Then 
\be
&& [g(\omega+\Delta)-g(\omega)] \omega \bar{\boldsymbol{\beta}} \\
&= &
\begin{multlined}[t]
\left(\omega^{\top} \omega\right)^{-1} \Delta^{\top} \omega \bar{\boldsymbol{\beta}} +\left[(\omega+\Delta)^{\top}(\omega+\Delta)\right]^{-1} \\ 
\left[\bI-\left(\omega^{\top} \omega+\Delta^{\top} \omega+\omega^{\top} \Delta+\Delta^{\top} \Delta\right)\left(\omega^{\top} \omega\right)^{-1}\right] \omega^{\top} \omega \bar{\boldsymbol{\beta}}
\end{multlined}
\\
&= & \left(\omega^{\top} \omega\right)^{-1} \Delta^{\top} \omega \bar{\boldsymbol{\beta}}-\left((\omega+\Delta)^{\top}(\omega+\Delta)\right)^{-1}\left(\Delta^{\top} \omega \bar{\boldsymbol{\beta}}+\omega^{\top} \Delta \bar{\boldsymbol{\beta}}+\Delta^{\top} \Delta \bar{\boldsymbol{\beta}}\right) \\
&= & 
\begin{multlined}[t]
\left[(\omega+\Delta)^{\top}(\omega+\Delta)\right]^{-1}\left(\Delta^{\top} \omega+\omega^{\top} \Delta+\Delta^{\top} \Delta\right)\left(\omega^{\top} \omega\right)^{-1} \Delta^{\top} \omega \bar{\boldsymbol{\beta}} \\
-\left[(\omega+\Delta)^{\top}(\omega+\Delta)\right]^{-1} \omega^{\top} \Delta \bar{\boldsymbol{\beta}}-\left[(\omega+\Delta)^{\top}(\omega+\Delta)\right]^{-1} \Delta^{\top} \Delta \bar{\boldsymbol{\beta}} .
\end{multlined}
\ee

Let's consider the smallest eigenvalue of $(\omega+\Delta)^{\top}(\omega+\Delta)$ as
\be
\lambdanew_{n}\left((\omega+\Delta)^{\top}(\omega+\Delta)\right)=\min _{z \in \mathbb{R}^{n}}\|(\omega+\Delta) z\|^{2}.
\ee

Note that
\be
\|\omega z\| \geq \sqrt{\lambdanew_{n}\left(\omega^{\top} \omega\right)} \geq \tau\|\omega\|_{op} \geqslant \tau \sqrt{n},
\ee
and
\be
\|\Delta z\| &\leq& \sqrt{2 n} \max \left\{\frac{\left\|\hat{\bu}-\bu\right\|}{\sqrt{n}}, \frac{\left\|\hat{\bv}-\bv\right\|}{\sqrt{n}}\right\} 
\leq \sqrt{2 n} \sqrt{\eta}.
\ee
Then we have 
\be
\lambdanew_{n}\left((\omega+\Delta)^{\top}(\omega+\Delta)\right) \geq \frac{1}{4} \tau^{2}\|\omega\|_{o p}^{2}.
\ee

Therefore,
\be
&& \|(g(\omega+\Delta)-g(\omega)) \omega \bar{\boldsymbol{\beta}}\| \\
&\leq & 4 \tau^{-4}\|\omega\|_{o p}^{-4} \cdot\left(2\|\Delta\|_{o p} \cdot\|\omega\|_{o p}+\|\Delta\|_{o p}^{2}\right)\|\Delta\|_{o p}\|\omega\|_{o p}\|\bar{\boldsymbol{\beta}}\|_{2} \\
&& + 4 \tau^{-2}\|\omega\|_{o p}^{-2}\left[\|\omega\|_{o p}\|\Delta \bar{\boldsymbol{\beta}}\|_{2}+\|\Delta\|_{o p}\|\Delta \bar{\boldsymbol{\beta}}\|_{2}\right] \\
&\leq & 4 \tau^{-4}\|\omega\|_{o p}^{-4} \cdot\left(2+\frac{1}{2} \tau\right)\|\omega\|_{o p}^{2} \cdot\|\Delta\|_{o p}^{2}\|\bar{\boldsymbol{\beta}}\|_{2} \\
&& + 4 \tau^{-2}\|\omega\|_{o p}^{-2}\left(1+\frac{1}{2} \tau\right)\|\omega\|_{o p}\|\Delta\|_{o p}\|\bar{\boldsymbol{\beta}}\|_{2} \\
&\leq & 4 \tau^{-4}\left(2+\frac{1}{2} \tau\right) 2 \eta \|\bar{\boldsymbol{\beta}}\|_{2} 
+ 4 \tau^{-2}\left(1+\frac{1}{2} \tau\right) \sqrt{2} \sqrt{\eta}\|\bar{\boldsymbol{\beta}}\|_{2}.
\ee

Now we turn to the second term in Inequality \eqref{eq:thm2-proof}.

\be
&& \|g(\omega+\Delta)-g(\omega)\|_{o p}\left\|\varepsilon_{y}\right\| \\
&\leq & \left[\tau^{-2}\|\omega\|_{op}^{-2}\|\Delta\|_{op}+4 \tau^{-2}\|\omega\|_{op}^{-2} \tau^{-2}\|\omega\|_{o p}^{-2}\left(2\|\Delta\|_{o p}\|\omega\|_{o p}+\|\Delta\|_{o p}^{2}\right)\right]\|\omega\|_{o p}\left\|\varepsilon_{y}\right\|_{2} \\
&\leq & \tau^{-2}\|\omega\|_{op}^{-1}\|\Delta\|_{op}\left[1+4 \tau^{-2}\|\omega\|_{o p}^{-1}\left(2+\frac{\tau}{2}\right)\right]\|\varepsilon_{y}\|_{2} \\
&\leq & \sqrt{\eta} \cdot \sqrt{2} \tau^{-2}\left\|\frac{\varepsilon_{y}}{\sqrt{n}}\right\|\left(1+\frac{8 \tau^{-2}+2 \tau^{-1}}{\sqrt{n}}\right) .
\ee

Under the high probability event that the bounds for $\hat{\bu}, \hat{\bv}$ in Proposition \ref{prop1} hold, we have

\be
&&\left\|g(\omega+\Delta)-g(\omega)\right\|_{op} \left\|\varepsilon_{y} \right\| 
\leq \sqrt{\eta} \sqrt{10} \tau^{-2}\left(1+\frac{8 \tau^{-2}+2 \tau^{-1}}{\sqrt{n}}\right) \sigma_{y}.
\ee

Now we turn to the third term in Inequality \eqref{eq:thm2-proof}. First note that,
\be
g(\omega) \cdot \varepsilon_y \sim N\left(\overrightarrow{0}, \sigma_{y}^{2} \cdot\left(\omega^{\top} \omega\right)^{-1}\right).
\ee
Let
$h=g(\omega) \cdot \varepsilon_{y}$,
then for $\xi > 0$, 

\be
\mathbb{E} e^{\xi\|h\|^{2}} & =& \int e^{\xi\left(\varepsilon_{y}^{\top} \omega\left(\omega^{\top} \omega\right)^{-2} \omega^{\top} \varepsilon_{y}\right)-\frac{1}{2 \sigma_{y}^{2}}\|\varepsilon_{y}\|^{2}} \frac{1}{(\sqrt{2 \pi} \sigma_{y})^{p+2}} d \varepsilon_{y} \\
& \leq& \int e^{\left(\xi \tau^{4} \cdot\|\omega\|_{op}^{-2}-\frac{1}{2 \sigma_{y}^{2}}\right)\left\|\varepsilon_{y}\right\|^{2}} \frac{1}{\left(\sqrt{2 \pi} \sigma_{y}\right)^{p+2}} d \varepsilon_{y} \\
&=&\left(\frac{1}{\sqrt{1-2 \sigma_{y}^{2} \xi \cdot \tau^{4}\|\omega\|_{op}^{-2}}}\right)^{p+2} \\
&\leq& \left(\frac{1}{\sqrt{1-2 \sigma_{y}^{2} \tau^{4} \cdot \frac{1}{n} \xi}}\right)^{p+2}; 
\ee
for $\xi<\frac{n}{2 \sigma_{y}^{2} \tau^{4}}$, 

\be
P\left(\|h\|^{2}>k\right) \leq \left(\frac{1}{\sqrt{1-2 \sigma_{y}^{2} \tau^{4} \cdot \frac{1}{n} \xi}}\right)^{p+2} e^{-\xi k}.
\ee

Let 
$\xi=\frac{n}{4 \sigma_{y}^{2} \cdot \tau^{4}}, \,
k=\frac{4 \sigma_{y}^{2} \cdot \tau^{4}}{n}\left(1+p/2+2 \log n \right)$, we have
\be
P\left(\left\|g(\omega) \cdot \varepsilon_{y}\right\|>\frac{2 \sigma_{y} \cdot \tau^{2}}{\sqrt{n}} \sqrt{1+p/2+2 \log n}\right)<\frac{1}{n^{2}}.
\ee

Combining the above results for the three terms in \eqref{eq:thm2-proof} gives that with probability at least $1-6 e^{-n}$,
\be
\|\hat{\boldsymbol{\beta}}-\boldsymbol{\beta}\| & \leq & \left(8 \tau^{-4}\left(2+\frac{1}{2} \tau\right) \eta+4 \sqrt{2} \tau^{-2}\left(1+\frac{1}{2} \tau\right) \sqrt{\eta}\right) \sqrt{\beta_{u}^{2}+\beta_{v}^{2}} \\
&& +\sqrt{\eta} \sigma_{y} \cdot \sqrt{10} \cdot \tau^{-2}\left(1+\frac{8 \tau^{-2}+2 \tau^{-1}}{\sqrt{n}}\right)+ 2 \sigma_{y} \tau^{2} \sqrt{\frac{1+ p/2 + 2 \log n}{n}}. 
\ee

Finally, Corollary \ref{thm2-cor}  directly follows from Theorem \ref{thm2} under the conditions \eqref{eq:thm2-cor-cond1}-\eqref{eq:thm2-cor-cond3}.

\end{proof}

\subsubsection{Proposition \ref{prop1}}
\label{sec:prop1}

\begin{proposition}
\label{prop1}
Suppose $\left(\bX, \bu, \bv\right)$ is full rank, then for any given $\delta\ge 0$, with probability at least $1-3 e^{-n}$, the following holds for all $(\hat{\bu}, \hat{\bv})$ such that $(\bX, \hat{\bu}, \hat{\bv})$ is full rank and $f_{1}(\hat{\bu}, \hat{\bv}) \leq f_{1}\left(\bu, \bv\right)+\delta$: 

\be
\left(1-\frac{\left\langle\hat{\bu}, \bu\right\rangle^{2}}{2n^{2}}-\frac{\left\langle\hat{\bv}, \bv\right\rangle^{2}}{2n^{2}}\right)
\leq \frac{30 \sigma_{a}^{2}}{\lambdanew\left(\done-d_{2}\right) \done n}+\frac{544 \sigma_{a}^{2}}{n\left(\done-d_{2}\right)^{2}}+\frac{2 \delta \sigma_{a}^{2}}{\lambdanew n^{2} \done\left(\done-d_{2}\right)}.
\ee
\end{proposition}

\begin{proof}
\label{prof:prop1}


\jcM{By Definitions \ref{defn4}-\ref{defn-f2} and the unified model \eqref{eq:model-supercent},}
$f_{1}(\hat{\bu}, \hat{\bv}) \leq f_{1} (\bu, \bv)+\delta$
gives 
$f_{2}(\hat{\bu}, \hat{\bv}) \leq f_{2} (\bu, \bv)+\delta.$
Therefore, 
\be
&& \E f_{2}(\hat{\bu}, \hat{\bv}) - \E f_{2} (\bu, \bv)  \leq \left[ f_{2} (\bu, \bv) - \E f_{2} (\bu, \bv) \right] - \left[ f_{2}(\hat{\bu}, \hat{\bv}) - \E f_{2}(\hat{\bu}, \hat{\bv}) \right] +\delta.
\ee

By Lemma \ref{lemma1} and Lemma \ref{lemma2}, 
\be
\left(1-\frac{\left\langle\hat{\bu}, \bu\right\rangle^{2}}{2n^{2}}-\frac{\left\langle\hat{\bv}, \bv\right\rangle^{2}}{2n^{2}}\right) & \leq &  \frac{\sigma_{a}^{2}}{\left(\done-d_{2}\right) \done \lambdanew n^{2}} \frac{6}{\sigma_{y}^{2}}\| \Tilde{\varepsilon} \|^{2} + 32\left(\frac{\| \bE\|_{op}}{n\left(\done-d_{2}\right)}\right)^{2} \\ 
&& +  \frac{2\|\bE\|_{op}^{2}}{n^{2} \done\left(\done-d_{2}\right)}+\frac{2 \sigma_{a}^{2} \delta}{\lambdanew \done\left(\done-d_{2}\right) n^{2}}.
\ee
Note that 
$\| \Tilde{\varepsilon} \| \sim \chi_{n-p}^{2}$.
By Lemma \ref{lemma3} and Lemma \ref{lemma4}, we have with probability at least $1-3 e^{-n}$, 
\be
\left(1-\frac{\left\langle\hat{\bu}, \bu\right\rangle^{2}}{2n^{2}}-\frac{\left\langle\hat{\bv}, \bv\right\rangle^{2}}{2n^{2}}\right) \leq \frac{30 \sigma_{a}^{2}}{\lambdanew\left(\done-d_{2}\right) \done n}+\frac{544 \sigma_{a}^{2}}{n\left(\done-d_{2}\right)^{2}}+\frac{2 \delta \sigma_{a}^{2}}{\lambdanew n^{2} \done\left(\done-d_{2}\right)}.
\ee

\end{proof}



\subsubsection{Technical Lemmas}
\label{sec:technical-lemmas}

We first present the following tail bounds that are frequently used in Sections \ref{sec:prop1} and \ref{sec:consistency}, and then provide the corresponding proofs in 
Sections \ref{prof:lemma1}-\ref{prof:lemma4}.  


\begin{lemma}
\label{lemma1}
For any $\bar{\bu} \in \mathbb{R}^{n}, \bar{\bv} \in \mathbb{R}^{n}$, such that $\|\bar{\bu}\|=\|\bar{\bv}\|=\sqrt{n}$, and $(\bX, \bar{\bu}, \bar{\bv})$ being full rank,
\be
 \E f_{2}(\bar{\bu}, \bar{\bv}) - \E f_{2}(\bu, \bv) & \geq & \frac{1}{\sigma_{y}^{2}}\left\| \bP_{(\bX, \bar{\bu}, \bar{\bv})^{\perp}}(\Tilde{\bu} \beta_{u} + \Tilde{\bv} \beta_{v})\right\|^{2}\\
&&  +  \frac{\lambdanew}{\sigma_{a}^{2}} n^{2} \left(1-\frac{\left\langle\bar{\bu}, \bu\right\rangle^{2}}{2n^{2}}-\frac{\left\langle\bar{\bv}, \bv\right\rangle^{2}}{2n^{2}}\right) \left(\done-d_{2}\right) \done.
\ee
\end{lemma}


\begin{lemma}
\label{lemma2}
For any $\bar{\bu} \in \mathbb{R}^{n}, \bar{\bv} \in \mathbb{R}^{n}$, such that $\|\bar{\bu}\|=\|\bar{\bv}\|=\sqrt{n}$, and $(\bX, \bar{\bu}, \bar{\bv})$ being full rank,
\be
&& \left[f_{2}\left(\bu, \bv\right) - \E f_{2}\left(\bu, \bv\right)\right]-\left[f_{2}(\bar{\bu}, \bar{\bv})-\E f_{2}(\bar{\bu}, \bar{\bv})\right] \\
&\leq & \frac{3}{\sigma_{y}^{2}}\| \Tilde{\varepsilon} \|^{2}+\frac{1}{2 \sigma_y^2}\left\| \bP_{(\bX, \bar{\bu}, \bar{\bv})^{\perp}}\left( \Tilde{\bu} \beta_{u} + \Tilde{\bv} \beta_{v} \right)\right\|^{2} +16 \frac{\lambdanew}{\sigma_{a}^{2}} \done \frac{\| \bE\|_{op}^{2}}{\done-d_{2}} \\
&& + \frac{\lambdanew n^{2} \done\left(\done-d_{2}\right)}{2 \sigma_{a}^{2}} \left(1-\frac{\left\langle \bu, \bar{\bu}\right\rangle^{2}}{2n^{2}}-\frac{\left\langle \bv, \bar{\bv}\right\rangle^{2}}{2n^{2}}\right) +\frac{\lambdanew}{\sigma_{a}^{2}}\|\bE\|_{op}^{2}.
\ee
\end{lemma}

\begin{lemma}
\label{lemma3}
For 
$\bE \in \mathbb{R}^{n \times n}$ where $\bE_{i j} \stackrel{\text{i.i.d}}{\sim} N(0, \sigma_{a}^{2})$, 
then for $t \geq 0$,

\be
P\left(\left\|\frac{\bE}{\sigma_{a}}\right\|_{op} \geq 2 \sqrt{n}+t\right) \leq 2 e^{-t^{2} / 2}.
\ee

\end{lemma}


\begin{lemma}
\label{lemma4}
Let $\chi_{n-p}^{2}$ follow $\chi^{2}$-distribution with degree of freedom $n-p$, then

\be
P\left(\chi_{n-p}^{2} \geq n-p + 2 \sqrt{n-p} \sqrt{x}+2 x\right) &\leq& e^{-x}, \\
\mathbb{E}\left(\chi_{n-p}^{2}\right) &=& n-p, \\
P\left(\chi_{n}^{2} \leq n-t\right) &\leq& e^{-t^{2} / 8 n}.
\ee

\end{lemma}

\paragraph{Proof of Lemma \ref{lemma1}}

\begin{proof}
\label{prof:lemma1}
$\forall\|\bar{\bu}\|_{2} = \|\bar{\bv}\|_{2}=\sqrt{n} $ and $(\bX, \bar{\bu}, \bar{\bv})$ full rank,  

\be
\bE f_{2}(\bar{\bu}, \bar{\bv}) &= &n-(p+2)+\frac{1}{\sigma_{y}^{2}}\left\|\bP_{(\bX, \bar{\bu}, \bar{\bv})^{\perp}} \left(\Tilde{\bu}_{1} \beta_{u} + \Tilde{\bv}_{1} \beta_{v} \right)\right\|^{2} \\
&& -  \frac{\lambdanew}{\sigma_{a}^{2}}\left[\left(\bar{\bu}^{\top}\left(\sum_{i=1}^{r} d_{i} \bu_{i} \bv_{i}^{\top}\right) \bar{\bv} / n\right)^{2}\right] -\lambdanew.
\ee

For simplicity of notation, denote 

\be
\bar{\bu}=a_{u} \bu+b_{u} \bu^{\perp},
\ee
where $\bu^{\perp} \perp \bu, \left\|\bu^{\perp}\right\|=\sqrt{n}$, and
\be
\bar{\bv}=a_{v} \bv+b_{v} \bv^{\perp},
\ee
where 
$\bv^{\perp} \perp \bv, \left\|\bv^{\perp}\right\|=\sqrt{n}$,
then
\be
&& \bE f_{2}(\bar{\bu}, \bar{\bv})- \bE f_{2}\left(\bu, \bv\right) \\
&\geq & \frac{1}{\sigma_{y}^{2}}\left\| \bP_{( \bX, \bar{\bu}, \bar{\bv})^{\perp}}\left(\Tilde{\bu}_{1} \beta_{u} + \Tilde{\bv}_{1} \beta_{v}\right)\right\|^{2} \\
&& - \frac{\lambdanew}{\sigma_{a}^{2}}\left[\left(a_{u} a_{v} \done n+b_{u} b_{v}\left(\bu^{\perp}\right)^{\top}\left(\sum_{i=2}^{r} d_{i} \bu_{i} \bv_{i}^{\top}\right) \bv^{\perp} / n\right)^{2}\right] +\frac{\lambdanew}{\sigma_{a}^{2}} \done^{2} n^{2} \\
&\geq & \frac{1}{\sigma_{y}^{2}}\left\| \bP_{( \bX, \bar{\bu}, \bar{\bv})^{\perp}}\left(\Tilde{\bu}_{1} \beta_{u} + \Tilde{\bv}_{1} \beta_{v}\right)\right\|^{2}
+\frac{\lambdanew}{\sigma_{a}^{2}} n^{2}\left[\done^{2}-\left(\left|a_{u} a_{v}\right| \done+\left|b_{u} b_{v}\right| d_{2}\right)^{2}\right] \\
&= & \frac{1}{\sigma_{y}^{2}}\left\| \bP_{( \bX, \bar{\bu}, \bar{\bv})^{\perp}}\left(\Tilde{\bu}_{1} \beta_{u} + \Tilde{\bv}_{1} \beta_{v}\right)\right\|^{2} \\
&& + \frac{\lambdanew}{\sigma_{a}^{2}} n^{2}\left[\done^{2}-\left(\left|a_{u} a_{v}\right| \done+\sqrt{\left(1-a_{u}^{2}\right)\left(1-a_{v}^{2}\right)} d_{2}\right)^{2}\right] \\
&\geq & \frac{1}{\sigma_{y}^{2}}\left\| \bP_{( \bX, \bar{\bu}, \bar{\bv})^{\perp}}\left(\Tilde{\bu}_{1} \beta_{u} + \Tilde{\bv}_{1} \beta_{v}\right)\right\|^{2} \\
&& + \frac{\lambdanew}{\sigma_{a}^{2}} n^{2}\left[ \done^{2} - \left(\frac{a_{u}^{2}+a_{v}^{2}}{2} \done+\frac{2-a_{u}^{2}-a_{v}^{2}}{2} d_{2}\right)^{2}\right] \\
&\geq & \frac{1}{\sigma_{y}^{2}}\left\| \bP_{( \bX, \bar{\bu}, \bar{\bv})^{\perp}}\left(\Tilde{\bu}_{1} \beta_{u} + \Tilde{\bv}_{1} \beta_{v}\right)\right\|^{2} \\
&& + \frac{\lambdanew}{\sigma_{a}^{2}} n^{2} \frac{\left(2-a_{u}^{2}-a_{v}^{2}\right)}{2}\left(\done-d_{2}\right)\left[\left(1+\frac{a_{u}^{2}+a_{v}^{2}}{2} \done+\frac{2-a_{u}^{2}-a_{v}^{2}}{2} d_{2}\right)\right] \\
&\geq & \frac{1}{\sigma_{y}^{2}}\left\| \bP_{( \bX, \bar{\bu}, \bar{\bv})^{\perp}}\left(\Tilde{\bu}_{1} \beta_{u} + \Tilde{\bv}_{1} \beta_{v}\right)\right\|^{2} 
 + \frac{\lambdanew}{\sigma_{a}^{2}} n^{2}  \frac{\left(2-a_{u}^{2}-a_{v}^{2}\right)}{2}\left(\done-d_{2}\right). \done
\ee
\end{proof}

\paragraph{Proof of Lemma \ref{lemma2}}

\begin{proof}
\label{prof:lemma2}
For $\forall \bar{\bu} \in \mathbb{R}^{n}, \bar{\bv} \in \mathbb{R}^{n}, \|\bar{\bu}\|=\|\bar{\bv}\|=\sqrt{n}$, and $(\bX, \bar{\bu}, \bar{\bv})$ full rank,
let 
\be
\Delta(\bar{\bu}, \bar{\bv})=\left[f_{2}\left(\bu, \bv\right)- \bE f_{2}\left(\bu, \bv\right)\right]-\left[f_{2}(\bar{\bu}, \bar{\bv})-\bE f_{2}(\bar{\bu}, \bar{\bv})\right].
\ee

Then
\be
\Delta & = & \frac{1}{\sigma_{y^{2}}}\left[\left\|\bP_{(\bX, \bu, \bv)^{\perp}} \Tilde{\varepsilon} \right\|^{2}-\left\|\bP_{(\bX, \bar{\bu}, \bar{\bv})^{\perp}} \Tilde{\varepsilon} \right\|^{2}\right] \\
&& -  \frac{2}{\sigma_{y}^{2}}\left\langle \bP_{(\bX, \bu, \bv)^{\perp}} \Tilde{\varepsilon}, \bP_{( \bX, \bar{\bu}, \bar{\bv})^{\perp}}\left(\Tilde{\bu}_{1} \beta_{u} + \Tilde{\bv}_{1} \beta_{v}\right)\right\rangle \\
&& + \frac{\lambdanew}{\sigma_{a}^{2}}\left(\bar{\bu}^{\top} \bE \bar{\bv}_{1} / n\right)^{2}-\frac{\lambdanew}{\sigma_{a}{ }^{2}}\left(\bu^{\top} \bE \bv / n\right)^{2} -\frac{2 \lambdanew}{\sigma_{a}^{2}}\left(n \done\right)\left(\bu^{\top} \bE \bv\right) / n \\
&& +  \frac{2 \lambdanew}{\sigma_{a}^{2}}\left(\bar{\bu}^{\top}\left(\sum_{i=1}^{r} d_{i} \bu_{i} \bv_{i}^{\top}\right) \bar{\bv} / n\right)\left(\bar{\bu}^{\top} \bE \bar{\bv} / n\right) \\
& \leq & \frac{1}{\sigma_{y}^{2}}\|\Tilde{\varepsilon}\|^{2}+\frac{2}{\sigma_{y}^{2}}\|\Tilde{\varepsilon}\| \cdot\left\|\bP_{(\bX, \bar{\bu}, \bar{\bv})^{\perp}}\left(\Tilde{\bu}_{1} \beta_{u} + \Tilde{\bv}_{1} \beta_{v}\right)\right\| +\frac{\lambdanew}{\sigma_{a}^{2}}\| \bE\|_{op}^{2} \\
&& + \frac{2 \lambdanew}{\sigma_{a}^{2}}\left[-\left(n \done\right)\left(\bu^{\top} \bE \bv\right) / n+ \left(n \done\right)\left(c_{1} c_{2} \bar{\bu}^{\top} \bE \bar{\bv}\right) / n \right. \\
&&+\left. \left(\bar{\bu}^{\top} \bE \bar{\bv} / n\right) \operatorname{tr}\left(\sum_{i=1}^{r} d_{i} \bu_{i} \bv_{i}^{\top}, \frac{\bar{\bu} \bar{\bv}^{\top}}{n}-\frac{\bu \bv^{\top} c_{1} c_{2}}{n}\right)\right] \\
& \leq & \frac{3}{\sigma_{y}^{2}}\left\|\Tilde{\varepsilon}\right\|^{2}+\frac{1}{2\sigma_y^2}\left\|\bP_{(\bX, \bar{\bu}, \bar{\bv})^{\perp}}\left(\Tilde{\bu}_{1} \beta_{u} + \Tilde{\bv}_{1} \beta_{v}\right)\right\|^{2} \\
&&+ \frac{2 \lambdanew}{\sigma_{a}^{2}}\|E\|_{o p} \cdot n \done \cdot 2\left\|\frac{\bu \bv^{\top}-\bar{\bu} \bar{\bv}^{\top} c_{1} c_{2}}{n}\right\|_{nuc}+\frac{\lambdanew}{\sigma_{a}^{2}}\|\bE\|_{op}^{2}, 
\ee
where 
\be
c_{1}&=&\operatorname{sign}\left(\left\langle\bar{\bu}, \bu\right\rangle\right)\\
c_{2}&=&\operatorname{sign}\left(\left\langle\bar{\bv}, \bv\right\rangle\right).
\ee

Note that 
\be
\left\|\frac{\bu \bv^{\top}-\bar{\bu} \bar{\bv}^{\top} c_{1} c_{2}}{n}\right\|_{nuc} & \leq & \sqrt{2}\left\|\frac{\bu \bv^{\top}-\bar{\bu} \bar{\bv}^{\top} c_{1} c_{2}}{n}\right\|_{F} \\
& =& \sqrt{2} \sqrt{1-\frac{c_{1}\left\langle \bu, \bar{\bu}\right\rangle \cdot c_{2}\left\langle \bv, \bar{\bv}\right\rangle}{n^{2}}} \\
& \leq & \sqrt{2} \sqrt{1-\frac{\left\langle \bu, \bar{\bu}\right\rangle^{2}}{2 n^{2}}-\frac{\left\langle \bv, \bar{\bv}\right\rangle^{2}}{2 n^{2}}}. 
\ee

Therefore,
\be
\Delta & \leq & \frac{3}{\sigma_{y}^{2}}\|\Tilde{\varepsilon}\|^{2}+\frac{1}{2}\left\|\bP_{(\bX, \bar{\bu}, \bar{\bv})^{\perp}}\left(\Tilde{\bu}_{1} \beta_{u} + \Tilde{\bv}_{1} \beta_{v}\right)\right\|^{2} \\
&& + \frac{4 \sqrt{2} \lambdanew}{\sigma_{a}^{2}} n \done\left(\frac{2 \sqrt{2}\|\bE\|_{op}^{2}}{n\left(\done-d_{2}\right)}+\frac{n\left(\done-d_{2}\right)}{8 \sqrt{2}}\left(1-\frac{\left\langle \bu, \bar{\bu}\right\rangle^{2}}{2 n^{2}}-\frac{\langle \bv, \bar{\bv}^{2}\rangle}{2 n^{2}}\right)\right) \\
&& +\frac{\lambdanew}{\sigma_{a}^{2}}\|\bE\|_{op}^{2} \\
&\leq & \frac{3}{\sigma_{y}^{2}}\|\Tilde{\varepsilon}\|^{2}+\frac{1}{2}\left\|\bP_{(\bX, \bar{\bu}, \bar{\bv})^{\perp}}\left(\Tilde{\bu}_{1} \beta_{u} + \Tilde{\bv}_{1} \beta_{v}\right)\right\|^{2} +16 \frac{\lambdanew}{\sigma_{a}^{2}} \done \frac{\|\bE\|_{op}^{2}}{\done-d_{2}} \\
&& + \frac{\lambdanew n^{2} \done\left(\done-d_{2}\right)}{2 \sigma_{a}^{2}}\left(1-\frac{\left\langle \bu, \bar{\bu}\right\rangle^{2}}{2 n^{2}}-\frac{\left\langle \bv, \bar{\bv}\right\rangle^{2}}{2 n^{2}}\right) +\frac{\lambdanew}{\sigma_{a}^{2}}\|\bE\|_{op}^{2}.
\ee

\end{proof}

\paragraph{Proof of Lemma \ref{lemma3}}

\begin{proof}
\label{prof:lemma3}
Let
\be
&& Z_{u, v}=\bu^{\top} \frac{\bE}{\sigma_{a}} \bv, \\
&& Y_{u, v}=\langle g, \bu\rangle+\langle h, \bv\rangle  \quad \text {, where } g, h \in N\left(0, \bI_{n}\right).
\ee
Clearly,
\be
\left\|\frac{\bE}{\sigma_{a}}\right\|_{op}=\sup_{\|\bu\|=\|\bv\|=1} Z_{u, v}.
\ee

For any pairs $(\hat{\bu}, \hat{\bv}),(\tilde{\bu}, \tilde{\bv})$ such that $\|\hat{\bu}\|=\|\hat{\bv}\|=\|\tilde{\bu}\|=\|\tilde{\bv}\|=1$,

\be
\mathbb{E}\left(\left(Z_{\hat{u}, \hat{v}}-Z_{\tilde{u}, \tilde{v}}\right)^{2}\right) & =& 2-2\langle\hat{\bu}, \tilde{\bu}\rangle\langle\hat{\bv}, \tilde{\bv}\rangle \\
\mathbb{E}\left(\left(Y_{\hat{u}, \hat{v}}-Y_{\tilde{u}, \tilde{v}}\right)^{2}\right) & =& \|\hat{\bu}-\tilde{\bu}\|^{2}+\|\hat{\bv}-\tilde{\bv}\|^{2} \\
& =& 4-2\langle\tilde{\bu}, \hat{\bu}\rangle-2\langle\hat{\bv}, \tilde{\bv}\rangle.
\ee

Therefore, 
\be
\mathbb{E}\left(\left(Z_{\hat{u}, \hat{v}}-Z_{\tilde{u}, \tilde{v}}\right)^{2}\right) \leq \mathbb{E}\left(\left(Y_{\hat{u}, \hat{v}}-Y_{\tilde{u}, \tilde{v}}\right)^{2}\right).
\ee

Let $w_{i}$ be a maximum $2^{-i}$-packing of $S^{n-1}$, then

\be
\mathbb{E}\left(\left\|\frac{\bE}{\sigma_{a}}\right\|_{op}\right) & =& \mathbb{E}\left(\sup _{\|\bu\|=\|\bv\|=1} \bu^{\top} \frac{\bE}{\sigma_{a}} \bv\right) \\
& \stackrel{(a)}{=} & \lim _{i \rightarrow+\infty} \mathbb{E}\left(\max_{\bu, \bv \in w_{i}} \bu^{\top} \frac{\bE}{\sigma_{a}} \bv\right) \\
& \stackrel{(b)}{\leq} & \lim_{i \rightarrow+\infty} \mathbb{E}\left(\max_{\bu, \bv \in w_{i}}\langle g, \bu\rangle + \langle h, \bv \rangle\right) \\
& =& \lim_{i \rightarrow+\infty} 2 \sqrt{n}=2 \sqrt{n}.
\ee
where 
(a) follows from $\left\|\frac{\bE}{\sigma_{a}}\right\|_{op} \leq \max _{\bu, \bv \in w_{i}} \bu^{\top} \frac{\bE}{\sigma_{a}} \bv+2^{1-i}\left\|\frac{\bE}{\sigma_{a}}\right\|_{op}$, and 
(b) follows from Sudakov-Fernique inequality.

Note that $\sup_{\|\bu\|=\|\bv\|=1} z_{u, v}$ is a 1-Lipschitz function of $\operatorname{vector} \left(\frac{\bE}{\sigma_{a}}\right)$, by Theorem 2.26 in \cite{wainwright2019high}, 
\be
P\left(\left|\sup_{\|\bu\|=\|\bv\|=1} Z_{u, v}- \mathbb{E} \left( \sup_{\|\bu\|=\|\bv\|=1} Z_{u, v}\right)\right| \geq t\right) \leq 2 e^{-\frac{t^{2}}{2}}.
\ee

Therefore 
\be
P\left( \sup \left\|\frac{\bE}{\sigma_{a}}\right\|_{op} \geq 2 \sqrt{n}+t\right) \leq 2 e^{-\frac{t^{2}}{2}}
\ee
for $t \geq 0$.

\end{proof}

\paragraph{Proof of Lemma \ref{lemma4}}

\begin{proof}
\label{prof:lemma4} 

We only need to prove that for any positive integer $n \geq 1$,

\be
&& P\left(\chi_{n}^{2} \geq n+2 \sqrt{n} \sqrt{x}+2 x\right) \leq e^{-x}, \\
&& \mathbb{E}\left(\chi_{n}^{2}\right)=n, \\
&& P\left(\chi_{n}^{2} \leq n-t\right) \leq 2 e^{-t^{2} / 8 n}.
\ee

Suppose $z_{1}, z_{2}, \cdots, z_{n} \stackrel{\text{i.i.d}}{\sim} N(0,1)$. 
Let $Y:=\sum_{i=1}^{n} z_{i}^{2} \sim \chi_{n}^{2}$. By definition of $\chi_{n}^{2}$-distribution, we have

\be
\mathbb{E}\left(\chi_{n}^{2}\right)=\mathbb{E}(Y)=n,
\ee
\be
 P\left(\chi_{n}^{2} \geq n+2 \sqrt{n} \sqrt{x}+2 x\right)&=&P(Y \geq n+2 \sqrt{n} \sqrt{x}+2 x) \\
&\leq& \inf_{\frac{1}{2}>\xi>0} \mathbb{E} e^{\xi(Y-n-2 \sqrt{n} \sqrt{x}-2 x)} \\
& =&\inf_{\frac{1}{2}>\xi>0} e^{-\frac{n}{2}\log^{1-2\xi}} e^{\xi(-n-2 \sqrt{n} \sqrt{x}-2 x)} \\
& \stackrel{(a)}{=} & e^{-\frac{n}{2} \log \frac{n}{n+2 \sqrt{n} \sqrt{x}+2 x}}+(-\sqrt{n} \sqrt{x}-x) \\
& =& e^{\frac{n}{2}\left(\log 1+2 \sqrt{\frac{x}{n}}+2 \frac{x}{n}-2 \sqrt{\frac{x}{n}}\right)-x} \\
& \stackrel{(b)}{\leq}& e^{-x}
\ee
where 
Step (a) follows from analysis derivative of the exponent w.r.t. $\xi$, 
and 
Step (b) follows from the fact $\log (1+2 t+2 t^{2}-2 t) \leq 0$ for $t \geq 0$.

\be
 P\left(\chi_{n}^{2} \leq n-t\right)&=&P(Y \leq n-t) \\
& \leq & \inf_{\xi>0} \mathbb{E} e^{-\xi Y+\xi(n-t)} \\
& =& \inf_{\xi>0} e^{-\frac{n}{2} \log 1+2 \xi}+\xi(n-t) \\
& =& e^{-\frac{n}{2} \log \frac{n}{n-t}+\frac{t}{2}} \\
& =& e^{\frac{n}{2} \log \frac{n-t}{n}+\frac{t}{2}+\frac{t^{2}}{\delta_n}-\frac{t^{2}}{\delta_n}}.
\ee

Note that

\be
&& \frac{n}{2} \log \frac{n-t}{n}+\frac{t}{2}+\frac{t^{2}}{8 n} \\
& = & \frac{n}{2}\left(\log (1-\frac{t}{n})+\frac{t}{n}+\frac{1}{4}\left(\frac{t}{n}\right)^{2}\right) \\
& \leq & \frac{n}{2} \sup_{1>t \geq 0} \log (1-t) +t+\frac{1}{4} t^{2} \\
& \leq & 0,
\ee
we have $P\left(\chi_{n}^{2} \leq n-t\right) \leq e^{-\frac{t^{2}}{8n}}$.

\end{proof}

\subsection{Proof of the theoretical results of two-stage }
\label{sec:proof-ts}

In this section, we prove that the asymptotic distribution of the two-stage estimators as stated in Theorem \ref{thm:two-stage-normality} and show the convergence rates of $\huts$ and $\hvts$ as in Proposition \ref{prop:two-stage-rate1}.

\subsubsection{Proof of Theorem \ref{thm:two-stage-normality}}
\label{sec:proof-thm1}


\begin{proof}
The naive two-stage procedure first 
estimates the centralities $\bu$ and $\bv$ by the leading left and right singular vectors, rescaled to have norm $\sqrt{n}$ and denoted as $\huts,\hvts$, from the SVD on the observed adjacency matrix $\bA$,
and then performs ordinary least square (OLS) regression of $\by$ on $\bX$ and $\huts,\hvts$, treating $\huts,\hvts$ as given covariates. It is, therefore, equivalent to solve the following two optimization problems sequentially,	
\begin{subequations}
\label{eq:two-stage-proof}
 \begin{align}[left ={\empheqlbrace}]
	(\hdts,\huts,\hvts) &:= \argmin_{d,\|\bu\|=\|\bv\|=\sqrt{n}}\|\bA - d\bu\bv^\top\|_F^2, \label{eq:ah-def2-proof} \\
	\hbbetats = ((\hbetaxts)^\top,\hbetauts,\hbetavts)^\top &:= \argmin_{\bbeta_x,\beta_u,\beta_v} \|\by - \bX\bbeta_x - \huts\beta_u - \hvts\beta_v\|_2^2. \label{eq:reg-obj-proof}
 \end{align}
\end{subequations}
Recall that $\bAp = \bUp \bDp \bVp^\top$ where $\bUp = (\bu_2, \ldots, \bu_r)$, $\bVp = (\bv_2, \ldots, \bv_r)$ and 
$\bDp = diag(d_2, \ldots, d_r)$. Then, $\bAp^\top \bu = 0$ and $\bAp \bv = \bm 0$.
The proof strategy is similar to that of Theorem \ref{thm:supercent-normality} in Section \ref{app:proof-thm2}. 

For the first stage, we minimize the objective function \eqref{eq:ah-def2-proof} and we obtain the following result for $\huts$ and $\hvts$:
\be
\left \{
\begin{aligned}
\duts & = \etauts + o\left(\etauts\right), \\
\dvts & = \etavts + o\left(\etavts\right), 
\end{aligned}
\right .
\label{eq:etas-ts-uv}
\ee
where
\be
\left(
  \begin{array}{c}
    \etauts \\
    \etavts \\
  \end{array}
\right)
&=&
\left(
\left(
  \begin{array}{cc}
    nd\bI & -\bAp \\
    (-\bAp)^\top & nd \bI \\
  \end{array}
\right)^{-1}
\left(
  \begin{array}{c}
   \bv^\top \otimes (\bI- \pu) \\
     \left(\bu^\top \otimes (\bI-\pv)\right)\bK \\
  \end{array}
\right)\vec1(\bE)
\right) \\
&\stackrel{def}{=}&
\left(
\left(
  \begin{array}{c}
    \bCts_{12} \\
    \bCts_{22} \\
  \end{array}
\right) \vec1(\bE)
\right), \label{2s:dudv}
\ee
$\frac{\|o\left(\etauts\right)\|}{\|\etauts\|} \overset{P}{\longrightarrow} 0$, and
$\frac{\|o\left(\etavts\right)\|}{\|\etavts\|} \overset{P}{\longrightarrow} 0$.

For the second stage, we plug in $\huts$ and $\hvts$ from the first stage into the objective function \eqref{eq:reg-obj-proof} of the second stage and then minimize the objective function.
For $\dbetauts$ and $\dbetavts$, we have
\be
\left \{
\begin{aligned}
\dbetau & = \etabetauts + o\left(\etabetauts\right), \\
\dbetav & = \etabetavts + o\left(\etabetavts\right), \\
\end{aligned}
\right .
\label{eq:etas-beta-ts}
\ee
where
{\footnotesize
\be
\left(
  \begin{array}{c}
    \etabetauts \\
    \etabetavts \\
  \end{array}
\right)
&=&
\left(
\bCuvi
\left(
 \begin{array}{c}
  \tu^\top \\
  \tv^\top \\
 \end{array}
\right)
\left(-\betau\bI_n~~ -\betav\bI_n ~~ \bI_n\right)
\left(
  \begin{array}{cc}
    \zero_{n\times n} & \bCts_{12} \\
    \zero_{n\times n} & \bCts_{22} \\
    \bI_n & \zero_{n\times n^2}
  \end{array}
\right)
\left(
  \begin{array}{c}
    \bepsilon \\
    \vec1(\bE) \\
  \end{array}
\right)
\right)  \\
&\stackrel{def}{=}&
\left(
\left(
  \begin{array}{cc}
    \bCts_{41} & \bCts_{42} \\
    \bCts_{51} & \bCts_{52} \\
  \end{array}
\right)
\left(
  \begin{array}{c}
    \bepsilon \\
    \vec1(\bE) \\
  \end{array}
\right)
\right)
\label{2s:dbetau-dbetav}
\ee
}
where explicitly 
\be
\left(
  \begin{array}{c}
    \bCts_{41}\\
    \bCts_{51}
  \end{array}
\right)
&=&
\bCuvi
 \left(
  \begin{array}{c}
    \tu^\top \\
  \tv^\top
  \end{array}
\right),
\ee
{\footnotesize
\be
\left(
  \begin{array}{c}
    \bCts_{42}\\
    \bCts_{52}
  \end{array}
\right)
&=&
\bCuvi
\left(
 \begin{array}{c}
  \tu^\top \\
  \tv^\top \\
 \end{array}
\right)
\left(-\betau\bI_n~~ -\betav\bI_n \right)
\left(
  \begin{array}{cc}
    nd\bI & -\bAp \\
    (-\bAp)^\top & nd \bI \\
  \end{array}
\right)^{-1}
\left(
  \begin{array}{c}
   \bv^\top \otimes (\bI- \pu) \\
     \left(\bu^\top \otimes (\bI-\pv)\right)\bK \\
  \end{array}
\right),
\ee
$\frac{|o\left(\etabetauts\right)|}{|\etabetauts|} \overset{P}{\longrightarrow} 0$, and
$\frac{|o\left(\etabetavts\right)|}{|\etabetavts|} \overset{P}{\longrightarrow} 0$.
}

For $\dbetaxts$, we have
$\dbetax = \etabetaxts + o\left(\etabetaxts\right)$
where $\left\|o\left(\etabetaxts\right) \right\| / \left\|\etabetaxts\right \|  \overset{P}{\longrightarrow} 0$ and
\be
\etabetaxts 
&=& 
\begin{multlined}[t]
    (\bX^\top\bX)^{-1}\bX^\top
 \left(
 -\betau\bI_n ~~ -\betav\bI_n ~~ -\bu ~~ -\bv ~~ \bI_n
\right)
\left(
  \begin{array}{cc}
    \zero_{n\times n} & \bCts_{12} \\
  \zero_{n\times n} & \bCts_{22} \\
    \bCts_{41} & \bCts_{52} \\
    \bCts_{51} & \bCts_{52} \\
    \bI_n & \zero_{n\times n^2}
  \end{array}
\right)
\left(
  \begin{array}{c}
    \bepsilon \\
    \vec1(\bE) \\
  \end{array}
\right) 
\end{multlined} \\
&\stackrel{def}{=}&
\left(
\left(
  \begin{array}{cc}
    \bCts_{31}, & \bCts_{32} \\
  \end{array}
\right)
\left(
  \begin{array}{c}
    \bepsilon \\
    \vec1(\bE) \\
  \end{array}
\right)
\right).
\label{2s:dbetax}
\ee

Finally, recall that we assume 
\be
\left(
  \begin{array}{c}
    \bepsilon \\
    \vec1(\bE) \\
  \end{array}
\right) \sim
N\left(\zero_{(n+n^2)\times 1}, \left(
                             \begin{array}{cc}
                               \sigma_y^2\bI_n & \zero_{n\times n^2} \\
                               \zero_{n^2\times n} & \sigma_a^2 \bI_{n^2} \\
                             \end{array}
                           \right)
\right).
\ee
Putting  \eqref{2s:dudv}, \eqref{2s:dbetau-dbetav} and \eqref{2s:dbetax} together, we have
\be
\left(
 \begin{array}{c}
  \huts-\bu\\
  \hvts-\bv\\
  \hbetaxts-\betax\\
  \hbetauts-\betau \\
  \hbetavts-\betav \\
 \end{array}
\right) = 
\left(
 \begin{array}{c}
 \etauts + o\left(\etauts\right) \\
 \etavts + o\left(\etavts\right) \\
 \etabetaxts + o\left(\etabetaxts\right) \\
 \etabetauts + o\left(\etabetauts\right) \\
 \etabetavts + o\left(\etabetavts\right) \\
 \end{array}
 \right )
\ee
where
\be
\left(
 \begin{array}{c}
 \etauts \\
 \etavts \\
 \etabetaxts \\
 \etabetauts \\
 \etabetavts\\
 \end{array}
 \right ) 
 &= &
\left(
 \begin{array}{cc}
  \bm{0}_{n\times n} & \bCts_{12} \\
  \bm{0}_{n\times n} & \bCts_{22} \\
  \bCts_{31} & \bCts_{32} \\
  \bCts_{41} & \bCts_{42} \\
  \bCts_{51} & \bCts_{52} \\
 \end{array}
\right)
\left(
  \begin{array}{c}
    \bepsilon \\
    \vec1(\bE) \\
  \end{array}
\right) \\
&= &
\bCts
\left(
  \begin{array}{c}
    \bepsilon \\
    \vec1(\bE) \\
  \end{array}
\right)\\
&\sim &  
N\Big(
\zero_{(2n+2+p)\times 1}, 
\bCts\left(
              \begin{array}{cc}
                \sigma_y^2\bI_n & \zero_{n\times n^2} \\
                \zero_{n^2\times n} & \sigma_a^2 \bI_{n^2} \\
              \end{array}
              \right){\bCts}^\top
\Big).
\ee



\end{proof}

\subsubsection{Proof of Proposition \ref{prop:two-stage-rate1}}

\begin{proof}
From Theorem \ref{thm:two-stage-normality}, we have
\be
\left(
  \begin{array}{c}
    \etauts \\
    \etavts \\
  \end{array}
\right)
&=&
\left(
\left(
  \begin{array}{cc}
    nd\bI & -\bAp \\
    (-\bAp)^\top & nd \bI \\
  \end{array}
\right)^{-1}
\left(
  \begin{array}{c}
   \bv^\top \otimes (\bI- \pu) \\
     \left(\bu^\top \otimes (\bI-\pv)\right)\bK \\
  \end{array}
\right)\vec1(\bE)
\right)  \\
&\stackrel{def}{=}&
\left(
\left(
  \begin{array}{c}
    \bCts_{12} \\
    \bCts_{22} \\
  \end{array}
\right) \vec1(\bE)
\right). \label{2s:dudv}
\ee
When $\bA_0$ is rank one, then $\bCts_{12} = (d n)^{-1} \bv^\top \otimes (\bI- \pu)$, $\bCts_{22} = (d n)^{-1} \left(\bu^\top \otimes (\bI-\pv)\right)\bK $ and  \\
 $\bCts_{32} = \left[ \opv \otimes \pu 
+ \pv \otimes \opu 
+ \pv\otimes \pu
 \right] $.
 Note that
\be
\frac{1}{n} \tr(\sigma_a^2 \bCts_{12} \bC_{12}^{ts\top} ) 
&=& \frac{1}{n} \frac{\sigma_a^2}{(dn)^2} 
\tr\left(\left( \bv^\top \otimes (\bI-\pu) \right) \left( \bv \otimes (\bI-\pu) \right) \right)\\
&=&\frac{1}{n}  \frac{\sigma_a^2}{(dn)^2}
\tr\left(\bv^\top \bv  \otimes (\bI-\pu) \right)\\
&=&\frac{1}{n}  \frac{\sigma_a^2}{(dn)^2}
\tr\left(n\bI-\bu\bu^\top) \right) \\
&=& \frac{\sigma_a^2}{d^2 n^2} (n-1).
\ee
Then for the rate of $\huts$,
\be
\frac{1}{n} 
\E \| \huts - \bu \|^2_2
&=& \left(\frac{\sigma_a^2}{d^2 n^2} (n-1)\right) \left(1+o(1)\right) = O(\kappa).
\label{2s:rate-u}
\ee

Similarly for the rate of $\hvts$,
\be 
 \frac{1}{n}  \tr(\sigma_a^2 \bCts_{22} \bC_{22}^{ts\top} ) 
&=& \frac{1}{n}  \frac{\sigma_a^2}{(dn)^2} 
\tr\left(\left( \bu^\top \otimes (\bI-\pv) \right) \left( \bu \otimes (\bI-\pv) \right) \right)\\
&=& \frac{\sigma_a^2}{d^2 n^2} (n-1)
\ee
where the first equality is due to $\bK \bK^\top = \bI$.
Hence
\be 
\frac{1}{n} \E \| \hvts - \bv \|^2_2 
&=& 
\left( \frac{\sigma_a^2}{d^2 n^2} (n-1) \right) \left(1+o(1)\right)
= O(\kappa).
\label{2s:rate-v}
\ee


\end{proof}

\subsubsection{Proof of Proposition \ref{prop:two-stage-rate2}}

\begin{proof}

We now prove the rate of $\hbetats$ of two-stage. 

\paragraph*{(1) Rate of $\hbetauts$ and $\hbetavts$. }

Recall that
\be
\left(
  \begin{array}{c}
    \etabetauts \\
    \etabetavts \\
  \end{array}
\right)
&=&
\bCuvi
\left(
  \begin{array}{c}
    \tu^\top \\
    \tv^\top
  \end{array}
\right) 
\Big[
\bepsilon - (\duts\betau+\dvts\betav)
\Big]\\
&=&
\bCuvi
\left(
  \begin{array}{c}
    \tu^\top \\
    \tv^\top
  \end{array}
\right)  
\Big[
\bepsilon - \frac{1}{dn}
\left(\betau \bv^\top\otimes \opu +
\betav \bu^\top\otimes \opv \bK
\right)
\vec1(\bE)
\Big] \\
&\stackrel{def}{=}&
\bCuvi
\left(
  \begin{array}{c}
    \tu^\top \\
    \tv^\top
  \end{array}
\right)  
\Big[
\bepsilon + \bB \vec1(\bE)
\Big] 
\label{2s:betaubetav-def1}
\\
&\stackrel{def}{=}&
\bD_1 \bepsilon + \bD_2 \vec1(\bE).
\ee

Note that
\be
\bD_1\bD_1^\top = 
\bCuvi
\left(
  \begin{array}{c}
    \tu^\top \\
    \tv^\top
  \end{array}
\right)  
\left(
  \begin{array}{cc}
    \tu &
    \tv
  \end{array}
\right) 
\bCuvi = \bCuvi
\label{2s:b1b1}
\ee
and
\be
\bD_2\bD_2^\top = 
\bCuvi
\left(
  \begin{array}{c}
    \tu^\top \\
    \tv^\top
  \end{array}
\right)  
\bB \bB^\top
\left(
  \begin{array}{cc}
    \tu &
    \tv
  \end{array}
\right) 
\bCuvi 
\label{2s:b2b2}
\ee
where
\footnotesize{
\be
\bB \bB^\top &= &
\frac{1}{(dn)^2} 
\Big[
\beta_u^2(\bv^\top \otimes \opu)(\bv \otimes \opu)
+ \beta_v^2(\bu^\top \otimes \opv)(\bu \otimes \opv) \\
&&
\quad \quad + \betau\betav(\bv^\top \otimes \opu)\bK^\top(\bu \otimes \opv) \\
&&
\quad \quad + \betau\betav(\bu^\top \otimes \opv)\bK(\bv \otimes \opu) 
\Big] \\
&=& \frac{1}{d^2n} \left[ \beta_u^2 \opu + \beta_v^2 \opv \right]
\label{eq:two-stage-aa}
\ee
}
since
$(\bv^\top \otimes \opu)\bK^\top(\bu \otimes \opv) 
= (\bv^\top \opv \otimes \opu \bu)\bK = 0  $
and
$(\bu^\top \otimes \opv)\bK(\bv \otimes \opu) 
=(\bu^\top \opu \otimes \opv \bv)\bK =0 $.

Plugging in \eqref{2s:b1b1}, \eqref{2s:b2b2} and \eqref{eq:two-stage-aa},
\be
Cov\left(
  \begin{array}{c}
    \dbetauts \\
    \dbetavts \\
  \end{array}
\right)
&\approx&
\sigma_y^2 \bD_1\bD_1^\top + \sigma_a^2 \bD_2 \bD_2^\top \\
&=& \sigma_y^2    \bCuvi \\
&&\vspace{1.5in}
+ \sigma_a^2 \frac{1}{d^2n} \bCuvi 
 \left(
  \begin{array}{c}
    \tu^\top \\
    \tv^\top
  \end{array}
\right) 
\left[ \beta_u^2 \opu + \beta_v^2 \opv \right]
\left(
  \begin{array}{cc}
    \tu &
    \tv
  \end{array}
\right) 
\bCuvi \\
&=& \sigma_y^2    
\bCuvi 
+ \sigma_a^2 \frac{1}{d^2n} \bCuvi 
\left(
\begin{array}{cc}
  \beta_v^2\tu^\top\opv\tu & 0\\
  0 & \beta_u^2\tv^\top\opu\tv \\
\end{array}
\right)
\bCuvi.
\label{2s:cov-betau-betav}
\ee
Therefore, we obtain the rate of $\hbetauts$ and $\hbetavts$
from the diagonal entries of $Cov\left(
  \begin{array}{c}
    \dbetauts \\
    \dbetavts \\
  \end{array}
\right)$ as
\footnotesize{
\be
\E (\hbetauts - \betau )^2 
&=& \frac{\sigma_y^2 }{c} \tv^\top\tv \\
& + & \frac{\sigma_a^2}{c^2} \frac{1}{d^2n} 
\left[
\beta_v^2 \tv^\top\tv\tu^\top\opv\tu\tv^\top\tv
+ \beta_u^2 \tu^\top\tv\tv^\top \opu \tv\tu^\top\tv
\right] (1+o(1)),\\
\E (\hbetavts - \betav )^2
&=& \frac{\sigma_y^2 }{c} \tu^\top\tu \\
& +& \frac{\sigma_a^2}{c^2} \frac{1}{d^2n}
\left[
\beta_u^2 \tu^\top\tu\tv^\top\opu\tv\tu^\top\tu
+ \beta_v^2 \tv^\top\tu\tu^\top \opv \tu\tv^\top\tu
\right] (1+o(1))
\ee
}
where $c = \tu^\top\tu\tv^\top\tv-(\tu^\top\tv)^2$.

\paragraph*{(2) Rate of $\hbetaxts$.} Recall that
\be
\dbetaxts &\approx& (\bX^\top\bX)^{-1}\bX^\top(\bepsilon-\bu\dbetauts-\duts\betau -\bv\dbetavts-\dvts\betav).
\label{2s:dbetax-2}
\ee
From \eqref{2s:betaubetav-def1}, we have
\be
\bepsilon - (\duts\betau+\dvts\betav) &\stackrel{def}{=}&
\bepsilon + \bB \vec1(\bE)
\ee
and
\be
\left(
  \begin{array}{c}
    \dbetauts \\
    \dbetavts \\
  \end{array}
\right)
&\approx&
\bCuvi
\left(
  \begin{array}{c}
    \tu^\top \\
    \tv^\top
  \end{array}
\right) 
\left[
\bepsilon + \bB \vec1(\bE)
\right].
\ee
Then we have
\be
\bu\dbetauts + \bv\dbetavts &=&
\left(
\begin{array}{cc}
  \bu & \bv
\end{array}
\right)
\left(
  \begin{array}{c}
    \dbetauts \\
    \dbetavts \\
  \end{array}
\right)
\approx
\left(
\begin{array}{cc}
  \bu & \bv
\end{array}
\right)
\bCuvi
\left(
  \begin{array}{c}
    \tu^\top \\
    \tv^\top
  \end{array}
\right) 
\left[
\bepsilon + \bB \vec1(\bE)
\right].
\label{2s:udbetau}
\ee

Let
\be
\tG =
\left(
\begin{array}{cc}
  \bu & \bv
\end{array}
\right)
\bCuvi
\left(
  \begin{array}{c}
    \tu^\top \\
    \tv^\top
  \end{array}
\right) 
\mbox{ and }
\bG = 
\left(
\begin{array}{cc}
  \bu & \bv
\end{array}
\right)
\bCuvi
\left(
  \begin{array}{c}
    \bu^\top \\
    \bv^\top
  \end{array}
\right).
\ee
Plugging \eqref{2s:betaubetav-def1} and \eqref{2s:udbetau} into \eqref{2s:dbetax-2}  yields
\be
\dbetaxts 
&\approx&
(\bX^\top\bX)^{-1}\bX^\top
\left[
\bI - 
\left(
\begin{array}{cc}
  \bu & \bv
\end{array}
\right)
\bCuvi
\left(
  \begin{array}{c}
    \tu^\top \\
    \tv^\top
  \end{array}
\right) 
\right]
(\bepsilon + \bB \vec1(\bE)) \\
&=&
(\bX^\top\bX)^{-1}\bX^\top
(\bI -\tG)
(\bepsilon + \bB \vec1(\bE))  \\
&\stackrel{def}{=}&
\bF_1\bepsilon + \bF_2 \vec1(\bE).
\ee

Hence, the variance-covariance matrix of $\dbetaxts$ is
\be
Cov\left(\dbetaxts\right) 
&\approx & \sigma_y^2 \bF_1 \bF_1^\top + \sigma_a^2 \bF_2 \bF_2^\top
\ee
where we will derive the explicit form of $\bF_1 \bF_1^\top$ and $\bF_2 \bF_2^\top $ in the following.

(a) $\bF_1 \bF_1^\top$.

Since
\be
(\bI -\tG)(\bI -\tG)^\top
= 
\bI - \tG - \tG^\top + \bG
\ee
and
\be
\left(
  \begin{array}{c}
    \tu^\top \\
    \tv^\top
  \end{array}
\right) \bX = \bm{0}
\mbox{ and thus }
\tG \bX = \bm{0},
\label{2s:tutvX}
\ee
consequently
\be
\bF_1 \bF_1^\top 
=(\bX^\top\bX)^{-1}
+ (\bX^\top\bX)^{-1}\bX^\top 
\bG
\bX(\bX^\top\bX)^{-1}.
\label{2s:f1f1}
\ee

(b) $\bF_2 \bF_2^\top$.

Recall \eqref{eq:two-stage-aa} where
\[
\bB \bB^\top = \frac{1}{d^2n} \left[ \beta_u^2 \opu + \beta_v^2 \opv \right].
\]
Plugging in we have,
\be
&& (\bI -\tG) \bB \bB^\top (\bI -\tG)^\top \\
&=& \frac{1}{d^2n}
\Bigg[
\beta_u^2 \opu + \beta_v^2 \opv 
- \tG \bB \bB^\top - \bB \bB^\top \tG^\top \\
&&\vspace{2in} 
+ 
\left(
\begin{array}{cc}
  \bu & \bv
\end{array}
\right) 
\bCuvi 
\left(
\begin{array}{cc}
  \beta_v^2\tu^\top\opv\tu & 0\\
  0 & \beta_u^2\tv^\top\opu\tv \\
\end{array}
\right)
\bCuvi 
\left(
  \begin{array}{c}
    \bu^\top \\
    \bv^\top
  \end{array}
\right)
\Bigg].
\ee

Together with \eqref{2s:f1f1} and using \eqref{2s:tutvX}, 
we obtain the variance-covariance matrix of $\dbetaxts$ as follows.
\be
&& Cov\left(\hbetaxts - \Bbetax\right)\\
&\approx &
\sigma_y^2
\left[
(\bX^\top\bX)^{-1}
+ (\bX^\top\bX)^{-1}\bX^\top
\left(
\begin{array}{cc}
	\bu & \bv
\end{array}
\right)
\bCuvi
\left(
 \begin{array}{c}
  \bu^\top \\
  \bv^\top
 \end{array}
\right)
\bX(\bX^\top\bX)^{-1}
\right]
\label{eq:two-stage-betax-exact-part1-proof}\\
\hspace{-1in}
&& 
+ \sigma_a^2
\frac{1}{d^2n}
(\bX^\top\bX)^{-1}\bX^\top
\Bigg[
\beta_u^2 \opu + \beta_v^2 \opv
\label{eq:two-stage-betax-exact-part2-proof}\\
\hspace{-1in}
&&
+
\left(
\begin{array}{cc}
	\bu & \bv
\end{array}
\right)
\bCuvi
\left(
\begin{array}{cc}
	\beta_v^2\tu^\top\opv\tu & 0\\
	0 & \beta_u^2\tv^\top\opu\tv \\
\end{array}
\right)
\bCuvi
\left(
 \begin{array}{c}
  \bu^\top \\
  \bv^\top
 \end{array}
\right)
\Bigg]
\bX(\bX^\top\bX)^{-1}.
\label{eq:two-stage-betax-exact-part3-proof}
\ee

\end{proof}

\end{document}